\def\etal{{\it et al.}}

\def\half{{\textstyle{1\over2}}}
\def\thalf{{\textstyle{3\over2}}}

\def\>{\rangle}
\def\<{\langle}

\def\rmb#1{{\bf #1}}

\def\beq{\begin{equation}}
\def\eeq{\end{equation}}

\def\vp{v^\prime}
\def\vvp{v\cdot v^\prime}

\def\g5{\gamma_5}
\def\ll{\Lambda} 

\def\lb{{\Lambda_b}}

\def\lcle{\Lambda_c^+ \to \ll e^+ \nu}

\def\lblc{\lb\to\ll_c}
\def\lble{\lb\to\ll_c e^- \nu}

\def\beqy{\begin{eqnarray}}
\def\eeqy{\end{eqnarray}}

\def\gursp{u^{\mu_1\dots\mu_j}(\vp)}
\def\gbursp{\bar u^{\mu_1\dots\mu_j}(\vp)}

\def\mdm{{\mu_1\dots\mu_j}}

\def\slash#1{#1 \hskip -0.5em / } 
\def\beqy{\begin{eqnarray}}
\def\eeqy{\end{eqnarray}}

\def\gursp{u^{\mu_1\dots\mu_j}(\vp)}
\def\gbursp{\bar u^{\mu_1\dots\mu_j}(\vp)}

\def\mdm{{\mu_1\dots\mu_j}}

\def\slash#1{#1 \hskip -0.5em / } 
\def\bll{\beta_{\lambda\lambda'}}
\def\bl{\beta_\lambda}
\def\blp{\beta_{\lambda'}}
\def\all{\alpha_{\lambda\lambda'}}
\def\al{\alpha_\lambda}
\def\alp{\alpha_{\lambda'}}
\def\ms{m_\sigma}
\def\ob{\Omega_b}
\def\omc{\Omega_c}
\def\omcs{\Omega_c^{(*)}}
\def\omq{\Omega_q}
\def\omqq{\Omega_Q}
\def\br{\beta_\rho}
\def\ar{\alpha_\rho}

\documentclass[aps,showpacs,floatfix]{revtex4}
\usepackage{graphicx,epsfig}
\newlength{\dinwidth}
\newlength{\dinmargin}
\begin{document}
\thispagestyle{empty}
\title{Semileptonic Decays of Heavy Omega Baryons in a Quark Model}
\author{Muslema Pervin$^{1,2}$, W. Roberts$^{1,3,4}$ and S. Capstick$^1$}
\affiliation{$^1$ Department of Physics, Florida State University, Tallahassee,
 FL 32306\\
$^2$ Physics Division,~Argonne National Laboratory, Argonne, IL-60439\\
$^3$ Department of Physics, Old Dominion University, Norfolk, VA
23529, USA\\
and\\
Thomas Jefferson National Accelerator Facility,
12000 Jefferson Avenue, Newport News, VA 23606, USA\\
$^4$ Office of Nuclear Physics,
Department of Energy, 19901 Germantown Road, Germantown, MD 20874}

\begin{abstract}
The semileptonic decays of $\Omega_c$ and $\Omega_b$ are treated in
the framework of a constituent quark model developed in a previous
paper on the semileptonic decays of heavy $\Lambda$ baryons. Analytic
results for the form factors for the decays to ground states and
a number of excited states are evaluated. For $\Omega_b$ to $\Omega_c$ the 
form factors obtained are shown to satisfy the relations predicted at 
leading order in the heavy-quark effective theory at the non-recoil
point. A modified fit of nonrelativistic and semirelativistic
Hamiltonians generates configuration-mixed baryon wave functions from
the known masses and the measured $\lcle$ rate, with wave functions
expanded in both harmonic oscillator and Sturmian bases. Decay rates
of $\ob$ to pairs of ground and excited $\omc$ states related by
heavy-quark symmetry calculated using these configuration-mixed wave
functions are in the ratios expected from heavy-quark effective
theory, to a good approximation. Our predictions for the semileptonic
elastic branching fraction of $\Omega_Q$ vary minimally within the
models we use. We obtain an average value of (84$\pm$ 2\%) for the
fraction of $\Omega_c\to \Xi^{(*)}$ decays to ground states, and
91\% for the fraction of $\Omega_c\to \Omega^{(*)}$ decays to
the ground state $\Omega$.  The elastic fraction of $\ob \to \omc$
ranges from about 50\% calculated with the two harmonic-oscillator
models, to about 67\% calculated with the two Sturmian models.
\flushright{JLAB-THY-06-480}
\end{abstract}
\maketitle
\setcounter{page}{1}

\section{Introduction and Motivation}

Semileptonic decay of hadrons are of interest for two basic reasons; they are
the primary source of information for the extraction of the
Cabbibo-Kobayashi-Maskawa (CKM) matrix elements
of the Standard Model from experiment, and the study of the semileptonic decays
of baryons provides information about their structure. In this manuscript, we
present results of a calculation of the form factors and rates of the
semileptonic decays of heavy $\Omega$ baryons ($\Omega_Q$) obtained using a
constituent quark model. This work is similar to a recently published
calculation of the semileptonic decays of heavy $\ll$ baryons~\cite{MWS}
($\Lambda_Q$) (hereinafter referred to as I). Although our motivation for the
present work is similar, it is briefly recapitulated here for completeness.

The heavy quark effective theory (HQET) \cite{HQET} has been a very powerful tool for
the extraction of the CKM matrix elements from data on the
semileptonic decay of mesons, especially for the decays of heavy
mesons to heavy mesons where a number of relations that simplify this
extraction are provided. For the semileptonic decays of a heavy baryon
to another heavy baryon, HQET makes predictions that are analogous to
those made for heavy-to-heavy meson decays: the six form factors that
describe the decays to the ground-state heavy baryons are replaced by
fewer form factors called Isgur-Wise functions; the
normalization of at least one of these Isgur-Wise functions is known at
the non-recoil point; corrections to this normalization first appear
at order $1/m_Q^2$; and corrections can be systematically estimated in
a $1/m_Q$ expansion. Note that for baryons, the number of Isgur-Wise
functions needed at leading order depends on the flavor-spin structure
of the parent and daughter baryons. For the semileptonic decays
$\Omega_Q\to \Omega_q$, only two functions are required to describe the
semileptonic decay in the heavy-quark limit. In the case of a heavy baryon
decaying to a light baryon, HQET makes predictions that are not as
powerful as in the heavy-to-heavy case. For example, for the
semileptonic decays $\Omega_Q\to \Omega$, the leading-order HQET
prediction is that the number of independent form factors decreases
from six to four for a daughter $\Omega$ with spin 1/2.

While HQET has been tremendously successful and useful in treating the
semileptonic decays of heavy hadrons, it has limitations. It is a
limit of QCD that applies only to hadrons containing heavy quarks, and
for the decays of such hadrons, it only predicts the relationships
among form factors, not their kinematic dependence. In addition, the
predictions of HQET are valid only as long as the energy of the
daughter hadron is much smaller than the mass of the heavy
quark. These limitations mean that the predictions of HQET must be
augmented by information arising from other approaches to hadron
structure.

While some work has been done in modeling the form factors for the
semileptonic decays of heavy baryons, to the best of our knowledge
little has been done in treating the decays to excited baryons. In I,
 the existing models of semileptonic decay
in both meson and baryon sectors have been discussed. We have
also outlined the procedure used to obtain the form factors and decay
rates for semileptonic decay of $\Lambda_Q$ baryons using a
non-relativistic quark model. Here we will focus only on the new
aspects relevant to the semileptonic decay of $\Omega_Q$.

Very little theoretical or experimental work has been published to date on the semileptonic
decay of $\Omega_Q$ baryons. Boyd and Brahm~\cite{omegab1} have used HQET to show that the
fourteen form factors that describe the decays $\ob \to \Omega_c e\bar\nu$ and $\ob \to
\Omega_c^*(J^P=3/2^+) e\bar\nu$ can be parametrized in terms of two nonperturbative functions at
leading order, and in terms of five additional nonperturbative functions and one dimensional
constant at order $1/m_c$. A $1/m_c$ expansion of the form factors for $\ob \to \omcs$ has also
been carried out by Sutherland~\cite{omegab2}, who treated the effect of the $\Omega_Q^* -
\Omega_Q$ mass splitting on the form factors evaluated at first order in $1/m_c$. A Bjorken sum
rule for the semileptonic decays of $\ob$ to the ground state and low-lying negative-parity
excited-state charmed baryons has been derived by Xu~\cite{omegab3}, again in the heavy quark
limit. To the best of our knowledge, there are no calculations performed outside of the HQET
framework, which motivates the present study.

There has recently been some progress in experiments dealing with
these decays. The CLEO-c collaboration~\cite{cleooc} has published
evidence for the observation of the decay $\Omega_c^0 \to \Omega^-
e^+\nu$, and have measured the product of the branching fraction and
cross section to be
$B(\Omega_c^0 \to \Omega^- e^+\nu)\cdot\sigma(e^+e^- \to
\Omega_cX)= 42.2\pm14.1\pm11.9$ fb.
The ARGUS~\cite{argus} and BELLE~\cite{belle} collaborations have also
seen evidence for the $\Omega_c^0 \to \Omega^- e^+\nu$ decay, but no
quantitative value for the branching fraction has yet been published.

The procedure we follow here to calculate the form factors and rates
for $\Omega_Q$ semileptonic decay is similar to that used for
$\Lambda_Q$ decays in I. In a quark model context the flavor part of
the $\Lambda_Q$ wave function is anti-symmetric under exchange of
quarks 1 and 2, which comprise the light diquark system. This requires
the spin-momentum part of the $\Lambda_Q$ wave function to be
anti-symmetric under exchange of the two light quarks to maintain the
appropriate Pauli exchange symmetry. On the other hand, $\Omega_Q$
baryons are described by a symmetric flavor wave function for the
light diquark system. As a result, the spin-momentum part of the wave
function has to be symmetric. This basic difference between the wave
functions of $\Lambda_Q$ and $\Omega$ makes calculations of the form
factors and the corresponding decay rates for their semileptonic
decays significantly different.

This manuscript is organized as follows: in Section II we discuss the
hadronic matrix elements and decay rates. Section III presents a brief
outline of heavy quark effective theory as it relates to the
$\Omega_Q$ decays that we discuss. In Section IV we describe the model
we use to obtain the form factors, including some description of the
Hamiltonian used to generate baryon wave functions. Our analytic
results are discussed and compared to HQET results in Section V, our
numerical results are given in Section VI, and Section VII presents
our conclusions and an outlook.  A number of details of the
calculation, including the explicit expressions for the form factors,
are shown in three Appendices.

\section{Matrix Elements and Decay Rates}

The transition matrix element for the semileptonic decay  $\omqq
\rightarrow \omq\ell\nu_\ell$ is written
\beq
T=\frac{G_F}{\sqrt{2}} V_{Qq} \overline{u_\ell} \gamma^\mu (1 - \gamma_5) 
u_{\nu_\ell}\<\omq(p', s')| J_{\mu}| \omqq(p, s)\>,
\eeq
where $G_F/\sqrt2=g^2/(8M_W^2)$ is the Fermi coupling constant, $M_W$ is the 
intermediate vector boson mass, $V_{Qq}$ is the CKM matrix element,
$\overline{u_{\ell}} \gamma^{\mu}(1 - \gamma_{5}) u_{\nu_\ell}$ is the lepton 
current, and $J_\mu=\overline{q}\gamma_\mu(1-\gamma_5)Q$ is the left handed current
between quarks $Q$ and $q$. The hadronic matrix element of $J_\mu$ is described in
terms of a number of form factors.   

For transitions between ground state 
$(J^P=1/2^+)$ baryons, the hadronic matrix elements of the vector 
($V_\mu\equiv\overline{q}\gamma_\mu Q$) and axial 
($A_\mu\equiv\overline{q}\gamma_\mu\gamma_5Q$) currents are
\begin{eqnarray}
\langle B_{q}(p', s')| V_{\mu}| B_{Q}(p, s)\rangle =
\overline{u}(p', s')\left(F_1(q^2) \gamma_{\mu} + F_2(q^2)
\frac{ p_\mu}{m_{B_Q}} +  F_3(q^2)\frac{ p'_\mu}{m_{B_q}}
\right) u(p,s),\label{vectorme}\\ 
\langle B_{q}(p', s')| A_{\mu}| B_{Q}(p, s)\rangle = 
\overline{u}(p', s')\left(G_1(q^2) \gamma_{\mu} + G_2(q^2)
\frac{ p_\mu}{m_{B_Q}} +  G_3(q^2)\frac{ p'_\mu}{m_{B_q}}\right)
\gamma_{5} u(p,s),\label{axialvme}
\end{eqnarray}
where the $F_i$ and $G_i$ are form factors which depend on the square  of the
momentum transfer $q=p-p^\prime$ between the initial and the final  baryons.
Similar expressions for transitions between $J^P=1/2^+$ ground states and final
state baryons with other spins and parities are given in I. If
$J\geq3/2$  these expressions involve a fourth pair of form factors $F_4$ and
$G_4$.

Expressions for the differential decay rates for semileptonic decays both
including and ignoring the mass of the final leptons are given in I. These expressions 
can be integrated to yield
the decay rates reported later in the present paper.

\section{Heavy Quark Effective Theory} 

In most applications of HQET, the aim has been to constrain the hadronic
uncertainties in the extraction of CKM matrix elements such as $V_{ub}$ and
$V_{cb}$. In this section, we take a different tack; we examine the predictions
of HQET for decays of a heavy $\Omega$ into a number of the allowed excited
heavy daughter baryons, with the aim
of comparing these predictions with the form factors that we obtain in our
model. 

\subsection{Structure of States and Parity Considerations}

In a heavy excited baryon, the light quark system has some total angular
momentum $j$, so that the total angular momentum of the baryon can be 
$J=j\pm 1/2$. These two states are degenerate because of the heavy quark spin
symmetry. It is useful to show explicitly the representation we use for these
two degenerate baryons. In the notation of Falk~\cite{Falk}, we write
$u^{\mu_1\dots\mu_j}_{j+1/2}(\vp)=\gursp-u^{\mu_1\dots\mu_j}_{j-1/2}(\vp)$,
with 
\beq \gursp=A^{\mu_1\dots\mu_j}(\vp)u_Q(\vp). 
\eeq 
Here, $u_Q(v)$ is the spinor of the heavy quark, with $v$ being the
four-velocity and $A^{\mu_1\dots\mu_j}(\vp)$
is a tensor that describes the spin-$j$ light quark system. This tensor is
symmetric in all of its Lorentz indices, meaning that $\gursp$ is also
symmetric in all its Lorentz indices. Both
$u^{\mu_1\dots\mu_j}_{j\pm 1/2}(\vp)$ satisfy the conditions 
\beqy 
\label{vmugmu}
\slash v^\prime\gursp&=&\gursp,\nonumber\\ 
\vp_{\mu_i}u^{\mu_1\dots\mu_i\dots\mu_j}&=&0,\,\,
g_{\mu_k\mu_l}{u}^{\mu_1\dots\mu_j}(v)=0, 
\eeqy 
where $\mu_k$ and $\mu_l$ indicate any pair of the indices $\mu_1\dots\mu_j$.
The state with $J=j+1/2$ also satisfies 
\beq \gamma_{\mu_i}u^{\mu_1\dots\mu_i\dots\mu_j}_{j+1/2}=0. 
\eeq 
Further details of the structure and properties of these tensors are given in
Falk's article~\cite{Falk}.

At this point, it is useful to discuss the parity of the states, which
is determined by the parity of the light component. A spin-$j$ light
quark component with parity $(-1)^j$ is said to have `natural' parity,
unnatural parity otherwise. The natural-parity light quark systems
therefore have $j^P=(2n)^+$ or $j^P=(2n+1)^-$, with $n$ a positive
integer or zero. The natural-parity light quark systems are
represented by tensors, while those with unnatural parity are
represented by pseudo-tensors. Since the parity of the baryon is that
of the light quark system, we may refer to the baryons as being
tensors or pseudo-tensors, with the understanding that this really
refers to the light-quark component of the baryon. It is thus
convenient to divide the decays we discuss into two classes, those in
which the daughter baryons are tensors, and those in which they are
pseudo-tensors.

\subsection{Heavy to Heavy $\Omega$ Transitions}
\label{hqetpredictions}

First, we note that the ground state of the $\Omega_Q$ has a symmetric
flavor wave function for the light diquark, and so has a spin wave
function that is also symmetric, corresponding to a total spin equal
to one in the light quark component of the wave function. This state
is therefore the spin-1/2 member of the lowest-lying
$(1/2^+,\,\,3/2^+)$ multiplet. The Falk representation of this state
is
\beq\label{falka}
\Omega_\nu(v)=\frac{1}{\sqrt{3}}\left(\gamma_\nu+v_\nu\right)\gamma_5 u(v),
\eeq
where $u(v)$ is a Dirac spinor. This state is a pseudotensor, and we
begin with a discussion of decays to other pseudotensor states.  We
are interested in the matrix element
\beq
{\cal A} =<\Omega_c^{(*)}(\vp,j)|\bar{c}\Gamma b|\Omega_b(v)>,
\eeq
where $c$ and $b$ are the heavy quark fields, and $\Gamma$ is an
arbitrary combination of Dirac matrices. With the use of HQET, we may
write this matrix element as
\beq \label{hqetme}
<\Omega_c^{(*)}(\vp,j)|\bar{c}\Gamma b|\Omega_b(v)>=\gbursp\Gamma \Omega^\nu 
M_{\mdm\nu},
\eeq
to leading order. Here, $M_{\mdm\nu}$ is the most general tensor that
we can construct, $\Omega^\nu$ is the Falk representation of
Eq.~(\ref{falka}), and $\gbursp$ is the analogous representation of
the daughter baryon. $M_{\mdm\nu}$ may not contain any factors of
$\vp_{\mu_i}$ or $g_{\mu_i\mu_j}$, and therefore takes the form
\beq\label{omegafirst}
M_{\mdm\nu}=\left(\eta_1^{(j)}g_{\mu_1\nu}+\eta_2^{(j)}v_{\mu_1}\vp_{\nu}\right)
v_{\mu_2}\dots v_{\mu_j}. 
\eeq 
Thus, two independent form factors, $\eta_{1,2}^{(j)}(\vvp)$ are needed to 
this order, regardless of the spin of the final baryon. 

Applying these results to the specific case of $j^P=0^-$, we find,
for $J^P=1/2^-$
\begin{equation} \label{hmsinglet}
F_1=\frac{w-1}{\sqrt{3}}\eta^{(0)}_2(w),\,\,F_2=G_2=0,\,\,
F_3=-G_3=\frac{2}{\sqrt{3}}\eta^{(0)}_2(w),\,\,
G_1=\frac{w+1}{\sqrt{3}}\eta^{(0)}_2(w), 
\end{equation}
where $w=v\cdot\vp$. In this case, the daughter baryon is a singlet, and has no Lorentz indices.
This means that the term in $g_{\mu\nu}$ is not present, and only the term in $\eta_2$
contributes to the matrix element.

When $j^P=1^+$, we find, for $J^P=1/2^+,$
\begin{eqnarray} \label{hpthpa}
F_1&=&G_1=\frac{1}{3}\left[w\eta_1^{(1)}+\left(w^2-1\right)\eta_2^{(1)}\right],\,\,
F_2=F_3=\frac{2}{3}\left[-\eta_1^{(1)}+\left(1-w\right)\eta_2^{(1)}\right],\nonumber\\
G_2&=&-G_3=-\frac{2}{3}\left[\eta_1^{(1)}+\left(1+w\right)\eta_2^{(1)}\right],
\end{eqnarray}
and for $J^P=3/2^+$,
\begin{eqnarray}\label{hpthpb}
F_1&=&-\frac{1}{\sqrt{3}}\left[\eta_1^{(1)}+\left(w-1\right)\eta_2^{(1)}\right],\,\,F_2=G_2=0,\,\,
F_3=-G_3=-\frac{2}{\sqrt{3}}\eta_2^{(1)},\,\,F_4=-G_4=-\frac{2}{\sqrt{3}}\eta_1^{(1)},\nonumber\\
G_1&=&-\frac{1}{\sqrt{3}}\left[\eta_1^{(1)}+\left(w+1\right)\eta_2^{(1)}\right].
\end{eqnarray}
In these two sets of equations $\eta^{(1)}_{1,2}$ are two universal
functions of the Isgur-Wise type. We note that there exist several
multiplets that have the same quantum numbers as this ground state
multiplet. The same is true in the case of $\Lambda_Q$ baryons, and
these excited $\Lambda_Q$s were identified as radial excitations of
the ground state. In the case of the $\Omega_Q$, such baryons are
indeed excitations of the ground state, but they are not necessarily
radial excitations. Some of these excitations are orbital
excitations. However, independent of whether the daughter baryon
belongs to the ground state $(1/2^+,\,\,3/2^+)$ or one of the excited
multiplets, the expressions above for the form factors are valid. The
explicit forms of the $\eta_i^{(1)}$ will depend on the details of the
structure of the daughter baryon. For the ground state we know that at
the non-recoil point, $\eta_1^{(1)}(v.v'=1)=-1$, while the
normalization of $\eta_2^{(1)}$ is not known.  The negative sign of
the normalization of $\eta_1^{(1)}$ arises because we have chosen a
positive sign for the $g_{\mu_1\nu}$ term in Eq.~(\ref{omegafirst}).

For $j^P=2^-$, we find for $J^P=3/2^-$, 
\begin{eqnarray}\label{thfhma}
F_1&=&\frac{1}{\sqrt{30}}\left[(2w-1)\eta_1^{(2)}+2\left(w^2-1\right)\eta_2^{(2)}\right],\,\,
F_2=-2\sqrt{\frac{2}{15}}\left[\eta_1^{(2)}+\left(w-1\right)\eta_2^{(2)}\right],\nonumber\\
F_3&=&-\sqrt{\frac{2}{15}}\left[\eta_1^{(2)}+2\left(w-1\right)\eta_2^{(2)}\right],\,\,
F_4=-\sqrt{\frac{2}{15}}(w-1)\eta_1^{(2)},\nonumber\\
G_1&=&\frac{1}{\sqrt{30}}\left[(2w+1)\eta_1^{(2)}+2\left(w^2-1\right)\eta_2^{(2)}\right],\,\,
G_2=-2\sqrt{\frac{2}{15}}\left[\eta_1^{(2)}+\left(w+1\right)\eta_2^{(2)}\right],\nonumber\\
G_3&=&\sqrt{\frac{2}{15}}\left[\eta_1^{(2)}+2\left(w+1\right)\eta_2^{(2)}\right],\,\,
G_4=\sqrt{\frac{2}{15}}(w+1)\eta_1^{(2)},
\end{eqnarray}
and for $J^P=5/2^-$,
\begin{eqnarray}\label{thfhmb}
F_1&=&-\frac{1}{\sqrt{3}}\left[\eta_1^{(2)}+\left(w-1\right)\eta_2^{(2)}\right],\,\,F_2=G_2=0,\,\,
F_3=-G_3=-\frac{2}{\sqrt{3}}\eta_2^{(2)},\,\,F_4=-G4=-\frac{2}{\sqrt{3}}\eta_1^{(2)},\nonumber\\
G_1&=&-\frac{1}{\sqrt{3}}\left[\eta_1^{(2)}+\left(w+1\right)\eta_2^{(2)}\right].
\end{eqnarray}
As with the previous example, the functions $\eta_{1,2}^{(2)}$ are
Isgur-Wise form factors common to both decays.

For $j^P=3^+$, we find for $J^P=5/2^+$,
\begin{eqnarray}\label{fhshpa}
F_1&=&\frac{1}{3\sqrt{7}}\left[(3w-2)\eta_1^{(3)}+3\left(w^2-1\right)\eta_2^{(3)}\right],\,\,
F_2=-2\sqrt{\frac{1}{7}}\left[\eta_1^{(3)}+\left(w-1\right)\eta_2^{(3)}\right],\nonumber\\
F_3&=&-\frac{2}{3\sqrt{7}}\left[\eta_1^{(3)}+3\left(w-1\right)\eta_2^{(3)}\right],\,\,
F_4=-\frac{4}{3\sqrt{7}}(w-1)\eta_1^{(3)},\nonumber\\
G_1&=&\frac{1}{3\sqrt{7}}\left[(3w+2)\eta_1^{(3)}+3\left(w^2-1\right)\eta_2^{(3)}\right],\,\,
G_2=-2\sqrt{\frac{1}{7}}\left[\eta_1^{(3)}+\left(w+1\right)\eta_2^{(3)}\right],\nonumber\\
G_3&=&\frac{2}{3\sqrt{7}}\left[\eta_1^{(3)}+3\left(w+1\right)\eta_2^{(3)}\right],\,\,
G_4=\frac{4}{3\sqrt{7}}(w+1)\eta_1^{(3)},
\end{eqnarray}
and for $J^P=7/2^+$,
\begin{eqnarray}\label{fhshpb}
F_1&=&-\frac{1}{\sqrt{3}}\left[\eta_1^{(3)}+\left(w-1\right)\eta_2^{(3)}\right],\,\,F_2=G_2=0,\,\,
F_3=-G_3=-\frac{2}{\sqrt{3}}\eta_2^{(3)},\,\,F_4=-G_4=-\frac{2}{\sqrt{3}}\eta_1^{(3)},\nonumber\\
G_1&=&-\frac{1}{\sqrt{3}}\left[\eta_1^{(3)}+\left(w+1\right)\eta_2^{(3)}\right]
\end{eqnarray}
The functions $\eta_{1,2}^{(3)}$ are Isgur-Wise form factors common to
both decays. The normalizations of $\eta_{1,2}^{(2,3)}$ are not known.

For the tensor decays, the matrix element again takes the form shown
in Eq.~(\ref{hqetme}), but $M_{\mdm\nu}$ must now be a
pseudotensor. The only form that we can write is
\beq
M_{\mdm\nu}=\tau^{(j)}(w) v_{\mu_2}\dots v_{\mu_j}\varepsilon_{\nu\mu_1\rho\lambda}
v^\rho v^{\prime\lambda}.
\eeq
When applied to the $1/2^+$ singlet daughter baryon, there is no way
to create this pseudotensor, so such amplitudes vanish at leading
order. For the other spin states, after some manipulation, we can
express the form factors in terms of the set of Isgur-Wise functions
$\tau^{(j)}(w)$.

For $j^P=1^-$, we find for the $1/2^-$ state
\beq\label{hmthma}
F_1=G_1=0,\,\,\,\, F_2=F_3=-G_2=G_3=-\frac{2}{3}\tau^{(1)},
\eeq
while for $3/2^-$ state, the form factors are
\beq\label{hmthmb}
F_2=G_2=0,\,G_3=-F_3=2F_1=-2G_1=-\frac{2}{\sqrt{3}}\tau^{(1)},\,
F_4=-\frac{2}{\sqrt{3}}(w-1)\tau^{(1)},\,
G_4=\frac{2}{\sqrt{3}}(w+1)\tau^{(1)}.
\eeq

For $j^P=2^+$, starting with $3/2^+$, the form factors are
\beqy\label{thfhpa}
F_1&=&\frac{1}{\sqrt{30}}(1-w)\tau^{(2)},\,\, F_2=-G_2=-2\sqrt{\frac{2}{15}}\tau^{(2)},
\,\, F_3=\sqrt{\frac{2}{15}}(w-2)\tau^{(2)},\,\, 
F_4=-G_4=\sqrt{\frac{2}{15}}(1-w^2)\tau^{(2)},\nonumber\\
G_1&=&\frac{1}{\sqrt{30}}(1+w)\tau^{(2)},\,\, G_3=-\sqrt{\frac{2}{15}}(w+2)\tau^{(2)}.
\eeqy
For the $5/2^+$ state, the form factors are
\beq\label{thfhpb}
F_2=G_2=0,\,G_3=-F_3=2F_1=-2G_1=-\frac{2}{\sqrt{3}}\tau^{(2)},\,
F_4=-\frac{2}{\sqrt{3}}(w-1)\tau^{(2)},\,
G_4=\frac{2}{\sqrt{3}}(w+1)\tau^{(2)}.
\eeq
The normalizations of none of the $\tau^{(i)}$ are known.

We do not present the predictions for the decays of $\Omega_Q$ to light $\Omega$
states, as the HQET predictions are not as useful as they are in the case of
heavy to light $\ll_Q$ decays. For instance, the decays of the ground state
$\Omega_Q$ to an $\Omega$ with $J^P=1/2^+$ are described in terms of four form
factors in HQET, instead of six in general. While this small simplification is
no doubt useful we do not pursue it here.

\section{The Model} 

\subsection{Wave Function Components}
\label{wfcomponents}

In our model, a baryon state has the form
\begin{eqnarray} |A_Q({\bf p},s)\rangle &=&
3^{3/4} \int d^3p_\rho d^3p_\lambda C^A \Psi_{A_Q}^S |q_1({\bf
p}_1,s_1)q_2({\bf p}_2,s_2)q_3({\bf p}_3,s_3)\rangle,\nonumber 
\end{eqnarray}  
where ${\bf p}_\rho= \frac{1}{\sqrt2}({\bf p}_1 - {\bf p}_2)$ and
${\bf p}_\lambda = \frac{1}{\sqrt6}({\bf p}_1 + {\bf p}_2 - 2{\bf
p}_3)$ are the Jacobi momenta, $C^A$ is the totally antisymmetric
color wave function, and $\Psi_{A_Q}^S = \phi_{A_Q} \psi_{A_Q}
\chi_{A_Q}$ 
is a symmetric combination of flavor, momentum and spin wave
functions. The flavor wave functions of $\Omega_Q$ and $\Xi$ are
\begin{eqnarray}
\phi_{\Omega_Q} =  ssQ,\,\,\,\, \phi_{\Xi^0} = ssu,\,\,\,\, \phi_{\Xi^-} = ssd,\nonumber
\end{eqnarray}
which are symmetric in quarks $1$ and $2$. The momentum-spin parts of
the wave functions must therefore be symmetric in quarks $1$ and $2$ to
keep the overall symmetry. The symmetric spin wave function
$\chi_{3/2}^S$, and the mixed symmetric spin wave functions
$\chi_{1/2}^\rho$, $\chi_{1/2}^\lambda$ are the usual eigenstates of
total spin made of three spin-$1/2$ quarks.

The momentum wave function for total $L=\ell_\rho+\ell_\lambda$ is
constructed from a Clebsch-Gordan sum of the wave functions of the two
Jacobi momenta ${\bf p}_\rho$ and ${\bf p}_\lambda$, and takes the
form
\begin{eqnarray}
\psi_{LMn_{\rho}\ell_{\rho}n_{\lambda}\ell_\lambda}({\bf p}_\rho, {\bf 
p}_\lambda) &=& 
\sum_m\langle LM|\ell_{\rho}m,\ell_\lambda M-m\rangle\psi_{n_\rho \ell_\rho m}
({\bf p}_\rho) \psi_{n_\lambda \ell_\lambda M-m}({\bf p}_\lambda).\nonumber
\end{eqnarray}
The momentum and spin wave functions are then coupled to
give symmetric wave functions corresponding to total spin $J$ and parity
$(-1)^{(l_\rho+l_\lambda)}$,
\begin{eqnarray}
\Psi_{JM}&=& \sum_{M_L}\< JM|LM_L, SM-M_L\>\psi_{LM_Ln_{\rho}
\ell_{\rho}n_{\lambda}\ell_\lambda}({\bf p}_\rho, {\bf p}_\lambda)\chi_{S}(M-M_L)\nonumber\\
&\equiv&\left[\psi_{LM_Ln_\rho\ell_\rho n_\lambda\ell_\lambda}({\bf p}_\rho, {\bf 
p}_\lambda)\chi_S(M-M_L)\right]_{J,M}.
\end{eqnarray}
The full wave function for a state $A$ is built from a linear 
superposition of such
components as 
\begin{equation}
\Psi_{A,J^PM}=\phi_A\sum_i \eta_i^A \Psi_{JM}^i.
\end{equation}
Here $\phi_A$ is the flavor wave function of the state $A$, and the
$\eta_i^A$ are coefficients that are determined by diagonalizing a
Hamiltonian in the basis of the $\Psi_{JM}^i$. For this calculation,
we limit the expansion in the last equation to components that satisfy
$N\le 2$, where $N=2(n_\rho+n_\lambda)+\ell_\rho+\ell_\lambda.$
Consistent with this is the fact that the states we discuss all
correspond to $N\le 2$.

The wave functions for $\Omega_Q$ with $J^P=1/2^+$ have the form
\begin{eqnarray}
\label{halfpwf}
\Psi^{\Omega_Q}_{1/2^+M}&=&\phi_{\Omega_Q}\left(\vphantom{\sum_i}
\left[\eta_1^{\Omega_Q}\psi_{000000}({\bf p}_\rho, {\bf 
p}_\lambda)
+\eta_2^{\Omega_Q}\psi_{001000}({\bf p}_\rho, {\bf 
p}_\lambda)
+\eta_3^{\Omega_Q}\psi_{000010}({\bf p}_\rho, 
{\bf p}_\lambda)\right]\chi_{1/2}^\lambda(M)\right.\nonumber \\
&+&\eta_4^{\Omega_Q}\psi_{000101}({\bf p}_\rho, 
{\bf p}_\lambda)\chi_{1/2}^\rho(M)
+\eta_5^{\Omega_Q}\left[\psi_{1M_L0101}({\bf p}_\rho, {\bf 
p}_\lambda)\chi_{1/2}^\rho(M-M_L)\right]_{1/2, M}\nonumber  \\
&+&\eta_6^{\Omega_Q}\left[\psi_{2M_L0200}({\bf p}_\rho, {\bf 
p}_\lambda)\chi_{3/2}^S(M-M_L)\right]_{1/2, M}\nonumber  \\
&+&\left.\eta_7^{\Omega_Q}\left[\psi_{2M_L0002}({\bf p}_\rho, {\bf p}_\lambda)
\chi_{3/2}^S(M-M_L)\right]_{1/2, M}\right).
\end{eqnarray}
The complete expressions for $\Omega_Q$ wave functions of different
spins and parities are given in Appendix~\ref{appendixA}.  A
simplified version of the model would truncate the expansion of the
wave functions, giving
\beq
\Psi_{\Omega_Q,1/2^+M} = \phi_\Omega \psi_{000000}({\bf p}_\rho, 
{\bf p}_\lambda)\chi_{1/2}^\lambda(M)
\eeq
for the ground state. There are a number of daughter baryons that have
an overlap with the ground state $\Omega_Q$ (in the spectator
approximation that we use), even when we limit the discussion to
states with $N\le 2$. There are three states with $J^P=1/2^+$, two
with $J^P=1/2^-$, two with $J^P=3/2^-$, four with $J^P=3/2^+$, one
with $J^P=5/2^-$, two with $J^P=5/2^+$, and one with $J^P=7/2^+$, all
of which occur in the $N\le 2$ bands. The single-component
representations of these states are
\begin{eqnarray}
\label{wavefunctionomega}
\Psi_{\Omega_Q,1/2^+M} &=& \phi_\Omega \psi_{000000}({\bf p}_\rho, 
{\bf p}_\lambda)\chi_{1/2}^\lambda(M),\nonumber\\
\Psi_{\Omega_Q,1/2^+_1M} &=& \phi_\Omega \psi_{000010}({\bf p}_\rho, 
{\bf p}_\lambda)\chi_{1/2}^\lambda(M),\nonumber\\
\Psi_{\Omega_Q,1/2^+_2M} &=& \phi_\Omega [\psi_{2M_L0002}({\bf p}_\rho, 
{\bf p}_\lambda)\chi_{3/2}^S(M-M_L)]_{1/2,M},\nonumber\\
\Psi_{\Omega_Q,1/2^-M} &=& \phi_\Omega [\psi_{1M_L0001}({\bf p}_\rho, 
{\bf p}_\lambda)\chi_{1/2}^\lambda(M-M_L)]_{1/2,M},\nonumber\\
\Psi_{\Omega_Q,1/2^-_1M} &=& \phi_\Omega [\psi_{1M_L0001}({\bf p}_\rho, 
{\bf p}_\lambda)\chi_{3/2}^S(M-M_L)]_{1/2,M},\nonumber\\
\Psi_{\Omega_Q,3/2^-M} &=& \phi_\Omega [\psi_{1M_L0001}({\bf p}_\rho, 
{\bf p}_\lambda)\chi_{1/2}^\lambda(M-M_L)]_{3/2,M},\nonumber\\
\Psi_{\Omega_Q,3/2^-_1M} &=& \phi_\Omega [\psi_{1M_L0001}({\bf p}_\rho, 
{\bf p}_\lambda)\chi_{3/2}^S(M-M_L)]_{3/2,M},\nonumber\\
\Psi_{\Omega_Q,5/2^-M} &=& \phi_\Omega [\psi_{1M_L0001}({\bf p}_\rho, 
{\bf p}_\lambda)\chi_{3/2}^S(M-M_L)]_{5/2,M},\nonumber\\
\Psi_{\Omega_Q,3/2^+M} &=& \phi_\Omega \psi_{000000}({\bf p}_\rho, 
{\bf p}_\lambda)\chi_{3/2}^S(M),\nonumber\\
\Psi_{\Omega_Q,3/2^+_1M} &=& \phi_\Omega \psi_{000010}({\bf p}_\rho, 
{\bf p}_\lambda)\chi_{3/2}^S(M),\nonumber\\
\Psi_{\Omega_Q,3/2^+_2M} &=& \phi_\Omega [\psi_{2M_L0002}({\bf p}_\rho, 
{\bf p}_\lambda)\chi_{3/2}^S(M-M_L)]_{3/2,M},\nonumber\\
\Psi_{\Omega_Q,3/2^+_3M} &=& \phi_\Omega [\psi_{2M_L0002}({\bf p}_\rho, 
{\bf p}_\lambda)\chi_{1/2}^\lambda(M-M_L)]_{3/2,M},\nonumber\\
\Psi_{\Omega_Q,5/2^+M} &=& \phi_\Omega [\psi_{2M_L0002}({\bf p}_\rho, 
{\bf p}_\lambda)\chi_{1/2}^\lambda(M-M_L)]_{5/2,M},\nonumber\\
\Psi_{\Omega_Q,5/2^+_1M} &=& \phi_\Omega [\psi_{2M_L0002}({\bf p}_\rho, 
{\bf p}_\lambda)\chi_{3/2}^S(M-M_L)]_{5/2,M},\nonumber\\
\Psi_{\Omega_Q,7/2^+M} &=& \phi_\Omega [\psi_{2M_L0002}({\bf p}_\rho, 
{\bf p}_\lambda)\chi_{3/2}^S(M-M_L)]_{7/2,M}.
\end{eqnarray}

A common choice for constructing baryon wave function is the harmonic
oscillator basis. One advantage of using this basis is that it
facilitates calculation of the required matrix elements. However, it
leads to form factors that fall off too rapidly at large values of
momentum transfer. We therefore also use the so-called Sturmian
basis~\cite{KP}. In this basis, form factors have multipole dependence
on $q^2$, which is what is expected experimentally.

The explicit wave functions in momentum space are
\begin{eqnarray}\label{hoa}
\psi^{\rm h.o.}_{nLm} ({\bf p})&=& \left[\frac{2\,n!}{\left(n + L +\half\right)!}
\right]^{\half} (i)^L(-1)^n \frac{1}{\alpha^{L+\thalf}} 
e^{-\frac{p^2}{(2\alpha^2)}}
L_n^{L+\half}(p^2/\alpha^2){\cal Y}_{Lm}({\bf p})
\end{eqnarray}
in the harmonic oscillator basis, and 
\begin{eqnarray} \label{sta}
\psi^{\rm St}_{nLm} ({\bf p}) &=& {2\left[ n! (n + 2L + 2)!\right]^{\half}
\over \left(n + L + \half\right)!}  (i)^L\frac{1}{\beta^{L+\thalf}}
\frac{1}{\left(\frac{p^2}{\beta^2} + 1\right)^{L+2}}
P_n^{\left(L +\thalf, L + \half\right)} \left(\frac{p^2 
-\beta^2}{p^2 
+\beta^2}\right) {\cal Y}_{Lm}({\bf p})\nonumber\\
\end{eqnarray}
in the Sturmian basis. The $L_n^\nu(x)$ are generalized Laguerre
polynomials and the $P_n^{(\mu,\nu)}(y)$ are Jacobi polynomials, with
$p=\left|{\bf p}\right|$.

\subsection{Hamiltonian}
\label{hamiltonian}

The phenomenological Hamiltonian we use has the form
\beq
\label{hamil}
H=\sum_{i=1}^3 K_i + C_{qqq}+ \sum_{i<j=1}^3 
\left({br_{ij}\over 2}-{2\alpha_{\rm Coul}\over3r_{ij}} +{2\alpha_{\rm hyp}\over 3 m_i 
m_j}{8\pi\over 3} \rmb{S}_i\cdot\rmb{S}_j\delta^3(\rmb{r}_{ij})\right),
\eeq
with $r_{ij}=\vert\rmb{r}_i-\rmb{r}_j \vert$. The spin independent
confining potential consists of a linear and a Coulomb component, and
the spin-dependent part of the potential takes the form of a contact
hyperfine interaction. Spin-orbit and tensor interactions are
neglected. We note here that $\alpha_{\rm Coul}$, $\alpha_{\rm hyp}$,
$b$, $C_{qqq}$, and $m_i$ are not fundamental, but are
phenomenological parameters obtained from a fit to the spectrum of
baryon states. In the sum $\sum_{i=1}^3K_i$ of the kinetic energies of
the quarks, each term has either the usual non-relativistic form given by
\beq
K_i= \left( m_i+\frac{p_i^2}{2m_i} \right),
\eeq
or a semi-relativistic form given by
\beq
K_i= \sqrt{p_i^2+m_i^2}.
\eeq

\subsection{Obtaining the Form Factors}
\label{obtaining}

\subsubsection{$\Omega_Q \to \Omega_q$}

Here, we illustrate the procedure we follow to obtain the form
factors, using the decay of the $B_Q$ to the ground state $B_q$ as an
example. We note here that $B_Q$ represents $\Omega_Q$ and $B_q$
refers to any of the $\Omega_q$ or $\Xi$ in their ground state. We
show the procedure only for the vector current matrix element from
Eq.~(\ref{vectorme}), with the assumption that the parent $B_Q$ is at
rest and the daughter $B_q$ has three momentum ${\bf p}$. The
left-hand side of Eq.~(\ref{vectorme}) is evaluated using the quark
model, after the operator $V_\mu=\bar{q} \gamma_\mu Q$ has been
reduced to its Pauli (non-relativistic) form. Specific values for the
index $\mu$ are chosen, as well as specific values of $s$ and
$s^\prime$. By making three sets of such choices, three equations for
the $F_i$ in terms of the quark-model matrix elements of three
operators are obtained. This system of equations is then solved to
obtain the expressions for the form factors. In the specific case at
hand, choosing $s=s^\prime=+1/2$ and $\mu=0$, for instance, leads to
\begin{eqnarray}
\langle B_q({\bf p},+)|\bar{q}\gamma_0 Q|B_Q(0,+)\rangle  
&=&  \int d^3p'_\rho d^3p'_\lambda d^3p_\rho
d^3p_\lambda  C^{A*} C^A  \Psi_{B_q}^{*S}(+)\nonumber\\
&\times&   \langle q'_1q'_2q|q^\dagger \gamma_0 Q|q_1q_2Q\rangle
\Psi_{B_Q}^S(+).\nonumber\\
\end{eqnarray}
where
\beq
\langle q'_1q'_2q|q^\dagger \gamma_0 Q|q_1q_2Q\rangle =  
\langle q'_1q'_2|q_1q_2\rangle \langle q|q^\dagger \gamma_0 Q|Q\rangle.\nonumber
\eeq
The matrix element $\langle q'_1q'_2|q_1q_2\rangle$ gives
$\delta$-functions in spin, momentum and flavor in the spectator
approximation. Using the $\delta$-functions in momentum and flavor,
the integral simplifies to
\begin{eqnarray}
\left< B_q({\bf p},+)\left|{\cal O}_0\right|
B_Q({\bf 0},+)\right> &=& \int d^3p_\rho d^3p_\lambda   
\psi_{B_q}^*({\bf p}'_\rho, {\bf p}'_\lambda) {\cal A}^{++}_{B_QB_q}({\cal O}_0)
\psi_{B_Q}({\bf p}_\rho, {\bf p}_\lambda),
\label{integral}
\end{eqnarray}
with ${\bf p}'_\rho = {\bf p}_\rho$, ${\bf p}'_\lambda = {\bf
p}_\lambda - 2\sqrt{3/2}\,m_\sigma {\bf p}/m_{B_q}$, 
where $m_\sigma$ is the mass of the light quark and ${\cal O}_0=q^\dagger
\gamma_0 Q$. It is useful for us to define
\begin{eqnarray}
{\cal A}^{s^\prime s}_{B_QB_q}({\cal O}_\mu)&=&\chi_{B_q}^\dag(s^\prime)\delta_{s'_1s_1}\delta_{s'_2s_2}
\langle q({\bf p}'_3,s'_3)\left|{\cal O}_\mu\right| Q({\bf p}_3,s_3)\rangle\chi_{B_Q}(s).
\end{eqnarray}
where ${\bf p}'_3={\bf p}-\sqrt{\frac{2}{3}}{\bf p}_\lambda$ and  ${\bf
p}_3=-\sqrt{\frac{2}{3}}{\bf p}_\lambda$.  We have
\begin{equation}
{\cal A}^{++}_{B_Q B_q}({\cal O}_\mu)=\frac{1}{3}\langle q({\bf p}'_3,\uparrow)\left|
{\cal O}_\mu\right| Q({\bf p}_3,\uparrow)\rangle+\frac{2}{3}\langle q({\bf p}'_3,\downarrow)\left|
{\cal O}_\mu\right| Q({\bf p}_3,\downarrow)\rangle,
\end{equation}
where both parent and daughter baryons are in the ground state.

After the spin matrix elements are evaluated, the momentum integrals
are performed using both bases for the momentum wave functions shown
earlier. The analytic results for the form factors for the
$\Omega_Q\to \Omega_q$ decays, evaluated using the truncated basis of
Eq.~(\ref{wavefunctionomega}), are given in
Appendix~\ref{appendixC}. For decays to excited states, the
calculation of the form factors is a little more involved, but the
basic idea is as outlined here.

\subsubsection{$\Omega_c \to \Omega^{(*)}$}

The calculation of form factors for $\Omega_c \to \Omega^{(*)}$ decays
is similar to that described above, up to a question of symmetry.  The
flavor wave functions of $\Omega_c$ and $\Omega$ are $\phi_{\Omega_c}=
ssc$ and $\phi_\Omega= sss$. In a semileptonic decay process the charm
quark of the $\Omega_c$ decays into an $s$ quark, which can be any of
the three $s$ quarks of the $\Omega$. Thus, we need to evaluate
\begin{eqnarray}
\frac{1}{\sqrt{3}}\left(\vphantom{\frac{1}{\sqrt{3}}}
\langle \Omega({\bf p},+)|{\cal O}_\mu|\Omega_Q({\bf 0},+)\rangle +
\langle \Omega({\bf p},+)|\{13\}{\cal O}_\mu|\Omega_Q({\bf 0},+)\rangle 
+\langle \Omega({\bf p},+)|\{23\}{\cal O}_\mu|\Omega_Q({\bf 0},+)\rangle\right)\nonumber.
\end{eqnarray}
The factor $\frac{1}{\sqrt{3}}$ comes from the normalization. The wave
functions for the $\Omega$ states are fully symmetric under
interchange of any of the quarks, so each of the permuted matrix
elements reproduces the one without permutations. The result is that
we must calculate
\begin{eqnarray}
\sqrt{3}\langle \Omega({\bf p},+)|{\cal O}_\mu|\Omega_Q({\bf 0},+)\rangle \nonumber,
\end{eqnarray}
to obtain the form factors for $\Omega_c\to\Omega$.

The procedure described in this subsection is relatively
straightforward to implement in the harmonic oscillator basis, largely
due to the fact that the Moshinsky rotations have been treated by a
number of authors, and are also fairly simple to calculate. In
particular, the fact that the `permuted' wave function can be written
in terms of a finite set of transformed wave function components is
another feature that makes the harmonic oscillator basis attractive
for calculations like these. In the Sturmian basis, however, the
permutation of particles requires an infinite sum of transformed wave
functions. This sum could be truncated at some point in a calculation
such as this. However, at this point we do not examine decays to
daughter $\Omega$'s in the Sturmian basis.

\section{Analytic Results and Comparison with HQET}
\label{analresults}

The analytic expressions that we obtain for the form factors are shown in
Appendix~\ref{appendixC}, for both the Sturmian and harmonic oscillator bases.
The results shown there are valid when the wave function for a
particular state is written as a single component, in either expansion basis.
As mentioned earlier, one of the advantages of the Sturmian basis is that it
leads to form factors that behave like multipoles in the kinematic variable,
and this is seen in the forms that we display. 

At this point, it is instructive
to compare, as far as possible, these analytic forms with the predictions of
HQET. While HQET does not give the explicit forms of the form factors, a number
of relationships among the form factors are expected, and any model should
reproduce these relationships. In what follows, we restrict our comparison to
the predictions that are valid at the non-recoil point, as we have ignored any
kinematic dependence beyond the Gaussian or multipole factors shown in
Appendix~\ref{appendixC}. In addition, we focus  on the predictions for
heavy to heavy transitions.

The quark model states we use are constructed in the coupling scheme
\beq
|J^P,L,S>=\left|\left[\left(\ell_\rho \ell_\lambda\right)_L \left(s_{12} s_3\right)_S\right]_J\right>,
\eeq
where the notation $(ab)_c$ means angular momentum $c$ is formed by
vector addition from angular momenta $a$ and $b$. The parity $P$ is
$(-1)^{\ell_\rho+\ell_\lambda}$, the total spin of the two light
quarks in the baryon is $s_{12}$, and $s_3$ is the spin of the third
quark, taken to be the heavy quark.

The HQET states are assumed to have the coupling scheme
\beq
|J^P,j>=\left|\left\{\left[\left(\ell_\rho \ell_\lambda\right)_L s_{12}\right]_{j} s_3\right\}_J\right>,
\eeq
where $j$ is the total spin of the light component of the baryon, so that $J=j\pm
1/2$. The states of one coupling scheme are linear combinations of the states of the second.
In particular, we find
\beqy
&&\left|\left\{\left[\left(\ell_\rho \ell_\lambda\right)_L s_{12}\right]_{j} s_3\right\}_J\right>
=(-1)^{1/2+s_{12}+L+J}\sqrt{2j+1}\nonumber\\
&&\times \sum_S\sqrt{2S+1}
\left\{\begin{array}{ccc}1/2&s_{12}&S\\ L& J & j\end{array}\right\}
\left|\left[\left(\ell_\rho \ell_\lambda\right)_L \left(s_{12} s_3\right)_S\right]_J\right>,
\eeqy
where $\left\{\begin{array}{ccc}1/2&s_{12}&S\\ L& J &
j\end{array}\right\}$ is a 6-J symbol.

For the states that we consider, the explicit expressions for the HQET
states in terms of the quark model states are
\beqy \label{mixedstates}
\left|1/2^-,j=1\right>&=&\sqrt{\frac{2}{3}}
\left|1/2^-,L=1,S=1/2\right>+\frac{1}{\sqrt{3}}\left|1/2^-,L=1,S=3/2\right>,\nonumber\\
\left|1/2^-,j=0\right>&=&-\frac{1}{\sqrt{3}}
\left|1/2^-,L=1,S=1/2\right>+\sqrt{\frac{2}{3}}\left|1/2^-,L=1,S=3/2\right>,\nonumber\\
\left|3/2^-,j=2\right>&=&\sqrt{\frac{5}{6}}
\left|3/2^-,L=1,S=1/2\right>+\frac{1}{\sqrt{6}}\left|3/2^-,L=1,S=3/2\right>,\nonumber\\
\left|3/2^-,j=1\right>&=&-\frac{1}{\sqrt{6}}
\left|3/2^-,L=1,S=1/2\right>+\sqrt{\frac{5}{6}}\left|3/2^-,L=1,S=3/2\right>,\nonumber\\
\left|3/2^+,j=2\right>&=&\frac{1}{\sqrt{2}}\left(
\left|3/2^+,L=2,S=1/2\right>+\left|3/2^+,L=2,S=3/2\right>\right),\nonumber\\
\left|3/2^+,j=1\right>&=&\frac{1}{\sqrt{2}}\left(
-\left|3/2^+,L=2,S=1/2\right>+\left|3/2^+,L=2,S=3/2\right>\right),\nonumber\\
\left|5/2^+,j=3\right>&=&\frac{\sqrt{7}}{3}
\left|5/2^+,L=2,S=1/2\right>+\frac{\sqrt{2}}{3}\left|5/2^+,L=2,S=3/2\right>,\nonumber\\
\left|5/2^+,j=2\right>&=&-\frac{\sqrt{2}}{3}
\left|5/2^+,L=2,S=1/2\right>+\frac{\sqrt{7}}{3}\left|5/2^+,L=2,S=3/2\right>.
\eeqy
For all of the quark model states shown on the r.h.s of these
equations, $S=1/2$ corresponds to spin wave function of the
$\chi^\lambda$ type. The form factors that describe transitions to
these states are shown in Appendix~\ref{hqetformfactors}.

Other states not shown above are single component states in both
representations, and these are 
\beqy
\left|1/2^+,j=1\right>&=&\left|1/2^+,L=0,S=1/2\right>,\nonumber\\
\left|3/2^+,j=1\right>&=&\left|3/2^+,L=0,S=3/2\right>,\nonumber\\
\left|1/2^+_1,j=1\right>&=&\left|1/2^+_1,L=0,S=1/2\right>,\nonumber\\
\left|3/2^+_1,j=1\right>&=&\left|3/2^+_1,L=0,S=3/2\right>,\nonumber\\
\left|1/2^+_2,j=1\right>&=&\left|1/2^+_2,L=2,S=3/2\right>,\nonumber\\
\left|5/2^-,j=2\right>&=&\left|5/2^-,L=1,S=3/2\right>.
\eeqy
The subscripts `1' denotes the first radially-excited copy of the
ground state multiplet. The `2' denotes an orbitally-excited state
with $J^P=1/2^+$. This state forms a $j=1$ multiplet with the second
$3/2^+$ state listed in Eq.~(\ref{mixedstates}).

We now examine the form factors of Appendix~\ref{hqetformfactors},
along with some of the form factors in Appendix~\ref{appendixC}, and
compare these with the predictions of HQET shown in
Section~\ref{hqetpredictions}. We begin with a discussion of the
decays to pseudotensor final states.

\subsection{$1/2^-$}

The HQET predictions for decays to this state are shown in
Eq.~(\ref{hmsinglet}), while the quark model form factors are shown in
Section~\ref{ffhmsinglet}. Noting that $w-1\approx{\cal
O}\left(1/m_q\right)$, the leading order predictions are that
$F_1=F_2=G_2=0$, and $F_3=-G_3=G_1$. The form factors of
Section~\ref{ffhmsinglet} satisfy these relations, and allow us to
identify
\beq
\eta_2^{(0)}=\frac{\ms}{\al} \left(\frac{\alpha_\lambda\alpha_{\lambda'}}{\alpha_{\lambda\lambda'}^2}\right)^{5/2}
\exp\left( -\frac{3 m^2_\sigma}{2m^2_{\Omega_q}}\frac{p^2}{\alpha_{\lambda\lambda'}^2}\right),\nonumber
\eeq
in the harmonic oscillator models, or
\beq
\eta_2^{(0)}=\frac{\ms}{\bl}\sqrt{2}\frac{\left(\frac{\beta_\lambda\beta_{\lambda'}}
{\beta_{\lambda\lambda'}^2}\right)^{5/2}}{\left[1+  \frac{3}{2}\frac{m^2_\sigma}{m^2_{\Omega_q}}
\frac{p^2}{\beta_{\lambda\lambda'}^2}\right]^3}
\eeq
in the Sturmian models. In these expressions, and in those that follow,
\beq
\all^2 =\frac{1}{2}(\al^2 + \alp^2)
\eeq 
and
\beq
\beta_{\lambda\lambda'} =\frac{1}{2}(\beta_\lambda + \beta_\lambda').
\eeq

\subsection{$(1/2^+,\,\,3/2^+)$}

The HQET predictions for decays to this pair of states are shown in
Eqs.~(\ref{hpthpa}) and (\ref{hpthpb}). The three multiplets
with these quantum numbers are discussed separately.

\subsubsection{Ground State}

The quark model form factors for the ground state doublet are shown in
Sections~\ref{ffhpgs} and~\ref{ffthpa}. Comparison of these form
factors with the predictions of HQET leads to
\beq
\eta_2^{(1)}=-\frac{1}{2}\eta_1^{(1)}
\eeq
at the non-recoil point, and allows us to identify
\beq
\eta_1^{(1)}=-\left(\frac{\alpha_\lambda\alpha_{\lambda'}}{\alpha_{\lambda\lambda'}^2}\right)^{3/2}
\exp\left( -\frac{3 m^2_\sigma}{2m^2_{\Omega_q}}\frac{p^2}{\alpha_{\lambda\lambda'}^2}\right),
\eeq
in the harmonic oscillator models, or
\beq
\eta_1^{(1)}=-\frac{\left(\frac{\beta_\lambda\beta_{\lambda'}}
{\beta_{\lambda\lambda'}^2}\right)^{3/2}}{\left[1+  \frac{3}{2}\frac{m^2_\sigma}{m^2_{\Omega_q}}
\frac{p^2}{\beta_{\lambda\lambda'}^2}\right]^3}
\eeq
in the Sturmian models. In the heavy quark limit, both forms yield the expected normalization at the
non-recoil point, namely $\eta_1^{(1)}(w=1)=-1$. It must be emphasized that the relationship between 
$\eta_1^{(1)}$ and $\eta_2^{(1)}$ given above is one that arises only in the context of the quark model.
In HQET, these two Isgur-Wise functions are {\it a priori} independent of each other. A more complete expression
of the relationship between $\eta_1^{(1)}$ and $\eta_2^{(1)}$ can be obtained by noting that
$G_2$ and $G_3$ for the $1/2^+$ final state both vanish at leading order in the quark model. This leads to
\beq
\eta_1^{(1)}(w)=-(1+w)\eta_2^{(1)}(w),
\eeq
valid at leading order in the heavy quark expansion.

\subsubsection{Radial Excitation}

The form factors for decays to the radially excited $(1/2^+,\,\, 3/2^+)$
multiplet are shown in Sections~\ref{ffhpa} and~\ref{ffthpb}. Comparison of
these form factors with the predictions of HQET again leads to 
\beq
\eta_2^{(1)}=-\frac{1}{2}\eta_1^{(1)},
\eeq
and allows us to identify
\beq
\eta_1^{(1)}=\sqrt{\frac{3}{8}} \frac{\al^2-\alp^2}{\al\alp}
\left(\frac{\alpha_\lambda\alpha_{\lambda'}}{\alpha_{\lambda\lambda'}^2}\right)^{5/2}
\exp\left( -\frac{3 m^2_\sigma}{2m^2_{\Omega_q}}\frac{p^2}{\alpha_{\lambda\lambda'}^2}\right),
\eeq
in the harmonic oscillator models, or
\beq
\eta_1^{(1)}=\frac{\sqrt{3}}{4} \frac{\bl^2-\blp^2}{\bl\blp}\frac{\left(\frac{\beta_\lambda\beta_{\lambda'}}
{\beta_{\lambda\lambda'}^2}\right)^{3/2}}{\left[1+  \frac{3}{2}\frac{m^2_\sigma}{m^2_{\Omega_q}}
\frac{p^2}{\beta_{\lambda\lambda'}^2}\right]^3}
\eeq
in the Sturmian models. As with decays to the ground state multiplet, the full relationship between 
$\eta_1^{(1)}$ and $\eta_2^{(1)}$ can be deduced to be 
\beq
\eta_1^{(1)}(w)=-(1+w)\eta_2^{(1)}(w),
\eeq
valid at leading order in the heavy quark expansion.

\subsubsection{Orbital Excitation}

The orbitally excited $(1/2^+,\,\,3/2^+)$ multiplet has a very different
structure from either of the two multiplets discussed previously, and the form
factors that are non-vanishing at leading order are different. For the ground
state and radially excited multiplets, $F_1$, $F_2$, $F_3$ and $G_1$ are the
non-vanishing form factors at leading order for the $1/2^+$ state, for
instance, and this pattern is repeated with the radially excited multiplet
[ignoring, for the moment, the fact that $\al^2-\alp^2\approx{\cal O}(1/m_q)$].
For the orbitally  excited states, whose form factors are shown in 
Sections~\ref{ffhpb} and~\ref{ffthpsl1}, the pattern is different, with $G_2$
and $G_3$ being the non-vanishing form factors for the $1/2^+$ state.

Comparing these quark model form factors with the leading order predictions of HQET allows us
to deduce that
\beqy
\eta_1^{(1)}&=&0,\nonumber\\
\eta_2^{(1)}&=&-\sqrt{\frac{27}{10}}\frac{\ms^2}{\al^2}
\left(\frac{\alpha_\lambda\alpha_{\lambda'}}{\alpha_{\lambda\lambda'}^2}\right)^{7/2}
\exp\left( -\frac{3 m^2_\sigma}{2m^2_{\Omega_q}}\frac{p^2}{\alpha_{\lambda\lambda'}^2}\right)
\eeqy
in the harmonic oscillator basis, or
\beqy
\eta_1^{(1)}&=&0,\nonumber\\
\eta_2^{(1)}&=&-\frac{9}{\sqrt{5}}\frac{\ms^2}{\bl^2}
\frac{\left(\frac{\beta_\lambda\beta_{\lambda'}}
{\beta_{\lambda\lambda'}^2}\right)^{7/2}}{\left[1+  \frac{3}{2}\frac{m^2_\sigma}{m^2_{\Omega_q}}
\frac{p^2}{\beta_{\lambda\lambda'}^2}\right]^4}
\eeqy
in the Sturmian basis.

\subsection{$(3/2^-,\,\,5/2^-)$}

The HQET predictions for this multiplet are shown in Eqs.~(\ref{thfhma}) and
(\ref{thfhmb}), while the quark model predictions for these states are shown in
Sections~\ref{fffhm} and~\ref{ffthmsl2}. Comparison of these two sets of
equations yields
\beq
\eta_1^{(2)}(w)=-(1+w)\eta_2^{(2)}(w)=-\sqrt{3}\frac{\ms}{\al}
\left(\frac{\alpha_\lambda\alpha_{\lambda'}}{\alpha_{\lambda\lambda'}^2}\right)^{5/2}
\exp\left( -\frac{3 m^2_\sigma}{2m^2_{\Omega_q}}\frac{p^2}{\alpha_{\lambda\lambda'}^2}\right)
\eeq
in the harmonic oscillator models, or
\beq
\eta_1^{(2)}(w)=-(1+w)\eta_2^{(2)}(w)=-\sqrt{6}\frac{\ms}{\bl}
\frac{\left(\frac{\beta_\lambda\beta_{\lambda'}}
{\beta_{\lambda\lambda'}^2}\right)^{5/2}}{\left[1+  \frac{3}{2}\frac{m^2_\sigma}{m^2_{\Omega_q}}
\frac{p^2}{\beta_{\lambda\lambda'}^2}\right]^3}\eeq
in the Sturmian models.

\subsection{$(5/2^+,\,\,7/2^+)$}

The HQET predictions for this multiplet are shown in Eqs.~(\ref{fhshpa}) and
(\ref{fhshpb}), while the quark model predictions for the $5/2^+$ state are 
shown in Section~\ref{fffhpsl3}. We have not calculated the form factors for the
$7/2^+$ state in our models. Comparison of the HQET predictions with the results
of the quark model calculation yields
\beq
\eta_1^{(3)}(w)=-(1+w)\eta_2^{(3)}(w)=-\frac{3}{\sqrt{2}}\frac{\ms^2}{\al^2}
\left(\frac{\alpha_\lambda\alpha_{\lambda'}}{\alpha_{\lambda\lambda'}^2}\right)^{7/2}
\exp\left( -\frac{3 m^2_\sigma}{2m^2_{\Omega_q}}\frac{p^2}{\alpha_{\lambda\lambda'}^2}\right)
\eeq
in the harmonic oscillator models, or
\beq
\eta_1^{(3)}(w)=-(1+w)\eta_2^{(3)}(w)=-3\sqrt{3}\frac{\ms^2}{\bl^2}
\frac{\left(\frac{\beta_\lambda\beta_{\lambda'}}
{\beta_{\lambda\lambda'}^2}\right)^{7/2}}{\left[1+  \frac{3}{2}\frac{m^2_\sigma}{m^2_{\Omega_q}}
\frac{p^2}{\beta_{\lambda\lambda'}^2}\right]^4}
\eeq
in the Sturmian models.

We now turn to a discussion of the decays to daughter baryons having tensor
light diquark. We note, first that there exists a $1/2^+,\,\,j=0$ singlet
state. At leading order, the form factors for decays to such a state vanish in
HQET. In the quark model, such a state can be constructed, but the overlap of
its wave function with that of the decaying parent baryon is zero to the
approximation to which we work, and is strongly suppressed beyond that. Thus we
do not have form factors for such a state. For the remaining decays of the tensor
type, there is a single Isgur-Wise type form factor.

\subsection{$(1/2^-,\,\,3/2^-)$}

The HQET predictions for this multiplet are shown in Eqs.~(\ref{hmthma}) and
(\ref{hmthmb}), while the quark model predictions are shown in 
Sections~\ref{ffhmsl1} and~\ref{ffthmsl1}. Comparison of these two sets yields
\beq
\tau^{(1)}=\sqrt{\frac{3}{2}}\frac{\ms}{\al}
\left(\frac{\alpha_\lambda\alpha_{\lambda'}}{\alpha_{\lambda\lambda'}^2}\right)^{7/2}
\exp\left( -\frac{3 m^2_\sigma}{2m^2_{\Omega_q}}\frac{p^2}{\alpha_{\lambda\lambda'}^2}\right)
\eeq
in the harmonic oscillator models, or
\beq
\tau^{(1)}=\sqrt{3}\frac{\ms}{\bl}
\frac{\left(\frac{\beta_\lambda\beta_{\lambda'}}
{\beta_{\lambda\lambda'}^2}\right)^{5/2}}{\left[1+  \frac{3}{2}\frac{m^2_\sigma}{m^2_{\Omega_q}}
\frac{p^2}{\beta_{\lambda\lambda'}^2}\right]^3}
\eeq
in the Sturmian models.

\subsection{$(3/2^+,\,\,5/2^+)$}

The HQET predictions for this multiplet are shown in Eqs.~(\ref{thfhpa}) and
(\ref{thfhpb}), while the quark model predictions are shown in 
Sections~\ref{ffthpsl2} and~\ref{fffhpsl2}. Comparison of these two sets yields
\beq
\tau^{(2)}=\sqrt{3}\frac{\ms^2}{\al^2}
\left(\frac{\alpha_\lambda\alpha_{\lambda'}}{\alpha_{\lambda\lambda'}^2}\right)^{5/2}
\exp\left( -\frac{3 m^2_\sigma}{2m^2_{\Omega_q}}\frac{p^2}{\alpha_{\lambda\lambda'}^2}\right)
\eeq
in the harmonic oscillator models, or
\beq
\tau^{(2)}=3\sqrt{2}\frac{\ms^2}{\bl^2}
\frac{\left(\frac{\beta_\lambda\beta_{\lambda'}}
{\beta_{\lambda\lambda'}^2}\right)^{5/2}}{\left[1+  \frac{3}{2}\frac{m^2_\sigma}{m^2_{\Omega_q}}
\frac{p^2}{\beta_{\lambda\lambda'}^2}\right]^3}
\eeq
in the Sturmian models.

\section{Numerical Results}

\subsection{Model Parameters, Mass Spectra and Wave Functions}
\label{parametersmw}
In Section~\ref{hamiltonian}, we introduced the two Hamiltonians we
diagonalize to obtain the baryon spectrum. They differ only in the
form chosen for the kinetic portion, one of which is non-relativistic
(NR), while the other is semi-relativistic (SR). In addition, we use
two different expansion bases to obtain the wave functions: the
harmonic oscillator (HO) basis, and the Sturmian (ST) basis. In the
following, the four spectra we obtain will be denoted HONR, HOSR, STNR
and STSR, in what should be an obvious notation.
\begin{center}
\begin{table}[h]
\caption{Hamiltonian parameters obtained from the four fits. In the
first column, HO refers to the harmonic oscillator basis, while ST
refers to the Sturmian basis. In the same column, NR and SR indicate
non-relativistic and semi-relativistic Hamiltonians, respectively. The
form of these Hamiltonians is described in
Section~\ref{hamiltonian}.\label{parameter1}}
\begin{tabular}{|l|ccccccccc|}
\hline
 model&$m_\sigma$ & $m_s$  & $m_c$  & $m_b$  & $b$& $\alpha_{\rm Coul}$ 
 &$\alpha_{\rm hyp}$ & $C_{qqq}$ & $\kappa$ \\
 &(GeV)&(GeV)&(GeV)&(GeV)&(GeV$^2$)& & &(GeV)&\\ \hline
 HONR & 0.39& 0.63& 1.90 & 5.30 & 0.16 &0.21& 1.18& -1.50 &0.73\\ 
 HOSR & 0.39& 0.55& 1.81 & 5.26 & 0.13 &0.15 & 0.86 & -1.11&0.70 \\
 STNR & 0.41& 0.63& 1.90 & 5.30 & 0.13 &0.23 & 0.34 & -1.40 &-\\ 
 STSR & 0.42& 0.60& 1.83 & 5.31 & 0.14 &0.08 & 0.25 & -1.40 &- \\ \hline
\end{tabular}
\end{table}
\end{center}

There are eight free parameters (nine in HO models) to be obtained for
each spectrum: four quark masses ($m_u=m_d$, $m_s$, $m_c$ and $m_b$),
and four parameters of the potential ($\alpha_{\rm hyp}$, $\alpha_{\rm
Coul}$, $b$ and $C_{qqq}$). These eight parameters are determined from
a `variational diagonalization' of the Hamiltonian. The variational
parameters are the wave function size parameters $\alpha_\rho$ and
$\alpha_\lambda$ of Eq.~(\ref{hoa}), or $\beta_\rho$ and
$\beta_\lambda$ of Eq.~(\ref{sta}). This variational diagonalization
is accompanied by a fit to the known spectrum. In this fit, the eight
parameters mentioned before are varied. In addition, it is important
to include any experimentally known information from semileptonic
decay rates into the fit of these parameters. At present, this information
 is limited to the decay rate for $\lcle$. The rationale here is
that the dynamics leading to the spectrum of states also play a
crucial role in the semileptonic processes we are studying. By
incorporating known semileptonic decay rates in our fit, we expect
that predictions for as yet unmeasured rates will be more robust. The
values we obtain for the Hamiltonian parameters are shown in
Table~\ref{parameter1}.

In I, we presented the values obtained for the fit parameters. In the
present manuscript, we have modified our variational diagonalization
procedure somewhat, so it is appropriate for us to show the resulting
values of the fit parameters again. The modification of this procedure is
best explained with a concrete example, that of the $1/2^+$
baryons. The wave functions for such states are expanded in terms of
the seven basis states of Eq.
(\ref{halfpwf}). When the Hamiltonian is diagonalized in this basis, we obtain
wave functions for seven states with $J^P=1/2^+$. The variational part
of the computation can be carried out by minimizing the energy of any
of these seven states. In I, we used the energy of the lowest-lying
state for this variation, and this led to the choices for $\al$ and
$\ar$ (or $\bl$ and $\br$) reported there.  In the present work, we no
longer use the lowest-lying state for the variational calculation, but
one of the excited states. This means that the values of the
wave-function size parameters, as well as of the parameters in the
Hamiltonian, in addition to the compositions of the states, are
modified from their values in I, sometimes significantly so. In our
current work we have chosen the third lowest-lying state for the
variational procedure.

One consequence of including the decay rate for $\lcle$ in the fit is
that the light ($u$ and $d$) quark masses we obtain are consistently
larger than conventional values. Other quark masses do not differ
significantly from those in our previous work. We also obtain somewhat
different values for $\alpha_{\rm Coul}$ and $\alpha_{\rm hyp}$, and
the strength of the hyperfine interaction in the two harmonic
oscillator models is larger than we reported in I. As a result, the
hyperfine splittings in the baryon spectrum are now quite well
reproduced in the HONR and HOSR models. We also note that the values
of $b$, the slope of the linear potential, remain similar to our
previous results, which tend to be smaller than in most published
studies of hadron spectra.  However, recent work by Barnes, Godfrey
and Swanson~\cite{godfrey} reports a value of $0.14$ for this
parameter, obtained by fitting a similar Hamiltonian to the
spectrum of heavy mesons.

In the two harmonic oscillator models we fit an
additional parameter $\kappa$, which appears in the $\lcle$ form
factors. We have discussed the origin of $\kappa$ in I. This parameter
was introduced in an {\it ad hoc} manner in the model of Isgur, Scora, Grinstein and
Wise (ISGW)~\cite{ISGW,ISGW1} to take into account ``relativistic
effects''. In the harmonic oscillator models, all form factors are
proportional to the exponential factor
\beqy
\exp\left( -\frac{3 m^2_\sigma}{2m^2_{B_q}}\frac{p^2}{\alpha_{\lambda\lambda'}^2}\right)\nonumber,
\eeqy
which ISGW modify to
\beqy
\exp\left( -\frac{3 m^2_\sigma}{2m^2_{B_q}}\frac{p^2}{\kappa^2\alpha_{\lambda\lambda'}^2}\right)\nonumber.
\eeqy
We include this parameter in our calculation of the HO form factors
and rates in part because the work we present is done in the same
spirit as the the work of ISGW, and such a parameter was found to be
necessary in Ref.~\cite{ISGW}. However, instead of choosing a
particular value, as was done in Ref.~\cite{ISGW}, we treat $\kappa$
as a free parameter constrained to lie between 0.7 and 1.0. The values
we obtain for $\kappa$ are shown in Table~\ref{parameter1} for both
HONR and HOSR models. We note, however, that we do not include this
parameter in the form factors for the decays of heavy $\Omega_Q$
baryons. This parameter is meant to mimic relativistic effects in the
spectator quarks in the decaying baryon, and such effects are expected
to be smaller for $s$ quarks than they are for $u$ and $d$ quarks.

The effect of the parameter $\kappa<1$ is to soften the form factors. 
It has been established that nonrelativistic or
semi-relativistic quark models using an oscillator basis tend to
underestimate the charge radius of light-quark systems such as the
proton, and that some part of this underestimation can be attributed
to relativistic effects in the evaluation of the electromagnetic
current~\cite{Hayne:1981zy}. A procedure similar to the inclusion of
the parameter $\kappa$ by ISGW was used by Foster and
Hughes~\cite{Foster:1983kn} to modify electromagnetic form factors of
light-quark systems calculated in a nonrelativistic quark model.
\begin{center}
\begin{table}[h]
\caption{Wave function size parameters, $\alpha_\rho$ and $\alpha_\lambda$, for
states of selected $J^P$ with spin-flavor symmetric light diquark, 
in different models. All values are in GeV.
\label{parameter3}}
\vspace{5mm}
\begin{tabular}{|l|l|cccc|}
\hline
$J^P$ &  model & $\Omega_b$ & $\Omega_c$ & $\Xi$ & $\Omega$  \\ 
&  & $\left(\alpha_\lambda,\,\,\,\,\alpha_\rho\right)$ &
$\left(\alpha_\lambda,\,\,\,\,\alpha_\rho\right)$ &
$\left(\alpha_\lambda,\,\,\,\,\alpha_\rho\right)$ &
$\left(\alpha_\lambda,\,\,\,\,\alpha_\rho\right)$  \\ \hline
$1/2^+$ &HONR &  (0.54, 0.42)& (0.49, 0.42) & (0.39, 0.42) & 0.42\\
$1/2^+$ &HOSR & (0.51, 0.48)& (0.49, 0.47) & (0.42, 0.44)& 0.42\\
$1/2^+$ & STNR &  (0.72, 0.42)& (0.67, 0.47) & (0.49, 0.65)&- \\
$1/2^+$ & STSR &  (0.72, 0.38)& (0.66, 0.44) & (0.63, 0.78) &-\\ \hline
$3/2^+$ &HONR &  -&(0.50, 0.42)  & (0.37, 0.40) & 0.40\\
$3/2^+$ &HOSR & -& (0.48, 0.47) & (0.36, 0.41)& 0.40\\
$3/2^+$ & STNR &  -& (0.74, 0.40) & (0.43, 0.71)&- \\
$3/2^+$ & STSR & -& (0.67, 0.43) & (0.63, 0.78) &-\\ \hline
$1/2^-$ & HONR &  -& (0.51, 0.43) & (0.38, 0.41) &0.42\\
$1/2^-$ & HOSR &  -& (0.48, 0.47) & (0.40, 0.42) &0.42 \\
$1/2^-$ &STNR & -&  (0.67, 0.47) & (0.61, 0.53)&- \\
$1/2^-$ &STSR & -& (0.65, 0.45) & (0.60, 0.75) &-\\ \hline
$5/2^-$ &HONR & -& (0.51, 0.43) & (0.37, 0.42) &0.42  \\
$5/2^-$ &HOSR & -& (0.48, 0.47) & (0.39, 0.42)& 0.42\\
$5/2^-$ &STNR & -&  (0.67, 0.47) & (0.61, 0.53)&- \\
$5/2^-$ &STSR & -& (0.65, 0.45) & (0.60, 0.75) &-\\ \hline
\end{tabular}
\end{table}
\end{center}

In carrying out our fits, we generally allow the values of
$\alpha_\rho$ to be different from $\alpha_\lambda$, as in I. The
exceptions occur in cases when the three quarks are identical, as they
are in the nucleon or the $\Omega$. In such cases, the variational
diagonalization automatically selects $\alpha_\rho=\alpha_\lambda$ in
the HO bases. Table~\ref{parameter3} shows some of the values we
obtain for the size parameters. The omitted parameters for the states
that are significant for this work are related to those
presented. For instance, for the $1/2^+_1$ states, the size parameters
are the same as for the $1/2^+$ states. Furthermore, since we do not
include a spin-orbit interaction in our Hamiltonian, the size
parameters for the $1/2^-$ and $3/2^-$ states are identical. We do not
show the size parameters for the $\Xi$ states with $J^P=5/2^+$ or
$7/2^+$ mainly because we find that semileptonic decay rates to these
states are very small. We also omit the size parameters for the
analogous $\Omega_Q$ states with $Q=c,\,\, s$.

\subsection{Mass Spectra}

Portions of the mass spectra we obtain using our four models are shown
in Tables~\ref{baryonspec2} and~\ref{baryonspec}. In these tables, the
first two columns identify the state and its experimental mass, while
the next four columns show the model masses that result from a fit of
the Hamiltonian parameters to those states whose experimental masses
are known. We note that for the $\Omega$ and $\Xi$ states, the
predicted masses are in satisfactory agreement with the available
experimental values, with little variation among the results from the
different models for these states.
\begin{center}
\begin{table}[h]
\caption{Baryon masses in GeV in the quark models we use. Hamiltonian 
parameters for each model are obtained from fits to the experimental masses 
where known; other masses shown are predictions of the models. The first two columns identify 
 the state and its experimental mass, while the next four columns 
show the masses that result from the models.
\label{baryonspec2}}
\vspace{5mm}
\begin{tabular}{|l|c|llll|}
\hline
State& Experimental Mass& HONR& HOSR& STNR& STSR\\ \hline
$\Xi(1/2^+)$&1.32&1.32&1.35&1.40&1.39\\
$\Xi(1/2^+_1(rad))$&-&2.03&1.79&2.06&1.83\\
$\Xi(1/2^+_2(orb))$&-&2.08&2.00&2.10&1.95\\
$\Xi(3/2^+)$&1.53&1.52&1.54&1.45&1.46\\ 
$\Xi(3/2^+_1(rad))$&-&2.16&2.08&2.10&2.12\\ 
$\Xi(3/2^+_2(orb))$&-&2.14&2.18&1.98&1.96\\ 
$\Xi(3/2^-)$&1.82&1.83&1.78&1.79&1.80\\
$\Xi(5/2^-)$&-&1.84&1.78&1.78&1.81\\ 
$\Xi(5/2^+)$&-&2.08&2.14&2.08&2.00\\ \hline
$\Omega(3/2^+)$&1.67&1.66&1.66&1.60&1.67\\ 
$\Omega(3/2^+_1 (rad))$&-&2.20&2.07&2.34&2.13\\ 
$\Omega(3/2^+_2(orb))$&-&2.23&2.11&2.24&2.14\\ 
$\Omega(3/2^-)$&-&1.95&1.84&1.88&1.88\\
$\Omega(5/2^-)$&-&1.95&1.89&1.89&1.89\\
$\Omega_c(1/2^+)$&2.70&2.69&2.72&2.73&2.71\\
$\Omega_c(1/2^+_1(rad))$&-&3.18&3.09&3.24&3.24\\
$\Omega_c(1/2^+_2(orb))$&-&3.25&3.17&3.24&3.26\\
$\Omega_c(3/2^+)$&-&2.77&2.78&2.75&2.73\\ 
$\Omega_c(3/2^+_1(rad))$&-&3.22&3.15&3.30&3.24\\
$\Omega_c(3/2^+_2(orb))$&-&3.24&3.18&3.23&3.26\\
$\Omega_c(1/2^-)$&3.00&3.00&2.97&3.00&3.02\\
$\Omega_c(5/2^-)$&-&3.02&2.99&3.01&3.02\\
$\Omega_b(1/2^+)$&-&6.08&6.13&6.08&6.14\\ \hline
\end{tabular}
\end{table}
\end{center}

In Table \ref{baryonspec}, we also present some of the masses of the nucleons and $\ll_Q$
states, mainly to show the improvement that has resulted from the
modified variational procedure. We have obtained a better spectrum for
almost all of the nucleons and $\Lambda_Q$ baryons, with significant
improvement in the $N(1440)$ and the $\Delta$ resonance model masses.
\begin{center}
\begin{table}[h]
\caption{Baryon masses in GeV fitted in the four quark models we use. The
first two columns identify the state and its experimental mass, while 
the next four columns show the masses that result from the models.
\label{baryonspec}}
\vspace{5mm}
\begin{tabular}{|l|c|llll|}
\hline
State& Experimental Mass& HONR& HOSR& STNR& STSR\\ \hline
$N(1/2^+)$&0.94&1.00&1.08&1.07&1.12\\
$N(1/2^+_1)$&1.44&1.68&1.56&1.76&1.58\\
$N(1/2^-)$&1.54&1.47&1.47&1.51&1.47\\
$N(3/2^+)$&1.72&1.72&1.76&1.77&1.73\\\hline 
$\Delta(3/2^+)$&1.23&1.24&1.32&1.20&1.20\\\hline 
$\Lambda(1/2^+)$&1.12&1.11&1.11&1.09&1.05\\
$\Lambda(1/2^+_1)$&1.60&1.74&1.63&1.61&1.59\\
$\Lambda(1/2^-)$&1.41&1.49&1.50&1.46&1.52\\
$\Lambda(3/2^+)$&1.89&1.85&1.74&1.73&1.81\\ \hline
$\Lambda_c(1/2^+)$&2.28&2.27&2.26&2.27&2.21\\
$\Lambda_c(1/2^-)$&2.59&2.63&2.60&2.60&2.66\\\hline
$\Lambda_b(1/2^+)$&5.62&5.62&5.62&5.62&5.62\\ \hline
\end{tabular}
\end{table}
\end{center}
\subsection{Wave Functions}
Significant mixing of wave function components occurs in many of the
$\Omega_Q$ and $\Xi$ states, for all flavors, particularly in the
Sturmian models. The mixing coefficients that result, along with
recalculated mixing coefficients for $N$ and $\Lambda_Q$ states, are
tabulated in Tables~\ref{mixingwf2} and~\ref{mixingwf}, for all four
models. In Table~\ref{mixingwf2}, we show the wave function
coefficients for the $1/2^+$, $3/2^+$ and $3/2^-$ states, in each flavor
sector, for each model for the $\Omega_Q$ and $\Xi$ baryons. The exceptions are
for $\Omega$, where we do not use the Sturmian basis. We do not present
the mixing for $1/2^-$ states because they have exactly the same
mixing coefficients as, and are degenerate with, the $3/2^-$ 
states. For other states we treat, such as $1/2^-_1$, $3/2^-_1$ and
$5/2^-$, the wave functions that result are single component wave
functions. The mixing shown in these tables complicates the extraction
of the form factors. However, in all numerical results that we show
for the form factors and the decay rates, this mixing is properly
taken into account.
\begin{center}
\begin{table}[h]
\caption{Mixing coefficients ($\eta_i$) of the lowest-lying $1/2^+, 3/2^+$ and
$3/2^-$ states of $\Xi$ and $\Omega_Q$, in different flavor sectors. The $\eta_i$ are defined in 
Appendix~\ref{appendixA}.\label{mixingwf2}}
\vspace{5mm}\begin{tabular}{|l|ccc|ccc|ccc|ccc|}
\hline
Baryon states&& HONR & & &HOSR & & STNR & & &STSR & & \\
&$\eta_1$& $\eta_2$& $\eta_3$& $\eta_1$& $\eta_2$& $\eta_3$ 
&$\eta_1$& $\eta_2$& $\eta_3$& $\eta_1$& $\eta_2$& $\eta_3$\\ \hline
$\Xi(1/2^+)$&0.970&0.100&0.198&0.962& 0.062&0.230
&0.969&-0.226&0.093&0.964& -0.256&0.058\\
$\Xi(3/2^+)$& 0.996&0.077&0.033&0.999& -0.009&0.038
& 0.947&-0.313&0.066&0.935& -0.334&-0.118 \\
$\Xi(3/2^-)$& 0.484&-&0.875&0.641& -&0.767
& -0.296&-&0.955&0.115& -&0.993\\
$\Omega_c(1/2^+)$& 0.976&0.093&0.189&0.980& -0.035&0.189
& 0.980&0.200&0.025&0.933& 0.361&$|\eta_3|<0.001$\\
$\Omega_c(3/2^+)$&0.995& 0.061&0.072& 0.993&-0.091&0.068
&0.964& 0.243&-0.107& 0.948&0.2317&-0.010\\
$\Omega_c(3/2^-)$&-0.234& -&0.997& -0.293&-&0.956
&-& -&1.00& -&-&1.00\\
$\Omega_b(1/2^+)$&0.985& 0.086&0.147&0.980& -0.09&0.173
&0.957& 0.291&0.006& 0.937&0.350&0.005\\ \hline
& & HONR & & &HOSR & & & & & & &\\
& $\eta_1$& $\eta_2$& $\eta_3$& $\eta_1$& $\eta_2$& $\eta_3$& & & & &&\\ \hline
$\Omega(3/2^+)$&0.996&0.063&0.063&0.998&0.042&0.042& & & & &&\\ \hline
\end{tabular}
\end{table}
\end{center}
\begin{center}
\begin{table}[h]
\caption{Mixing coefficients ($\eta_i$) of the lowest-lying $1/2^+$  states of N and $\Lambda_Q$ in different flavor sectors. The $\eta_i$ are defined in 
Appendix~\ref{appendixA}.
\label{mixingwf}}
\begin{tabular}{|l|ccc|ccc|ccc|ccc|}
\hline
Baryon states&& HONR &&& HOSR&  && STNR&& &STSR &\\
& $\eta_1$& $\eta_2$& $\eta_3$& $\eta_1$& $\eta_2$& $\eta_3$& $\eta_1$& $\eta_2$
& $\eta_3$& $\eta_1$& $\eta_2$& $\eta_3$\\ \hline
$N(1/2^+)$&0.959&0.095&0.246&0.949& 0.067&0.272&-&-&-&-&-&-\\
$\Lambda(1/2^+)$&0.976&0.186&0.026&0.933&0.289&-0.089& 0.963&0.219& 0.154
&0.869&0.478&0.129\\
$\Lambda_c(1/2^+)$&0.976&0.186&0.103& 0.939&0.337&0.006&0.962&0.224&0.158
&0.876&0.447&0.179\\
$\Lambda_b(1/2^+)$&0.977&$$0.185&0.106& 0.929&0.368&0.024&0.951&0.227&0.206
 &0.861&0.438&0.259\\ \hline
\end{tabular}
\end{table}
\end{center}
\subsection{Form Factors and Decay Rates: $\Omega_Q$}

\subsubsection{$\Omega_b\rightarrow \Omega_c^{(*)}$}

The form factors at the non-recoil point for the $\Omega_b\rightarrow
\Omega_c^{(*)}$ decays calculated using all four models are shown in 
Table~\ref{ffob1}. We show only form factors for the decays to
final states with $J^P=1/2^+$ and $3/2^+$. These states constitute the elastic
channels. We note that the form factor values at the non-recoil points
are very close to each other in all our models.

It is instructive to compare our form factor values at the non-recoil
point with the HQET predictions for decays to the ground state
doublet. For the state with $J^P =1/2^+$, the HQET prediction is 
$F_1 = G_1 = \eta_1^{(1)}/3$. With the known normalization condition on
$\eta_1^{(1)}$, this means that we expect $F_1=G_1=-1/3$ at the
non-recoil point. Our results for $G_1$ are very close to this, but
those for $F_1$ are not. The deviations from the HQET predictions can
easily be traced to the presence of $1/m_q$ and $1/m_Q$ terms in the
quark model results for $F_1$, while no such terms exist in the model
predictions for $G_1$. If such terms are ignored in $F_1$ then the
HQET prediction is indeed satisfied in the quark models. Similarly,
HQET predicts that $F_2$ and $F_3$ should each have the value of 2/3
at the non-recoil point. The model results agree with this prediction
for $F_2$ but not for $F_3$, and the differences can again be traced
to the presence of $1/m_q$ terms in the quark model results. The HQET
predictions for $G_2$ and $G_3$ are less easily interpreted, but
comparison of those predictions with the quark model calculations
suggests that the Isgur-Wise function $\eta_2^{(1)}$ is normalized to
1/2 at the non-recoil point. We have already raised this point in
Section~\ref{analresults} where we compare the quark model predictions
for the form factors with those of HQET.

For the state with $J^P=3/2^+$, HQET predicts that
$F_4=-G_4=2/\sqrt{3}$. This is well satisfied by the model predictions
for $G_4$, but the model predictions for $F_4$ include $1/m_q$
contributions. Assuming that $\eta_2^{(1)}$ is indeed normalized to
1/2 at the non-recoil point, the HQET prediction is then that $G_1=0$
and $F_3=-G_3=-1/\sqrt{3}$. The quark model results for $G_3$ are
close to the HQET prediction, but those for $F_3$ deviate from this
prediction because of the presence of $1/m_q$ terms in the quark model
results.

\begin{center}
\begin{table}[h]
\caption{The form factors for $\Omega_b$ transitions to the ground-state
$\Omega_c$ multiplet with $J^P=(1/2^+,\,\,3/2^+)$, calculated at the non-recoil point, in the four models used here.
\label{ffob1}}
\vspace{5mm}\begin{tabular}{|l|l|lllclllc|}
\hline
spin & model & $F_1$\,\,\, & $F_2$\,\,\,  & $F_3$ \,\,\, & $F_4$ \,\,\, & 
$G_1$ \,\,\,  & $G_2$ \,\,\, & $G_3$ \,\,\, & $G_4$\,\,\,  \\ \hline
$1/2^+$ &HONR & -0.48 & 0.58 & 0.86 & -&-0.32 & 0.13 & -0.03  & - \\
$1/2^+$ &HOSR & -0.47& 0.56 & 0.87 & -&-0.32& 0.12 & -0.02  & - \\
$1/2^+$ & STNR & -0.47 & 0.63 & 0.83 & -&-0.33& 0.11 & -0.04  & - \\
$1/2^+$ & STSR &-0.47 & 0.62 & 0.83 & -&-0.33 & 0.11 & -0.04  & - \\ \hline
$3/2^+$ & HONR & 0.80 & 0.0 & -0.80 & 1.61 &0.0 & -0.17 & 0.63 & -1.12 \\
$3/2^+$ &HOSR & 0.78& 0.0 & -0.78 & 1.58 &0.0 & -0.16 & 0.62 & -1.12 \\
$3/2^+$ & STNR & 0.83& 0.0 & -0.83 & 1.65 &0.0 & -0.18 & 0.64 & -1.15 \\
$3/2^+$ &STSR &0.82 & 0.0 & -0.82 & 1.64 &0.0 & -0.19 & 0.63 & -1.14 \\ \hline
\end{tabular}
\end{table}
\end{center}

\begin{figure}[h]
\centerline{\hspace*{0.35in}\epsfig{file=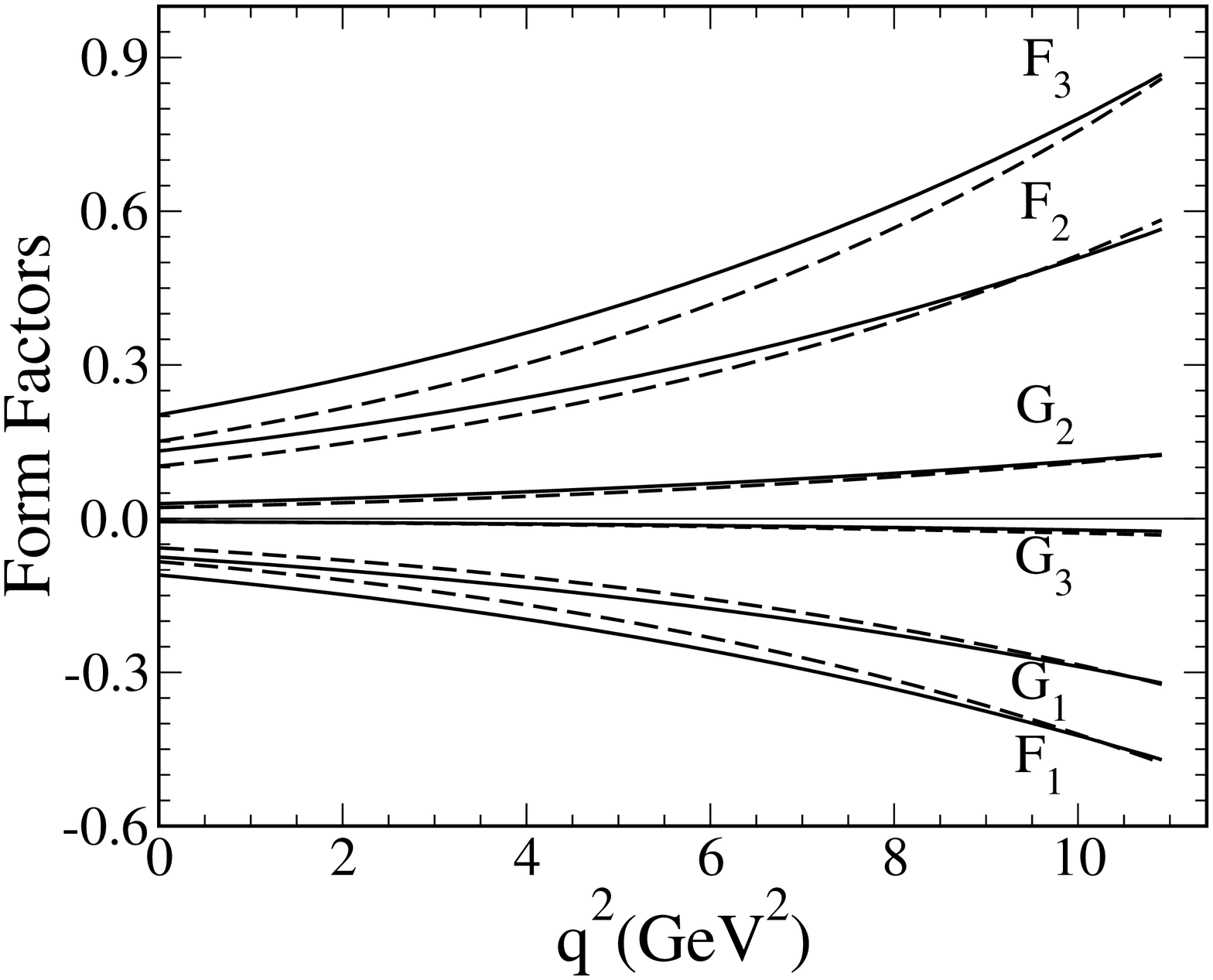,width=9.8cm}
\hspace*{-0.25in}\epsfig{file=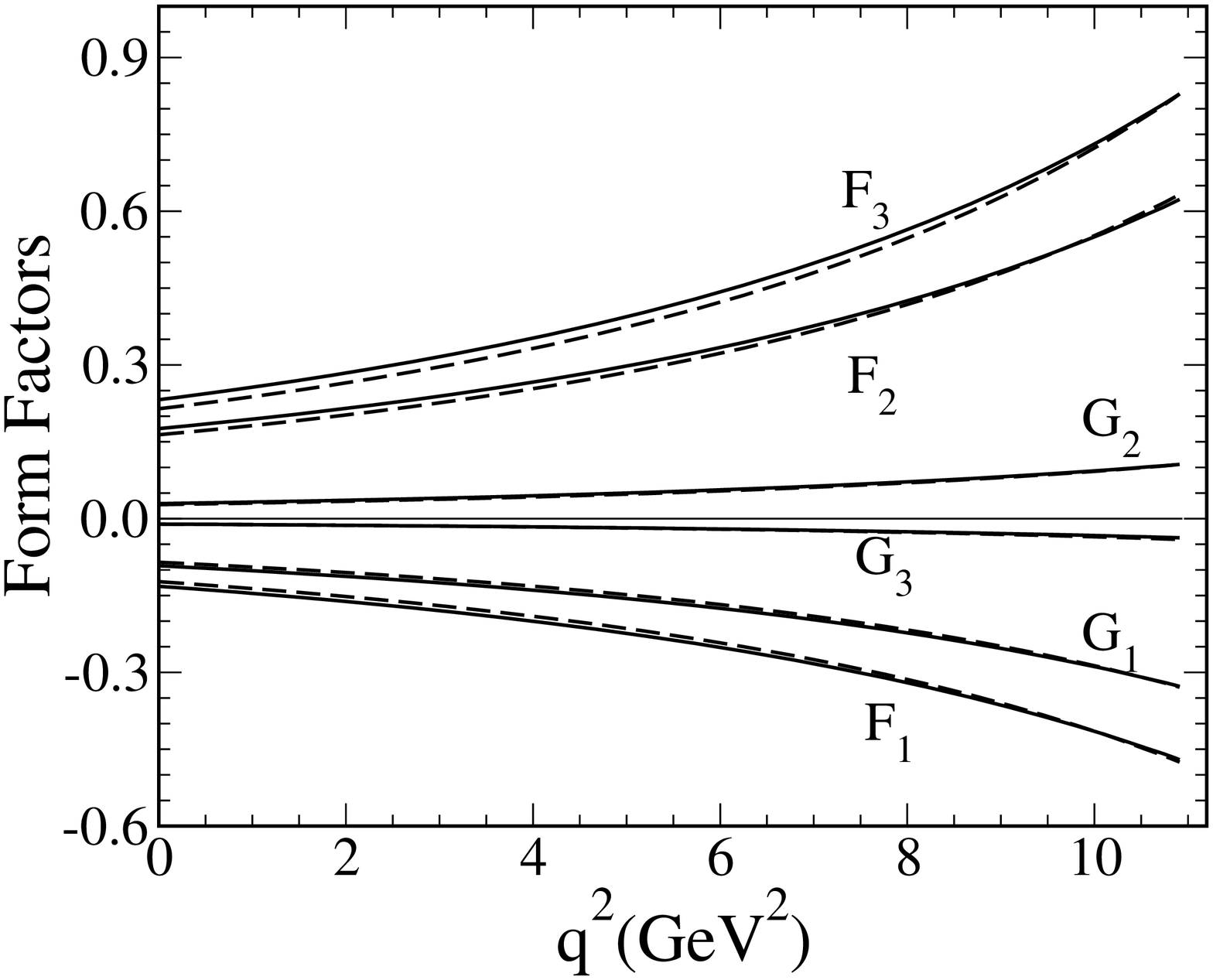,width=9.8cm}}
\caption{Form factors for $\Omega_b\to\Omega_c(1/2^+)$ obtained using harmonic
oscillator wave functions (left panel, HOSR and HONR models) and Sturmian wave 
functions (right panel, STSR and STSR models). In each panel, the solid curves 
arise from the semi-relativistic version of the model, while the dashed curves 
arise from the non-relativistic version.
\label{ffomega1}}
\end{figure}

Figure~\ref{ffomega1} shows the $q^2$ dependence of the form factors for the
elastic transition $\Omega_b\to\Omega_c(1/2^+)$ calculated in the HONR and
HOSR models on the left, and in the STSR and STNR models on the right. In each
panel, the solid curves arise from the SR version of the model, while the
dashed curves are from the NR version. Here we
note that the form factors calculated using the Sturmian wave functions have
slopes near the non-recoil point that are similar to those calculated using the
harmonic oscillator wave functions. This is due to the fact that we have similar 
mixing patterns for the $\ob$ and $\omc$ ground state wave functions in all models as
can be seen in Table~\ref{mixingwf2}. 
\begin{figure}[h]
\centerline{\epsfig{file=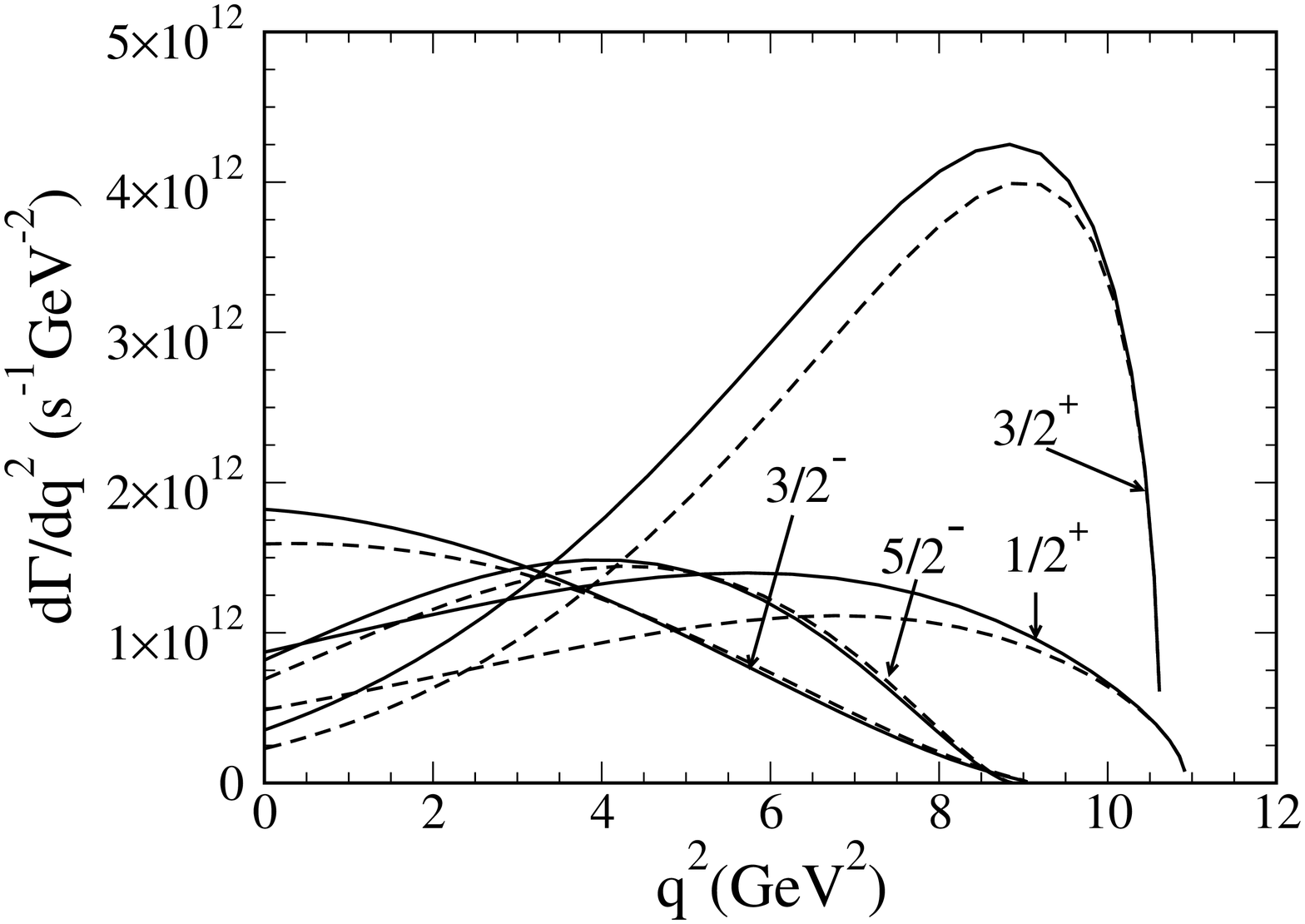,width=9.5cm}\hspace{-0.25in}
\epsfig{file=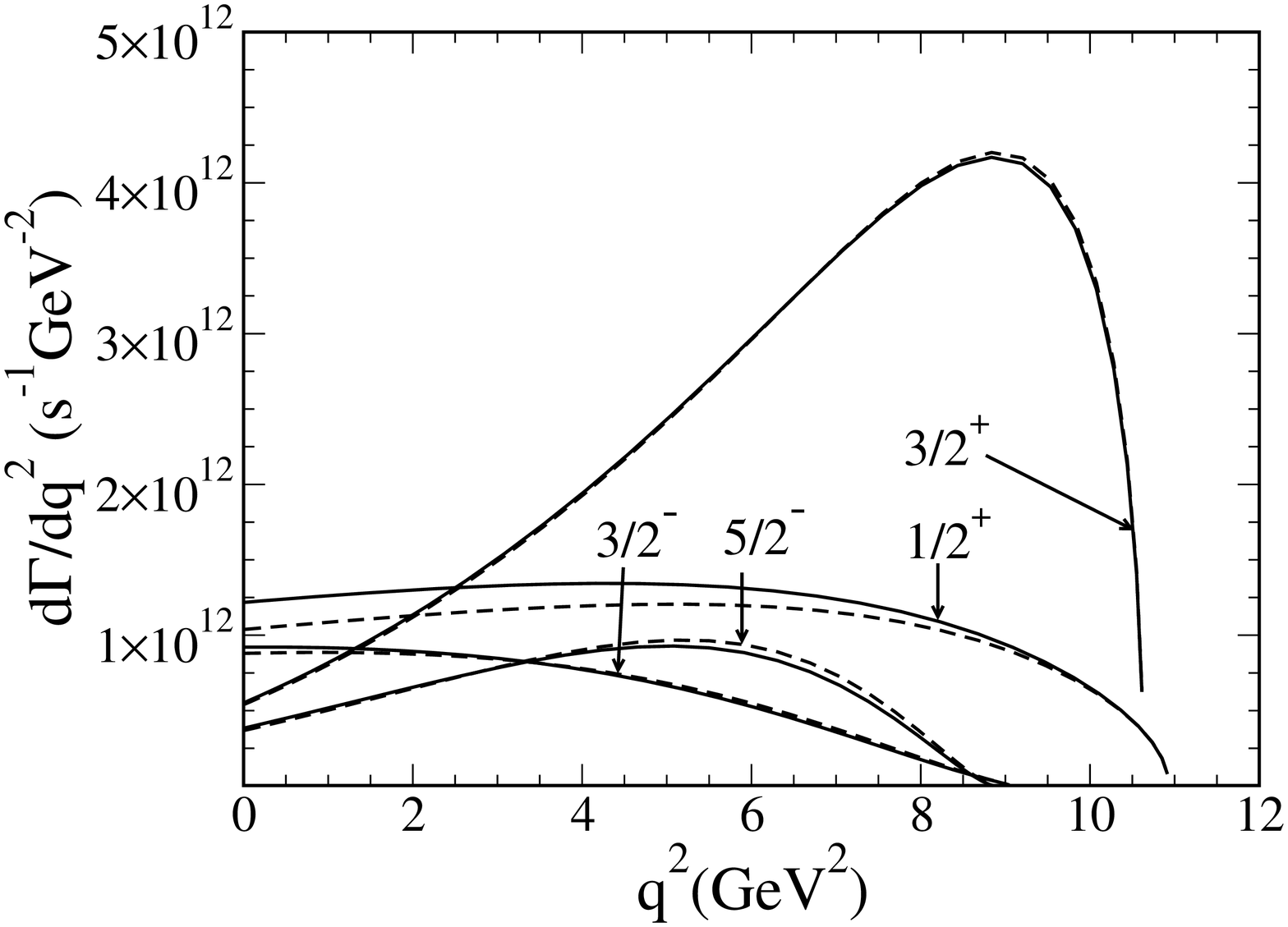,width=9.5cm}}
\vskip -.3in
\centerline{\epsfig{file=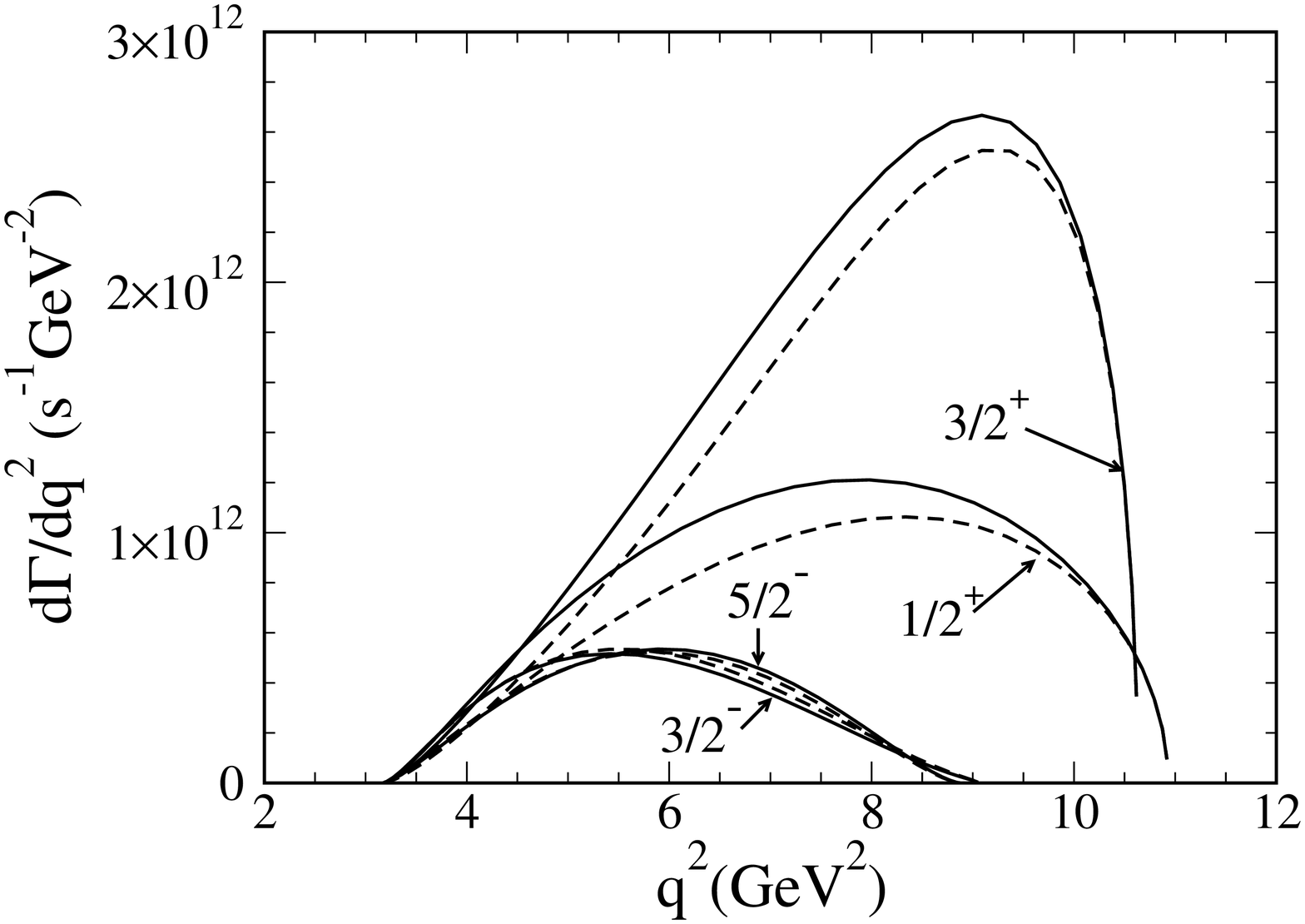,width=9.5cm}\hspace{-0.25in}
\epsfig{file=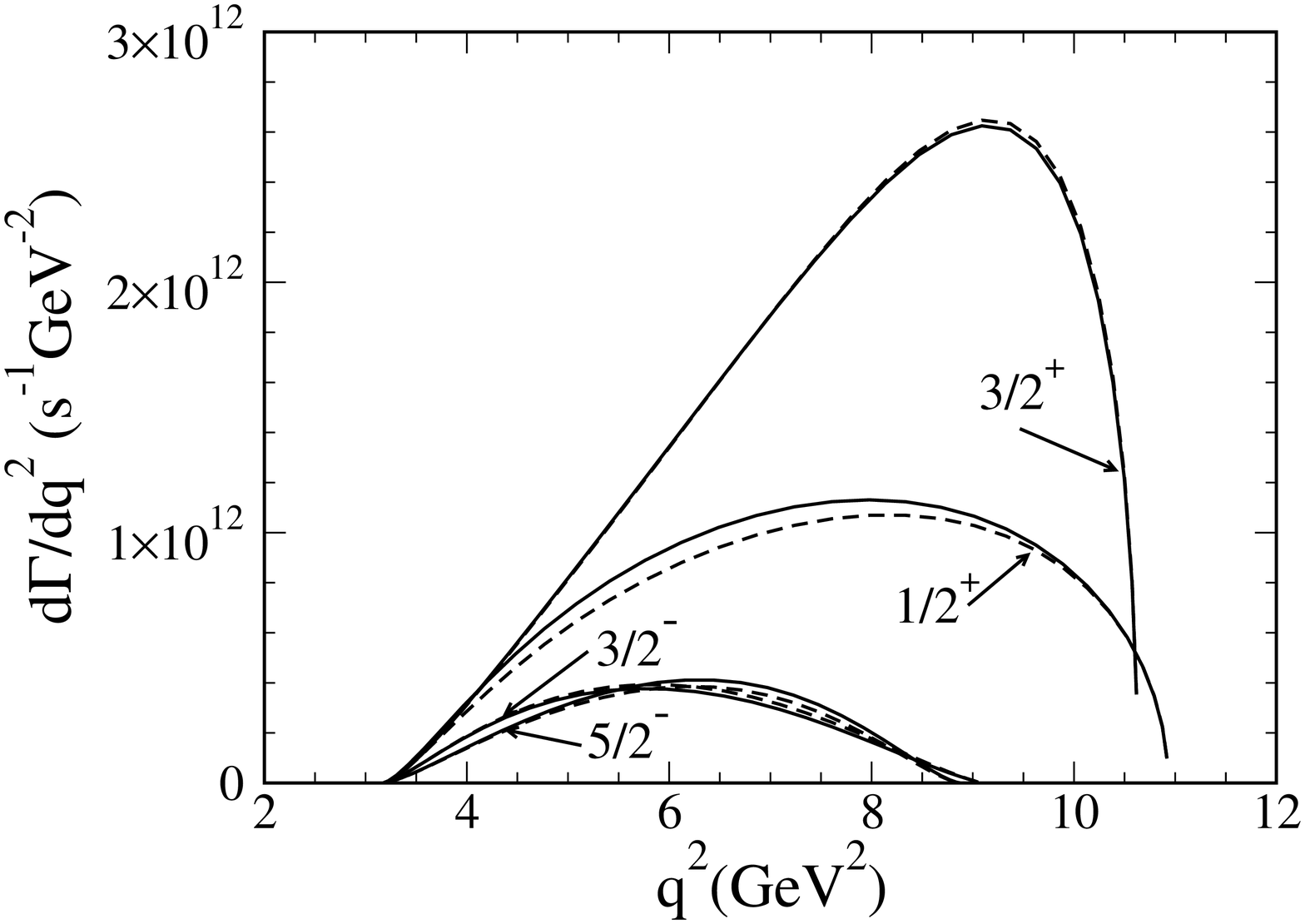,width=9.5cm}}
\caption{The differential decay rates for selected
$\Omega_b\to\Omega_c^{(*)}$ transitions, in the various models
that we use. The panels on the left arise from the two versions of the
harmonic oscillator model, while those on the right are from the
Sturmian models. The upper panels are for
$\Omega_b\to\Omega_c^{(*)}\ell\bar\nu_\ell$, where $\ell$ is $e^-$ or
$\mu^-$. The lower panels are for
$\Omega_b\to\Omega_c^{(*)}\tau\bar\nu_\tau$. The curves are for final
states with $J^P=3/2^+$, $1/2^+$, $3/2^-$ and $5/2^-$. In each panel, the solid 
curves arise from  
the semi-relativistic version of the model, while the dashed curves arise from 
the non-relativistic version. \label{decayrate7}}
\end{figure}

The differential decay rates $d\Gamma/dq^2$ obtained in the
four models for different final states in
$\ob\rightarrow\omcs\ell\bar{\nu}_\ell$, with $\ell= e^-,\mu^-$, are
shown in the upper panels of Figure~\ref{decayrate7}. For these rates,
we use $|V_{cb}|=0.041$. In these figures, we only show the
differential rates for the dominant decays to the two elastic
channels, with $J^P=3/2^+$ and $J^P=1/2^+$, and for two orbital
excitations, the states with $J^P=3/2^-$ and $5/2^-$. We have also
examined the differential decay rates to the $1/2^-$, $1/2^-_1$, and
$3/2^-_1$ orbitally excited states, as well as to the radially excited
states $1/2^+_1$ and $3/2^+_1$ (notations defined in
Section~\ref{wfcomponents}). We have found that the branching fraction
for the radially excited states (not shown in Table~\ref{rateob})
are small, whereas the branching fraction for the decays to the
orbitally excited states are not insignificant, as shown in
Table~\ref{rateob}. The lower panels of Figure~\ref{decayrate7} show
the differential decay rates of $\ob$ decaying to the same $\omc$
final states as in the upper panels, but with a $\tau$ lepton in the final 
state.

In Table~\ref{rateob} we show the integrated decay rates obtained for
the selected final states in the four quark models we use. The first
part of this table shows the rate with a vanishing lepton mass, while
the second part shows the rate when the final lepton is a $\tau$. The
last two rows of the first part of the table present the total decay
rate and the ratio of the elastic to the total semileptonic
decay rate. The integrated rates for the elastic decay modes $(1/2^+,
3/2^+)$ obtained in all models are similar. However, the two Sturmian
models predict somewhat smaller rates for decays into the inelastic
$\omc$ channels. As a result the branching fraction for the elastic
decay mode is smaller in the HO models than in the Sturmian models. If
we consider the two HO models alone, the predicted elastic branching
fraction is 49.5$\pm1.5\%$. The corresponding prediction from the
Sturmian models is 67.5$\pm0.5\%$. Thus, the two HO models are
consistent with each other, and the two ST models are consistent with
each other, but the HO and ST models are in disagreement. Both sets of
models predict that the elastic decay processes dominate the $\ob$
semileptonic decay but do not saturate it; there is some significant
branching fraction to the inelastic channels.

\begin{center}
\begin{table}[h]
\caption{Integrated decay rates for $\Omega_b\to\omcs$
 in units of $10^{10}s^{-1}$, for a selection of $\Omega_c$ states in the
four models we consider. The upper portion of the table presents the decay 
rates obtained
with massless leptons (electron or muon), while the lower portion
corresponds to the rates with a massive $\tau$ lepton in the final state.
\label{rateob}}
\vspace{5mm}
\begin{tabular}{|l|cc|cc|}
\hline
 &\multicolumn{4}{c|}{$\Omega_b \to \Omega^{(*)}_c\ell^-\bar\nu_\ell$} \\ \hline
Spin & $\Gamma$(HONR) & $\Gamma$(HOSR) 
& $\Gamma$(STNR) & $\Gamma$(STSR) \\ \hline
$3/2^+$ & $1.68$ &  $1.93$ & 2.01& 2.00 \\ \hline
$1/2^+$ & $0.71$ &  $0.94$ & 0.87  & 0.96 \\ \hline
$3/2^-$ & $0.69$ &  $0.71$ &0.43& 0.43\\ \hline
$3/2^-_1$ & $0.44$ &$0.47$&  0.23& 0.22  \\ \hline
$5/2^-$ & $0.70$ &  $0.72$ & 0.46& 0.45\\ \hline
$1/2^-$ & $0.32$ &  $0.33$ &0.20 & 0.20\\ \hline
$1/2^-_1$ & $0.44$ &$0.48$& 0.07 & 0.07\\ \hline
total      &4.98& 5.58   &4.27 & 4.33\\ \hline
$\Gamma_{\Omega_c(1/2^+,3/2^+)}/\Gamma_{\rm total}$&0.48&0.51&0.67&0.68\\ \hline
&\multicolumn{4}{c|} {$\Omega_b\to \Omega^{(*)}_c\tau^-\bar\nu_\tau$}\\ \hline
Spin & $\Gamma$(HONR) & $\Gamma$(HOSR) & $\Gamma$(STNR) & $\Gamma$(STSR)\\ \hline
$3/2^+$ &$0.81$ &  $0.89$   & 0.89 &0.88\\ \hline
$1/2^+$ &$0.42$ &  $0.49$ &  0.45&0.49\\ \hline
$3/2^-$ & $0.15$ & $0.14$ &  0.10 &0.11\\ \hline 
$3/2^-_1$ &$0.09$ &$0.10$ &   0.07& 0.07\\ \hline
$5/2^-$ & $0.14$ & $0.14$ &   0.11& 0.10 \\ \hline
$1/2^-$ &$0.07$ &  $0.06$ &  0.05&0.05 \\ \hline
$1/2^-_1$& $0.08$ &$0.09$  &  0.02&0.02\\ \hline
total & 1.76& 1.91     & 1.69&1.72\\ \hline
\end{tabular}
\end{table}
\end{center}

We may also compare our decay rates with HQET predictions. As
discussed in Section~\ref{hqetpredictions}, there are a number of
pairs of degenerate states, such as states with $J^P=(1/2^+, 3/2^+)$,
$J^P=(1/2^-, 3/2^-)$ and $J^P=(3/2^-_1, 5/2^-)$. In the heavy quark
limit, the ratio between the rates of the ground state heavy baryon
decaying to the states in the first two degenerate pairs is expected
to be $1:2$, and for the third pair it is $2:3$.  In other words, for
example, we expect the rate for $\ob(1/2^+) \to
\omc(3/2^+)$ to be twice as large as the rate for
$\ob(1/2^+)\to\omc(1/2^+)$. 
The expected pattern of the rates coming from the HQET prediction is
reflected in all of our quark model calculations, as can be seen in
Table~\ref{rateob}. Departures from these predictions are due to
$1/m_Q$ and $1/m_q$ corrections, and the fact that, for instance, the
$3/2^-$ state shown in the table is not exactly the state in the
$(1/2^-, 3/2^-)$ multiplet, but contains some admixture of the $3/2^-$
state from the $(3/2^-_1, 5/2^-)$ multiplet.

In Table~\ref{rateob} we have shown only the rates to the
$\omcs$ states which have a significant branching fraction. However, we have calculated
the rates of $\ob$ decaying to the majority of states listed in 
Eq.~\ref{wavefunctionomega}, and have found that the rates for states not
shown in Table~\ref{rateob} are small (of the order of 1\% of the total rate
that we have obtained).

\subsubsection{$\Omega_c (\Omega_b)\rightarrow \Xi^{(*)}$}

\begin{figure}[h]
\centerline{\epsfig{file=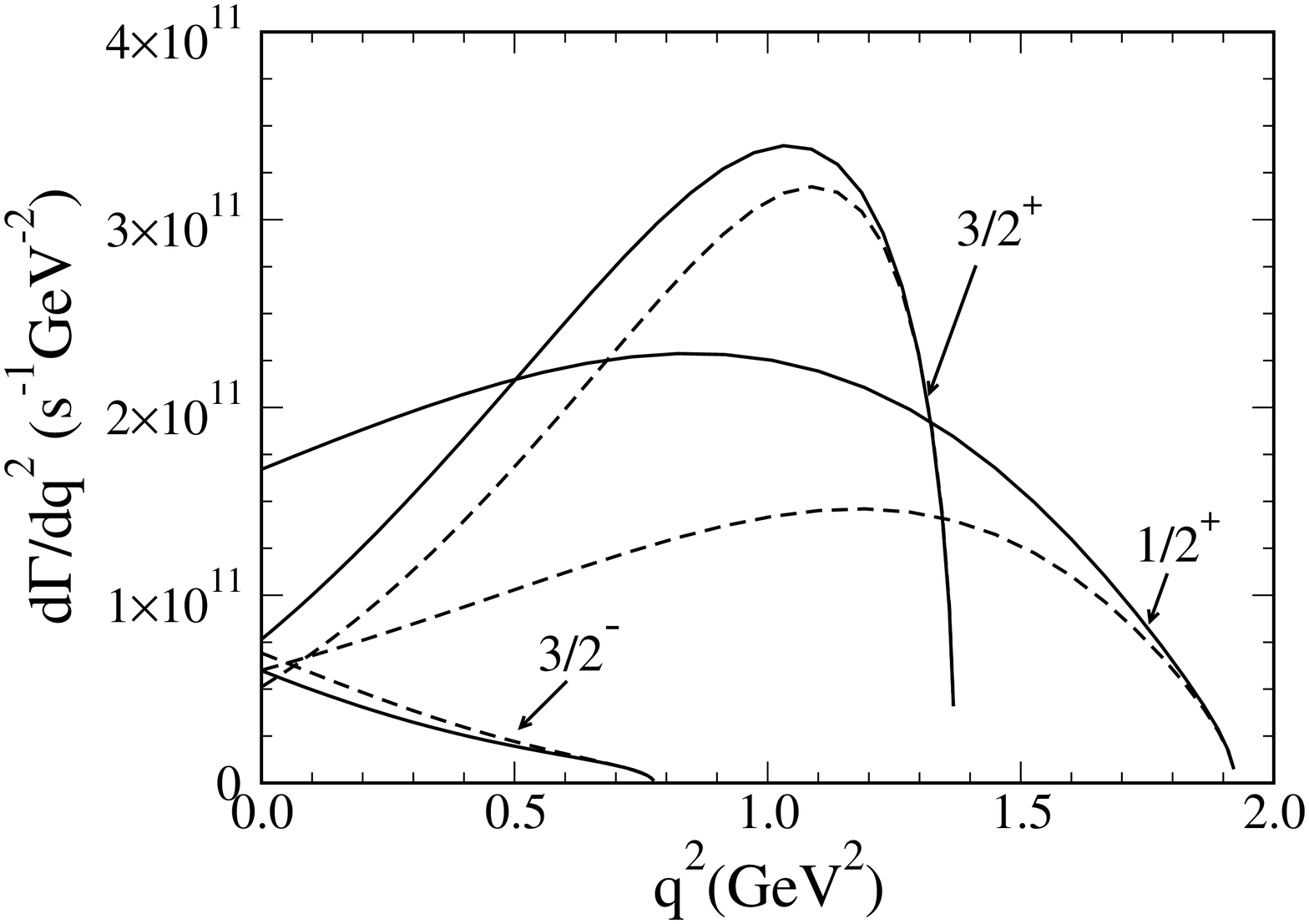,width=9.8cm}\hspace{-0.25in}
\epsfig{file=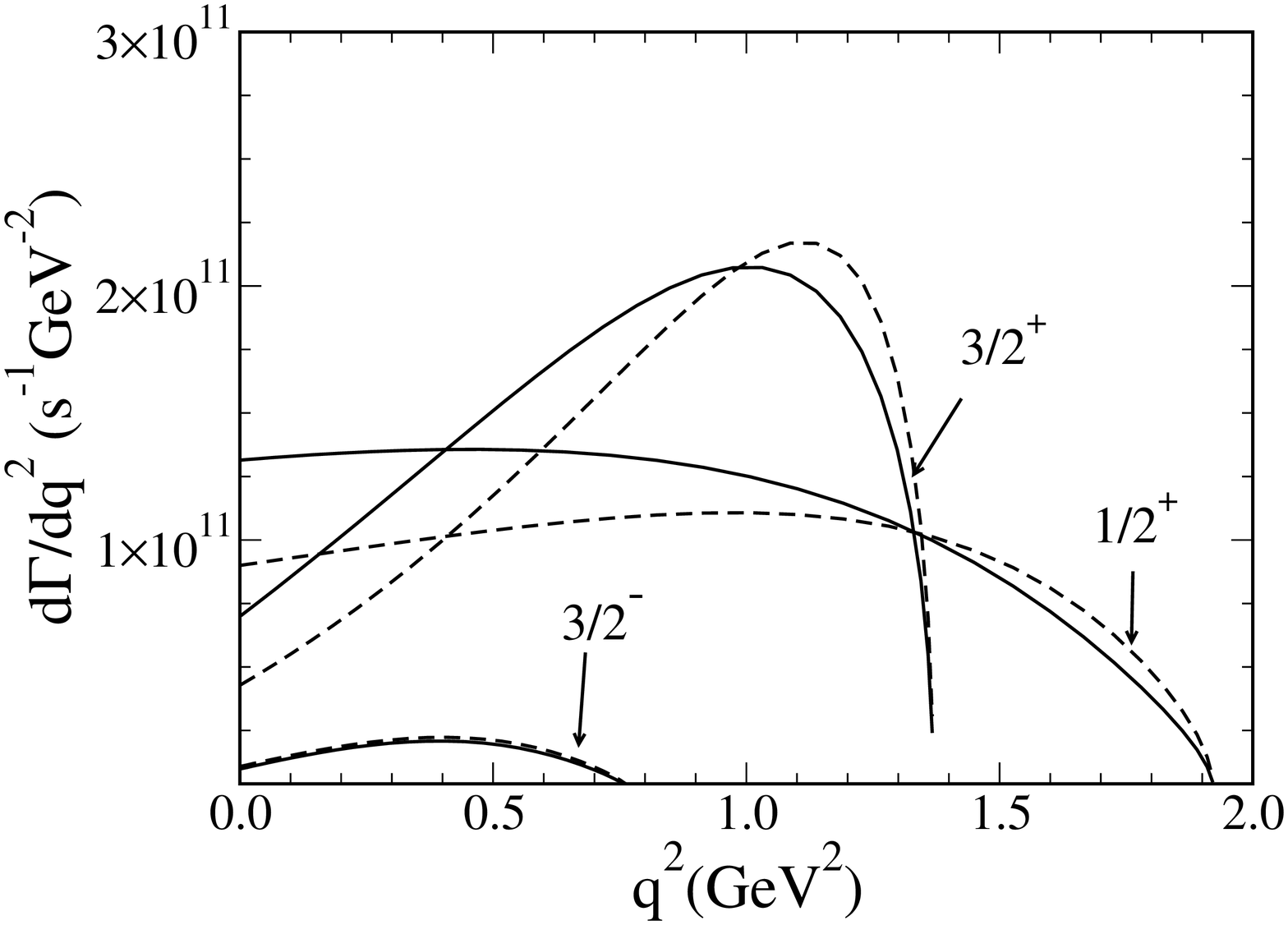,width=9.8cm}}
\caption{The differential decay rates for selected 
$\Omega_c\to\Xi^{(*)}$ transitions, in the various models
that we use. The panel on the left arises from the
harmonic oscillator models, while the one on the right arises from the
Sturmian models. 
They key to the curves is as in Fig. \ref{decayrate7}.\label{decayrate9}}
\end{figure}

The differential decay rates $d\Gamma/dq^2$ for $\omc$ decaying to
the dominant final states of $\Xi^{(*)}$ are shown as functions of
$q^2$ in Figure~\ref{decayrate9}. Here, we use $|V_{cs}|=0.224$. The left panel shows the
differential decay rates obtained in the HONR and HOSR models, and the
right panel shows the results from the STNR and STSR models. In both
harmonic oscillator and Sturmian models we see significant differences
between the non-relativistic and semi-relativistic
predictions, with the most significant difference occuring between the HOSR and
HONR predictions for the $1/2^+$ final state. Although the different size parameters obtained in
various models play a role in these differences, we should also note
here that this particular decay rate is very sensitive to the value of
$m_\sigma$ ($\sigma= s$ in this decay). Our fitted values for the
strange quark mass in the different models, shown in
Table~\ref{parameter1}, show significant variation. We have seen
similar differences between NR and SR model differential decay rates
for the $\Lambda_c \to n$ decay mode presented in I.
\begin{center}
\begin{table}[h]
\caption{Integrated decay  rates for $\Omega_c\to\Xi^{(*)}$ in units of
$10^{10}s^{-1}$, for selected $\Xi$ states in the four models we consider.
\label{rateoc}}
\vspace{5mm}\begin{tabular}{|l|ll|ll|}
\hline
Spin & $\Gamma$(HONR) & $\Gamma$(HOSR)
 & $\Gamma$(STNR) & $\Gamma$(STSR) \\ \hline
$3/2^+$ & $0.51$ &  $0.59$ & 0.36& 0.39 \\ \hline
$1/2^+$ & $0.39$ &  $0.65$ & 0.34  & 0.39\\ \hline
$3/2^-$ & $0.05$ &  $0.05$ &  0.03& 0.02\\ \hline
$5/2^-$ & $0.03$ &  $0.03$ &  0.04& 0.03 \\ \hline
$1/2^-$ & $0.04$ &  $0.04$ & 0.05 & 0.05\\ \hline
$1/2^-_1$ & $0.08$ &$0.11$ &0.02 & 0.02 \\ \hline
total & 1.10& 1.47 & 0.84 & 0.90  \\ \hline
$\Gamma_{\Xi(1/2^+,3/2^+)}/\Gamma_{\rm total}$ & 0.82 & 0.84& 0.83 & 0.87 \\ \hline
\end{tabular}
\end{table}
\end{center}
The integrated decay rates for $\omc$ decaying to various states of
$\Xi$ are shown in Table~\ref{rateoc} in the four different models we
use. The last two rows give the total decay rate and the branching
fraction to the `elastic decay' channels. The two elastic decay modes
dominate the decay, with a small fraction of $\omc$ decays going to
the orbitally excited $\Xi^{(*)}$ states. The rates obtained in the
Sturmian models are smaller than those obtained in the harmonic oscillator
models. Nevertheless, the predicted elastic fraction does not depend
strongly on the models we use.

\begin{center}
\begin{table}
\caption{Integrated decay rates for $\Omega_b\to\Xi^{(*)}$ in
units of $10^{12}|V_{ub}|^2s^{-1}$, for a selection of $\Xi$ states in the four
models we consider. The last row of the top portion of the table shows the 
`elastic fraction' (evaluated using the lowest lying $1/2^+$ and $3/2^+$ states) obtained in our
model, where the decays shown in
the table are assumed to saturate the semileptonic decays.
\label{rateobc}}
\vspace{5mm}\begin{tabular}{|l|cc|cc|}
\hline
 &\multicolumn{4}{c|}{$\Omega_b \to \Xi^{(*)}\ell^-\bar\nu_\ell$} \\ \hline
Spin & $\Gamma$(HONR) & $\Gamma$(HOSR)& $\Gamma$(STNR) & $\Gamma$(STSR) \\ \hline
$3/2^+$ & $2.90$ &  $3.74$ &  2.16& 3.60 \\ \hline
$1/2^+$ & $0.82$ &  $1.46$  & 1.09 &1.78\\ \hline
$3/2^-$ &  $2.17$ & $2.73$   &3.80 &3.96 \\ \hline
$3/2^-_1$& $2.25$ & $4.15$ & 2.04& 2.03\\ \hline
$5/2^-$ & $2.15$ &  $3.73$  & 3.27&3.20 \\ \hline
$1/2^-$ & $0.64$ &  $0.87$ & 1.33 & 1.39 \\ \hline
$1/2^-_1$ & $1.15$ &  $1.80$ & 0.35 & 0.37 \\ \hline
total & 12.08& 18.48 & 14.04 & 16.33 \\ \hline
$\Gamma_{\Xi(1/2^+,3/2^+)}/\Gamma_{\rm total}$ & 0.31& 0.28 & 0.23 & 0.33\\ \hline
&\multicolumn{4}{c|} {$\Omega_b\to \Xi^{(*)}\tau^-\bar\nu_\tau$}\\ \hline
$3/2^+$ & $2.42$ &  $3.10$  & 1.76 &2.83  \\ \hline
$1/2^+$ & $0.89$ &  $1.53$  & 1.06 &1.61 \\ \hline
$3/2^-$ & $1.86$ &  $2.32$  &2.88 &2.99 \\ \hline
$3/2^-_1$ &$1.86$ &$3.36$   & 1.39&1.35 \\ \hline
$5/2^-$ & $1.62$ &  $2.75$  & 2.33& 2.25 \\ \hline
$1/2^-$ & $0.59$ &  $0.77$  & 1.04&1.07\\ \hline
$1/2^-_1$ & 1.09 &  $1.67$  & 0.30&0.30\\ \hline
Total& 10.33 & 15.50 & 10.76 & 12.4 \\\hline
\end{tabular}
\end{table}
\end{center}

In Table~\ref{rateobc} the integrated decay rates of $\ob$ to various
states of $\Xi$ are presented, along with the rates for the $\ob \to
\Xi^{(*)}\tau^-\bar\nu_\tau$ in the second part of the table. 
We note that due to the large phase space available, a significant
fraction of $\ob$ is predicted to decay semileptonically to various
excited $\Xi$ states. Apart from a few decay modes, such as
$\Xi(1/2_1^-)$, the integrated decay rates of $\ob$ to the various
states of $\Xi$ vary minimally within the models we use.

\subsubsection{$\Omega_c\rightarrow \Omega^{(*)}$}

Table~\ref{rateoc2} shows the integrated rates for $\omc$ decaying to
$\Omega$ in our two harmonic oscillator models. The decay to the
ground state ($J^P=3/2^+$) almost saturates the decay, providing
91\% of the total decay rate. The lack of phase space
suppresses the decays to 'inelastic' channels.

\begin{center}
\begin{table}[h]
\caption{Integrated decay rates for $\Omega_c\to\Omega^{(*)}$ in
units of $10^{11}s^{-1}$, for a selection of $\Omega$ states in the two HO
models we consider. The last row shows the `elastic fraction' obtained
in our model, assuming the decays shown in the table saturate the
semileptonic decays.
\label{rateoc2}}
\vspace{5mm}
\begin{tabular}{|l|ll|}
\hline
Spin & $\Gamma$(HONR) & $\Gamma$(HOSR)\\ \hline
$3/2^+$ & $4.74$ &  $5.39$\\ \hline
$3/2^-$ & $0.31$ &  $0.33$\\ \hline
$1/2^-$ & $0.18$ &  $0.19$\\ \hline
total & 5.23& 5.91\\ \hline
$\Gamma_{\Omega(3/2^+)}/\Gamma_{\rm total}$ & 0.91& 0.91\\ \hline
\end{tabular}
\end{table}
\end{center}

\subsection{Form Factors and Decay Rates: $\Lambda_Q$, revisited}
\label{lrevisit}
One consequence of the modified variational procedure mentioned in
Section~\ref{parametersmw}, coupled with fitting to the decay rate for
$\lcle$, is that most of the decay rates for $\Lambda_Q$ have been
modified. In this section, we briefly discuss the new results for
$\ll_Q$ decays, and compare them with those presented in I. We also
discuss the effects of $\kappa$ on these decay rates.

\begin{center}
\begin{table}[h]
\caption{Integrated decay rates for $\Lambda_c\to\Lambda^{(*)}$ in
units of $10^{11}s^{-1}$, in the models we consider. The row
labeled `total' is obtained by adding all calculated exclusive decay
rates. The fourth and fifth columns show the $\kappa$-affected
integrated rates in the HO models. Integrated rates from our previous
work are also shown in parentheses for comparison. The last row 
shows the
`elastic fraction' estimated in our models, assuming that the exclusive
channels calculated saturate the semileptonic decays of the $\Lambda_c$. In each
row, the numbers in parentheses are from I \cite{MWS}.
\label{ratelc}}
\vspace{5mm}\begin{tabular}{|l||l|l||l|l||l|l||l|}
\hline
$J^P$ & $\Gamma$(HONR) & $\Gamma$(HOSR) & $\Gamma$(HONR$\kappa)$ & $\Gamma$(HOSR$\kappa)$
& $\Gamma$(STNR) & $\Gamma$(STSR) & Expt.~\protect{\cite{CLEO}}\\ \hline
$1/2^+$  & $1.80$ (2.10) &  $2.16$  (2.36) & $1.07$ &  $1.31$& 1.32 (0.79)  & 1.44 (1.11) & 1.05$\pm$0.35\\ \hline
total & 2.00 (2.36)& 2.37 (2.73) & 1.22& 1.47 &1.53 (0.97) & 1.60 (1.31) & - \\ \hline
$\Gamma_\Lambda/\Gamma_{\rm total}$  & 0.90 (0.89) & 0.91 (0.86) & 0.88 & 0.89& 0.86 (0.81) & 0.90 (0.85) & 1.0
(assumed) \\ \hline
\end{tabular}
\end{table}
\end{center}

Table~\ref{ratelc} shows the integrated decay rates for $\Lambda_c\to\Lambda^{(*)}$
obtained in the various models we use. The second and third columns of
this table show the rates we obtain in the two harmonic oscillator
models. Note that the rates obtained in the HO models are somewhat
improved from those presented in I, but are still larger than the
experimentally measured rate by about a factor of two, even though the
rate is included in the fit. The fourth and fifth columns show the
rates we obtain when $\kappa$ is included in the harmonic oscillator
form factors. The second row in the table shows the rates we obtain
with the modified variational procedure, while the numbers in parentheses in
that row are the corresponding rates presented in I. In the Sturmian models, the elastic 
rate is
larger than in I, resulting in elastic fractions that are similar to
those obtained in the HO models in the present work. As a result, the
overall model dependence of the elastic fraction has decreased
significantly. One of the reasons for this improvement is that with
the new minimization scheme both harmonic oscillator and Sturmian
model wave functions for different baryon states have similar mixing
coefficients, as can be seen in Tables~\ref{mixingwf2} and
\ref{mixingwf}.

The third row presents the total rate we estimate for
$\Lambda_c^+\to\Lambda^{(*)}e^+\nu$ (the numbers in parentheses are from I). 
Examination of these rates indicates
that the model dependence in the $\lcle$ rate has decreased compared
with I. We have obtained an average value of
$1.28^{+0.16}_{-0.21}\times10^{11}s^{-1}$ for the $\lcle$ rate in the
Sturmian models and the two $\kappa$-modified HO models, consistent
with the value reported by the CLEO collaboration~\cite{CLEO}. In the
last row of Table~\ref{ratelc} the `elastic fractions' obtained in
the various models are shown. Our new calculations predict smaller
inelastic branching fractions (9 to 14\%) than reported in I (11 to
19\%). The elastic fraction, averaged over the Sturmian models and the
$\kappa$-modified harmonic-oscillator models, is found to be
$0.88\pm0.02$.

\begin{center}
\begin{table}[h]
\caption{Rates for $\Lambda_b\to \Lambda_c^{(*)}$
decays in units of $10^{10}s^{-1}$. The fourth and fifth columns show
the $\kappa$-affected integrated rates in the HO models. Integrated rates from our previous
work are also shown in parentheses for comparison. The row
labeled `total' is obtained by adding all calculated exclusive decay
rates, while the row with the branching fractions assumes that the
exclusive channels calculated saturate the semileptonic decays of the
$\Lambda_b$. The elastic fraction reported by the DELPHI 
collaboration~\cite{delphi}
(third row of numbers, eighth column) is actually
$\Gamma(\Lambda_b\to\Lambda_c\ell\bar\nu_\ell)/[
\Gamma(\Lambda_b\to\Lambda_c\ell\bar\nu_\ell)+
\Gamma(\Lambda_b\to\Lambda_c\pi\pi\ell\bar\nu_\ell)]$. The errors on both
numbers from DELPHI are statistical and systematic, respectively.
\label{ratelb}}
\vspace{5mm}\begin{tabular}{|l||c|c||c|c||c|c||c|}
\hline
$J^P$ & $\Gamma$(HONR) & $\Gamma$(HOSR)& $\Gamma$(HONR$\kappa)$ & $\Gamma$(HOSR$\kappa)$
 & $\Gamma$(STNR) & $\Gamma$(STSR)&$\Gamma_{\rm DELPHI}$~\cite{delphi}  \\ \hline
$1/2^+$ & $3.50$ (4.60) &  $4.14$ (5.39) & $1.81$ &  $2.29$ & 2.39 (1.47)  & 2.52 (2.00) & $4.07^{+0.90+1.30}_{-0.65-0.98}$\\ \hline
Total ($\Lambda_c^{(*)}\ell^-\bar\nu_\ell$) & 4.83 (5.95) & 5.21 (6.82) & 2.49 & 2.89& 3.31 (2.36) &
3.04 (2.88) &-\\ \hline
$\Gamma_{\Lambda_c}/\Gamma_{\rm total}$  & 0.72 (0.76) & 0.79 (0.79) & 0.73 & 0.79& 0.72 (0.62) & 0.83 (0.69) &
$0.47^{+0.10+0.07}_{-0.08-0.06}$ \\ \hline

\end{tabular}
\end{table}
\end{center}

The integrated rates for $\Lambda_b\to \Lambda_c^{(*)}$ are shown in Table~\ref{ratelb}. Here
we again present both the results from this work and from I for
comparison. We note that the HO and ST models now provide predictions
for both the elastic $\lblc$ decay and the total decay rates that are
more consistent with each other. It is intriguing that, for the decays
of both the $\Lambda_b$ and $\Lambda_c$, the results obtained in the
$\kappa$-modified HO models are consistent with the results obtained
in the ST models.

The elastic fraction for the $\Lambda_b\to \Lambda_c^{(*)}$ decay obtained in the various
models is shown in the last row of Table~\ref{ratelb}. In the
present work this fraction is slightly less model-dependent than in I. 
The average value for the $\lble$ rate obtained using the
two Sturmian models and the two $\kappa$-modified HO models is
$2.25^{+0.27}_{-0.44} \times10^{10}s^{-1}$. These four models predict
an average value of $0.77\pm0.05$ for the elastic fraction.

\begin{center}
\begin{table}[h]
\caption{Decay rates of 
$\Lambda_b\to N^{(*)+}\ell^-\bar\nu_\ell$ in units of
$10^{12}|V_{ub}|^2s^{-1}$. The fourth, fifth, eighth and
ninth columns show the $\kappa$-affected integrated rates in the HO
models. The row labeled `total' is obtained by summing all the
exclusive decay rates shown in the table. Integrated rates from our previous
work are also shown in parentheses for comparison. The first four columns of
numbers are for decays with a muon or electron in the final state,
while the last four columns are for decays with a $\tau$ in the final
state. Also shown are the rates for $\Lambda_c\to
N^{(*)0}\ell^+\nu_\ell$ in units of $10^{10}s^{-1}$.
\label{ratelbp}}
\vspace{5mm}\begin{tabular}{|l||c|c|c|c||c|c|c|c|}
\hline
 &\multicolumn{4}{c|}{$\Lambda_b\to N^{(*)+}\ell^-\bar\nu_\ell$}&
\multicolumn{4}{c|}
 {$\Lambda_b\to N^{(*)+}\tau^-\bar\nu_\tau$} \\ \hline
$J^P$ & $\Gamma$(HONR) & $\Gamma$(HOSR)& $\Gamma$(HONR$\kappa)$ & $\Gamma($HOSR$\kappa)$ 
& $\Gamma$(HONR) & $\Gamma$(HOSR) & $\Gamma$(HONR$\kappa)$ & $\Gamma$(HOSR$\kappa)$ \\
\hline
$1/2^+$ &  $2.38$ (4.55) &  $3.53$ (7.55)& $0.90$ &  $1.19$ & 2.12 (4.01) & 3.11 (6.55)& $0.81$ &  $1.06$\\\hline
$1/2^+_1$& $0.44$ &  0.85&   $0.49$ &  $0.65$ & 0.35 & 0.64 &$0.12$ &  $0.22$\\ \hline
$1/2^-$ &  $1.18$ &  1.14&   $0.45$ &  $0.42$ & 0.95 & 0.85 & $0.38$ &  $0.35$   \\ \hline
$3/2^-$ &  $0.87$ & $0.73$&  $0.20$ &  $0.15$ & 0.64 & 0.46& $0.17$ &  $0.06$ \\ \hline
$3/2^+$ &  $0.52$ &  0.28&   $0.11$ &  $0.06$ & 0.36 & 0.17& $0.09$ &  $0.05$   \\ \hline
$5/2^+$ &  $1.29$ & $0.44$&  $0.18$ &  $0.06$ & 0.69 & 0.18& $0.13$ &  $0.04$ \\ \hline
Total & 6.68 (12.25) & 6.97 (21.31) &  2.33 & 2.53 & 5.11 (9.00) &5.41 (15.53) & 1.70 & 1.78 \\ \hline
 &\multicolumn{4}{c|}{$\Lambda_c\to N^{(*)0}\ell^+\nu_\ell$} &-&-&- &-\\\hline
 $1/2^+$ & $0.68$ &  $0.84$ &0.31 &0.38&- &-&- &-\\ \hline
$1/2^-$ & $0.03$ &  $0.03$  &0.02& 0.02 &- &-&- &-\\ \hline
\end{tabular}
\end{table}
\end{center}

In Table~\ref{ratelbp} we present the integrated rates for
$\Lambda_b\to N^{(*)+}\ell^-\bar\nu_\ell$ obtained in the two HO
models. Integrated rates for $\Lambda_b\to N^{(*)+}\tau^-\bar\nu_\tau$
are also shown in this table. From the integrated rates for all
allowed decay modes to ground state nucleons and N$^*$ states shown in
Table~\ref{ratelbp}, we predict that a significant
fraction of these decays are to excited nucleons, due in part to the
large phase space available. We also present $\Lambda_b\to p$ rates
and total decay rates from I in parentheses for comparison. We note that, as in the
two previous decay processes, the modified variational procedure leads
to results in the two HO models that are more consistent with each other.

\section{Conclusion and outlook}

A constituent quark model calculation of semileptonic decays of
$\Omega_Q$, which has several intriguing features, has been described
in the preceding sections. Analytic results for the form factors for
the decays to ground states and excited states with a selection of quantum
numbers are evaluated, and compared to HQET predictions. For
$\Omega_b\to\Omega_c$ transitions, the HQET relationships among the
form factors detailed in Appendix~\ref{hqetformfactors} in the
$(1/2^+,\,3/2^+)$, $(1/2^-,\,3/2^-)$, $(3/2^-,\,5/2^-)$, and
$(3/2^+,\,5/2^+)$ doublets are respected in our predicted form factors,
at leading order.

These form factors depend on the size parameters of the initial and
final baryon wave functions, and so a fit is made to the spectrum of
the states treated here. Two model Hamiltonians are used, with either
a nonrelativistic or semi-relativistic kinetic energy term, and with
Coulomb and spin-spin contact interactions, as we did before in our
work on $\Lambda_Q$ decay. For the present work we have modified our
variational procedure from that used in I, and we have incorporated
the $\lcle$ rate in the fit to the spectrum. As a result our size
parameters are somewhat different from those shown in I, and our four
baryon spectra are improved from those shown in I.

As in I, the wave functions are expanded in either a harmonic
oscillator or Sturmian basis, up to second-order polynomials, and our
numerical results for form factors and rates are calculated using the
resulting mixed wave functions. Four sets of predictions are made for
form factors and rates, with wave functions, size parameters and
mixing coefficients arising from fits using both the non-relativistic
and semi-relativistic Hamiltonians, and using the two different
bases. The variation among these predictions can be used to assess the
model dependence in the results we obtain.

At present there exist few quantitative measurements of the
semileptonic decay rates of $\Omega_Q$. The CLEO
collaboration~\cite{cleooc} has measured $B(\Omega_c^0 \to
\Omega^- e^+\nu)\cdot\sigma(e^+e^- \to \Omega_cX)= 42.2\pm14.1\pm11.9$
fb.  In addition the ARGUS~\cite{argus} and BELLE~\cite{belle}
collaborations have reported the $\omc^0 \to \Omega^-$ decay as
`seen'. However, the branching fraction for this decay has not yet
been measured, which means that a comparison of our model predictions
with experiment is not possible at this time. Nevertheless, it is
instructive to examine our predictions for the various integrated
rates for $\ob$ and $\omc$ decays in light of HQET predictions for
these decays. There are a number of pairs of degenerate states, such
as states with $J^P=(1/2^+, 3/2^+)$, $J^P=(1/2^-, 3/2^-)$,
$J^P=(3/2^-_1, 5/2^-)$, for which HQET predicts the relationships
between the form factors as well as the ratio of rates of the
degenerate pairs. According to this prediction, the rate for the $\ob
\to
\omc(3/2^+)$  is expected to be twice as large as the rate for $\ob \to
\omc(1/2^+)$, and the rate for the $J^P=3/2^-$ final state  is also expected to
be twice as large as the rate for the $J^P=1/2^-$ final state
decay. In the same way the ratio of the rates for the final states of
the $(3/2^-_1,\,5/2^-)$ doublet is expected to be 2:3. The states we
obtained in our spectral fits are not exactly the HQET states, but
appear to be close approximations to such states, because these
relationships among the rates of the degenerate pairs are well
respected in all our quark model calculations for the semileptonic
decays of $\ob$.

Our predictions for the semileptonic elastic branching fraction of
$\Omega_Q$ vary minimally within the models we use. We obtain an
average value of (84$\pm$ 2\%) for the elastic fraction of
$\Omega_c\to \Xi^{(*)}$. For the case of $\Omega_c\to \Omega^{(*)}$,
91\% of all decays are elastic. The two HO models give results that
are consistent with each other (48\%, 51\%) for the elastic fraction
of $\ob \to \omcs$. The two Sturmian models also give results that are
consistent with each other (67\% and 68\%), but which are somewhat
different from the fraction predicted by the HO models.

We have recalculated the $\ll_Q$ rates in all four models. Following
ISGW, we have also modified the results of the HO models by including
a parameter $\kappa$ in the exponential factor that multiplies all HO
form factors.  As a result the model dependence of the $\ll_Q$ rates
and the elastic rates have been decreased significantly. We have
obtained an average value of $1.28^{+0.16}_{-0.21}
\times10^{11}s^{-1}$ for the $\lcle$ decay rate in the Sturmian and
$\kappa$-modified HO models, and the elastic fraction for the decays
of the $\Lambda_c$ is $0.88\pm0.02$. For $\lble$ our models predict an
average value of $2.25^{+0.27}_{-0.44}\times10^{10}s^{-1}$, and the
predicted elastic fraction is $0.77\pm0.05$.

As noted in I, the work presented in this manuscript can be extended
in a number of directions. We can apply our model to the description
of the semileptonic decays of the light baryons, although these are
already successfully described by Cabibbo theory. Essentially all
experimentally accessible observables for these decays have been
measured, and it will be interesting to see if our model, constructed
with no special reference to chiral symmetry or current algebra, can
describe the results of these measurements.

We have not examined the predictions of our model for the many
polarization observables which can, in principle, be measured in
semileptonic decays. In addition, the rare decays of heavy baryons,
such as $\ob\to\Omega$ can easily be treated in the framework that we
have developed. Such processes, along with their meson analogs, are
used in searches for physics beyond the Standard Model. However, the
interpretation of the measured rates depend strongly on estimates of
the form factors involved (in much the same way that extraction of CKM
matrix elements depends on the form factors that describe semileptonic
decays). Finally, if factorization in some form is valid, the
semileptonic form factors calculated in the manuscript may also be
useful in the description of nonleptonic weak decays.

It may also be possible to systematically improve the quark model used
in the present calculation. An obvious first step is the
implementation of full symmetrization of the spatial wave functions in
the Sturmian basis, which would allow calculation of results for
decays to final state nucleons in this basis.

We also plan to modify and expand all our baryon spectrum codes to
make predictions for baryons containing quarks with three different
masses. One advantage of this modification is that it will allow us to
examine the semileptonic decays of $\Xi_Q$. This study will be
interesting, as some of the $\Xi_Q$ states have an antisymmetric
($\Lambda_Q$-like) light diquark, while some have a symmetric
($\Omega_Q$-like) light diquark~\cite{pdgxi}.

\section*{Acknowledgements}
Helpful discussions with Dr. J. Pickarewicz and Dr. L. Reina are gratefully acknowledged. 
This research is supported by the U.S. Department of Energy under contracts 
DE-FG02-92ER40750 (M.P. and S.C.) and DE-FG05-94ER40832 (W.R.).

Notice: This manuscript has been authored by The Southeastern Universities Research
Association, Inc. under Contract No. DE-AC05-84150 with the U.S. Department of Energy.
The United States Government retains and the publisher, by accepting the article for
publication, acknowledges that the United States Government retains a non-exclusive,
paid-up, irrevocable, world wide license to publish or reproduce the published form of
this manuscript, or allow others to do so, for United States Government purposes.

\appendix

\section{Wave Functions}
\label{appendixA}

The wave function components for states that we consider are shown here. These
components are valid for both the $\Omega_Q$ and $\Xi$ states that are treated in this
manuscript.  For $J^P=1/2^+$, wave functions are expanded as 
\begin{eqnarray}
\Psi^{\Omega_Q}_{1/2^+M}&=&\phi_{\Omega_Q}\left(\vphantom{\sum_i}
\left[\eta_1^{\Omega_Q}\psi_{000000}({\bf p}_\rho, {\bf 
p}_\lambda)
+\eta_2^{\Omega_Q}\psi_{001000}({\bf p}_\rho, {\bf 
p}_\lambda)
+\eta_3^{\Omega_Q}\psi_{000010}({\bf p}_\rho, 
{\bf p}_\lambda)\right]\chi_{1/2}^\lambda(M)\right.\nonumber \\
&+&\eta_4^{\Omega_Q}\psi_{000101}({\bf p}_\rho, 
{\bf p}_\lambda)\chi_{1/2}^\rho(M)
+\eta_5^{\Omega_Q}\left[\psi_{1M_L0101}({\bf p}_\rho, {\bf 
p}_\lambda)\chi_{1/2}^\rho(M-M_L)\right]^{1/2}_M\nonumber  \\
&+&\eta_6^{\Omega_Q}\left[\psi_{2M_L0200}({\bf p}_\rho, {\bf 
p}_\lambda)\chi_{3/2}^S(M-M_L)\right]^{1/2}_M\nonumber  \\
&+&\left.\eta_7^{\Omega_Q}\left[\psi_{2M_L0002}({\bf p}_\rho, {\bf p}_\lambda)
\chi_{3/2}^S(M-M_L)\right]^{1/2}_M\right),
\end{eqnarray}

For $\Omega_Q$ states with $J^P=3/2^+$, the expansion is
\begin{eqnarray}
\Psi^{\Omega_Q}_{3/2^+M}&=&\phi_{\Omega_Q}\left(\vphantom{\sum_i}
\left[\eta_1^{\Omega_Q}\psi_{000000}({\bf p}_\rho, {\bf p}_\lambda)
+\eta_2^{\Omega_Q}\psi_{001000}({\bf p}_\rho, {\bf p}_\lambda)
+\eta_3^{\Omega_Q}\psi_{000010}
({\bf p}_\rho, {\bf p}_\lambda)\right]\chi_{3/2}^S(M)\right.\nonumber\\
&+&\eta_4^{\Omega_Q}\left[\psi_{1M_L0101}
({\bf p}_\rho, {\bf p}_\lambda)\chi_{1/2}^\rho(M-M_L)\right]^{3/2}_M\nonumber\\
&+&\eta_5^{\Omega_Q}\left[\psi_{2M_L0200}
({\bf p}_\rho, {\bf p}_\lambda)\chi_{3/2}^S(M-M_L)\right]^{3/2}_M\nonumber\\
&+&\eta_6^{\Omega_Q}\left[\psi_{2M_L0200}
({\bf p}_\rho, {\bf p}_\lambda)\chi_{1/2}^\lambda(M-M_L)\right]^{3/2}_M\nonumber\\
&+&\eta_7^{\Omega_Q}\left[\psi_{2M_L0101}
({\bf p}_\rho, {\bf p}_\lambda)\chi_{1/2}^\rho(M-M_L)\right]^{3/2}_M\nonumber\\
&+&\eta_8^{\Omega_Q}\left[\psi_{2M_L0002}
({\bf p}_\rho, {\bf p}_\lambda)\chi_{3/2}^S(M-M_L)\right]^{3/2}_M\nonumber\\
&+&\left.\eta_9^{\Omega_Q}\left[\psi_{2M_L0002}
({\bf p}_\rho, {\bf p}_\lambda)\chi_{1/2}^\lambda(M-M_L)\right]^{3/2}_M\right),
\end{eqnarray}
where $\left[\psi_{LM_Ln_\rho\ell_\rho n_\lambda\ell_\lambda}({\bf p}_\rho,
{\bf  p}_\lambda)\chi_{S}(M-M_L)\right]^J_M$ is a shorthand notation that
denotes the Clebsch-Gordan sum $\sum_{M_L}\< JM|LM_L, SM-M_L\>
\psi_{LM_Ln_\rho\ell_\rho n_\lambda\ell_\lambda}({\bf p}_\rho, {\bf 
p}_\lambda)\chi_{S}(M-M_L)$.

For $J^P=1/2^-$ and $3/2^-$, the expansion is
\begin{eqnarray} \label{nthm}
\Psi^{\Omega_Q}_{J^-M}&=&\phi_{\Omega_Q}\left(
\eta_1^{\Omega_Q}\left[\psi_{1M_L0100}({\bf p}_\rho, {\bf
p}_\lambda)\chi_{1/2}^\rho(M-M_L)\right]^J_M\right.\nonumber\\
&&+\eta_2^{\Omega_Q}\left[\psi_{1M_L0001}({\bf p}_\rho, 
{\bf p}_\lambda)
\chi_{3/2}^S(M-M_L)\right]^J_M\nonumber\\
&&+\left.\eta_3^{\Omega_Q}\left[\psi_{1M_L0001}({\bf 
p}_\rho, 
{\bf p}_\lambda)\chi_{1/2}^\lambda(M-M_L)\right]^J_M\right),
\end{eqnarray}
where $J$ can take the values 1/2 or 3/2.

For $J^P=5/2^+$, the expansion is 
\begin{eqnarray}
\Psi^{\Omega_Q}_{5/2^+M}&=&\phi_{\Omega_Q}\left(\vphantom{\sum_i}
\eta_1^{\Omega_Q}\psi_{2M_L0101}({\bf p}_\rho, {\bf 
p}_\lambda)\chi_{1/2}^\rho(M-M_L)
+\eta_2^{\Omega_Q}\left[\psi_{2M_L0200}
({\bf p}_\rho, {\bf p}_\lambda)\chi_{3/2}^S(M-M_L)\right]^{5/2}_M\right.\nonumber\\
&+&\eta_3^{\Omega_Q}\left[\psi_{2M_L0200}
({\bf p}_\rho, {\bf p}_\lambda)\chi_{1/2}^\lambda(M-M_L)\right]^{5/2}_M
+\eta_4^{\Omega_Q}\left[\psi_{2M_L0002}
({\bf p}_\rho, {\bf p}_\lambda)\chi_{3/2}^S(M-M_L)\right]^{5/2}_M\nonumber\\
&+&\left.\eta_5^{\Omega_Q}\left[\psi_{2M_L0002}({\bf p}_\rho, 
{\bf p}_\lambda)\chi_{1/2}^\lambda(M-M_L)\right]^{5/2}_M\right)
\end{eqnarray}

For $J^P=5/2^-$, the wave function is
\begin{equation} \label{nfhm}
\Psi^{\Omega_Q}_{J^-M}=\phi_{\Omega_Q}
\left[\psi_{1M_L0001}({\bf p}_\rho, {\bf
p}_\lambda)\chi_{3/2}^S(M-M_L)\right]^{5/2}_M,
\end{equation}

No other states are expected to have significant overlap with the decaying 
ground-state $\Omega_Q$ in the spectator approximation that we use.

\section{Quark Model Form Factors}
\label{appendixC}

In this section, we present the analytic expressions we obtained for the form
factors, assuming single component wave functions. We list
 both harmonic oscillator and Sturmian form factors
together, ordered by the spin and parity of the daughter baryon, beginning with the
ground state. We place the form factors for each spin and parity in a separate
subsection.

For most of the $J^P$ we treat, there is usually more than one example of the
state: the exception is $5/2^-$, for which there is a single state up to $N=2$.
We therefore distinguish among the different states with the same $J^P$ by
presenting their quark model quantum numbers, in the notation 
$|n_S,L,n_\rho,\ell_\rho,n_\lambda,\ell_\lambda>$. Here $n_S$ takes values between 1
and 3, with 1 denoting total quark spin 1/2, with spin wave function of type
$\chi_\rho$, 2 denoting total quark spin 1/2, with spin wave function of type
$\chi_\lambda$, and 3 denoting total quark spin 3/2.
\newpage
\subsection{$1/2^+,|200000>$}
\label{ffhpgs}
\subsubsection{Harmonic Oscillator Form Factors}
\begin{eqnarray}
F_1 &=&  -I_H \left[1 +\frac{m_\sigma}{\alpha^2_{\lambda\lambda'}}\left(\frac{\alpha^2_{\lambda'}}{m_q}
+\frac{\alpha^2_{\lambda}}{m_Q}\right)\right],\nonumber\\
F_2 &=& 2 I_H  \left[1 -\frac{m_\sigma}{\alpha^2_{\lambda\lambda'}}\left(\frac{\alpha^2_{\lambda'}}
{2m_q}-\frac{\alpha^2_{\lambda}}{m_Q}\right)\right],\nonumber\\
F_3 &=& 2I_H \left[1 +\frac{m_\sigma}{\alpha^2_{\lambda\lambda'}}\left(\frac{\alpha^2_{\lambda'}}{m_q}
-\frac{\alpha^2_{\lambda}}{2m_Q}\right)\right],\nonumber\\ 
G_1 &=& - I_H,\nonumber\\
G_2 &=& I_H\frac{m_\sigma}{m_q}\frac{\alpha^2_{\lambda'}}{\alpha^2_{\lambda\lambda'}},\nonumber\\
G_3 &=& -I_H \frac{m_\sigma}{m_Q}\frac{\alpha^2_{\lambda}}{\alpha^2_{\lambda\lambda'}},\nonumber
\end{eqnarray}
where
\begin{eqnarray}
I_H =\frac{1}{3}\left(\frac{\alpha_\lambda\alpha_{\lambda'}}{\alpha_{\lambda\lambda'}^2}\right)^{3/2}
\exp\left( -\frac{3m^2_\sigma}{2m^2_{\Omega_q}}\frac{p^2}{\alpha_{\lambda\lambda'}^2}\right).\nonumber
\end{eqnarray}
$\all^2 =\frac{1}{2}(\al^2 + \alp^2)$, and $m_\sigma$ is the mass of
the light quark.
\subsubsection{Sturmian Form Factors}
\begin{eqnarray}
F_1 &=& - I_S \left[1 +\frac{m_\sigma}{\beta_{\lambda\lambda'}}\left(\frac{\beta
_{\lambda'}}{m_q}+ \frac{\beta_\lambda}{m_Q}\right)\right],\nonumber\\
F_2 &=& 2 I_S  \left[1 -\frac{m_\sigma}{\beta_{\lambda\lambda'}}\left(\frac{\beta
_{\lambda'}}{2m_q}-\frac{\beta_\lambda}{m_Q}\right)\right],\nonumber\\
F_3 &=& 2 I_S  \left[1 +\frac{m_\sigma}{\beta_{\lambda\lambda'}}\left(\frac{\beta
_{\lambda'}}{m_q}- \frac{\beta_\lambda}{2m_Q}\right)\right],\nonumber\\ 
G_1 &=&  -I_S,\nonumber\\
G_2 &=&  I_S\frac{m_\sigma\beta_{\lambda'}}{m_q\beta_{\lambda\lambda'}},\nonumber\\
G_3 &=& -I_S\frac{m_\sigma\beta_{\lambda}}{m_Q\beta_{\lambda\lambda'}},\nonumber
\end{eqnarray}
where
\begin{eqnarray}
I_S = \frac{1}{3}\frac{\left(\frac{\beta_\lambda\beta_{\lambda'}}
{\beta_{\lambda\lambda'}^2}\right)^{3/2}}{\left[1+  \frac{3}{2}\frac{m^2_\sigma}{m^2_{\Omega_q}}
\frac{p^2}{\beta_{\lambda\lambda'}^2}\right]^2},\nonumber
\end{eqnarray}
and $\beta_{\lambda\lambda'} =\frac{1}{2}(\beta_\lambda + \beta_\lambda')$.
\subsection{$1/2^+_1,|200010>$}
\label{ffhpa}
\subsubsection{Harmonic Oscillator Form Factors}
\begin{eqnarray}
F_1 &=&  -I_H \frac{1}{2\alpha_{\lambda}\alpha_{\lambda'}}\left[\left(\alpha^2_{\lambda}
-\alpha^2_{\lambda'}\right)
+\frac{m_\sigma}{3\alpha^2_{\lambda\lambda'}}\left(\frac{\alpha^2_{\lambda'}(7\alpha^2_{\lambda} 
-3\alpha^2_{\lambda'})}{m_q}+\frac{\alpha^2_{\lambda}(3\alpha^2_{\lambda} -7\alpha^2_{\lambda'})}{m_Q}
\right)\right],\nonumber\\
F_2 &=& I_H \frac{1}{\alpha_{\lambda}\alpha_{\lambda'}}\left[\left(\alpha^2_{\lambda} -\alpha^2_{\lambda'}\right)
-\frac{m_\sigma}{3\alpha^2_{\lambda\lambda'}}\left(\frac{\alpha^2_{\lambda'}(7\alpha^2_{\lambda} 
-3\alpha^2_{\lambda'})}{2m_q}-\frac{\alpha^2_{\lambda}(3\alpha^2_{\lambda} -7\alpha^2_{\lambda'})}{m_Q}
\right)\right],\nonumber\\
F_3 &=& I_H \frac{1}{\alpha_{\lambda}\alpha_{\lambda'}}\left[\left(\alpha^2_{\lambda} -\alpha^2_{\lambda'}\right)
+\frac{m_\sigma}{3\alpha^2_{\lambda\lambda'}}\left(\frac{\alpha^2_{\lambda'}(7\alpha^2_{\lambda} 
-3\alpha^2_{\lambda'})}{m_q}-\frac{\alpha^2_{\lambda}(3\alpha^2_{\lambda} -7\alpha^2_{\lambda'})}{2m_Q}
\right)\right],\nonumber\\ 
G_1 &=& - I_H\frac{\left(\alpha^2_{\lambda} -\alpha^2_{\lambda'}\right)}{2\alpha_{\lambda}\alpha_{\lambda'}},\nonumber\\
G_2 &=&
I_H\frac{m_\sigma}{6m_q}\frac{\alpha_{\lambda'}(7\alpha^2_{\lambda}-3\alpha^2_{\lambda'})}
{\alpha_\lambda\alpha^2_{\lambda\lambda'}},\nonumber\\
G_3 &=& -I_H
\frac{m_\sigma}{6m_Q}\frac{\alpha_{\lambda}(3\alpha^2_{\lambda}-7\alpha^2_{\lambda'})}
{\alpha_{\lambda'}\alpha^2_{\lambda\lambda'}},\nonumber
\end{eqnarray}
where
\begin{center}
\begin{eqnarray}
I_H =\frac{1}{\sqrt{6}}\left(\frac{\alpha_\lambda\alpha_{\lambda'}}{\alpha_{\lambda\lambda'}^2}\right)^{5/2}
\exp\left( -\frac{3m^2_\sigma}{2m^2_{\Omega_q}}\frac{p^2}{\alpha_{\lambda\lambda'}^2}\right).\nonumber
\end{eqnarray}
\end{center}
\subsubsection{Sturmian Form Factors}
\begin{eqnarray}
F_1 &=&  -I_S \frac{1}{2\beta_{\lambda}\beta_{\lambda'}}\left[(\beta^2_{\lambda} -\beta^2_{\lambda'})
+\frac{2m_\sigma}{3}\left(\frac{\beta_{\lambda'}(5\beta_{\lambda} 
-3\beta_{\lambda'})}{m_q}+\frac{\beta_{\lambda}(3\beta_{\lambda} -5\beta_{\lambda'})}{m_Q}
\right)\right],\nonumber\\
F_2 &=& I_S \frac{1}{\beta_{\lambda}\beta_{\lambda'}}\left[(\beta^2_{\lambda} -\beta^2_{\lambda'})
-\frac{2m_\sigma}{3}\left(\frac{\beta_{\lambda'}(5\beta_{\lambda}-3\beta_{\lambda'})}{2m_q}-
\frac{\beta_{\lambda}(3\beta_{\lambda} -5\beta_{\lambda'})}{m_Q}\right)\right],\nonumber\\
F_3 &=& I_S \frac{1}{\beta_{\lambda}\beta_{\lambda'}}\left[(\beta^2_{\lambda} -\beta^2_{\lambda'})
+\frac{2m_\sigma}{3}\left(\frac{\beta_{\lambda'}(5\beta_{\lambda} -3\beta_{\lambda'})}{m_q}-
\frac{\beta_{\lambda}(3\beta_{\lambda} -5\beta_{\lambda'})}{2m_Q}\right)\right],\nonumber\\ 
G_1 &=& - I_S\frac{(\beta^2_{\lambda} -\beta^2_{\lambda'})}{2\beta_{\lambda}\beta_{\lambda'}},\nonumber\\
G_2 &=& I_S\frac{m_\sigma}{3m_q}\frac{(5\beta_{\lambda} -3\beta_{\lambda'})}{\beta_{\lambda}},\nonumber\\
G_3 &=& -I_S\frac{m_\sigma}{3m_Q}\frac{(3\beta_{\lambda}-5\beta_{\lambda'})}{\beta_{\lambda'}},\nonumber
\end{eqnarray}
where
\begin{eqnarray}
I_S = \frac{1}{2\sqrt{3}}\frac{\left(\frac{\beta_\lambda\beta_{\lambda'}}
{\beta_{\lambda\lambda'}^2}\right)^{5/2}}{\left[1+  \frac{3}{2}\frac{m^2_\sigma}{m^2_{\Omega_q}}
\frac{p^2}{\beta_{\lambda\lambda'}^2}\right]^3}.\nonumber
\end{eqnarray}
\subsection{$1/2^+_2, |320002>$}
\label{ffhpb}
\subsubsection{Harmonic Oscillator Form Factors}
\begin{eqnarray}
F_1 &=&  -I_H\ms \left(\frac{1}{m_q}-\frac{1}{m_Q}\right),\nonumber\\
F_2 &=&I_H\frac{\ms}{2} \left(\frac{1}{m_q}-\frac{1}{m_Q}\right),\nonumber\\
F_3 &=& I_H\frac{\ms}{2} \left(\frac{1}{m_q}-\frac{1}{m_Q}\right),\nonumber\\
G_1 &=&0,\nonumber\\
G_2&=&I_H\frac{\ms}{\al}\left[\frac{18m_\sigma}{5\alpha_\lambda}-
\frac{\al}{2}\left(\frac{4}{m_q}+\frac{3}{m_Q}\right)\right],\nonumber\\
G_3 &=&-I_H \frac{\ms}{\al} \left(\frac{18m_\sigma}{5\alpha_\lambda}+
\frac{\al}{2m_Q}\right),\nonumber
\end{eqnarray}
where
\begin{eqnarray}
I_H =-\frac{\sqrt{10}}{3\sqrt{3}}\left(\frac{\alpha_\lambda\alpha_{\lambda'}}
{\alpha_{\lambda\lambda'}^2}\right)^{7/2}\exp\left( -\frac{3m^2_\sigma}{2m^2_{\Omega_q}}
\frac{p^2}{\alpha_{\lambda\lambda'}^2}\right).\nonumber
\end{eqnarray}
\subsubsection{Sturmian Form Factors}
\begin{eqnarray}
F_1 &=&-I_S\frac{\ms\bll}{2\bl}\left(\frac{1}{m_q}-\frac{1}{m_Q}\right),\nonumber\\
F_2 &=&I_S\frac{\ms\bll}{4\bl}\left(\frac{1}{m_q}-\frac{1}{m_Q}\right),\nonumber\\
F_3 &=&I_S\frac{\ms\bll}{4\bl}\left(\frac{1}{m_q}-\frac{1}{m_Q}\right) ,\nonumber\\
G_1 &=& 0,\nonumber\\
G_2 &=&I_S\frac{\ms}{\bl}\left[\frac{27\ms}{5\bl}-\frac{\bll}{4}\left(\frac{4}{m_q}+\frac{3}{m_Q}\right)\right],\nonumber\\
G_3 &=&-I_S\frac{\ms}{\bl} \left(\frac{27\ms}{5\bl}+\frac{\bll}{4m_Q}\right),\nonumber
\end{eqnarray}
where
\begin{eqnarray}
I_S = -\frac{4\sqrt{5}}{9}\frac{\left(\frac{\beta_\lambda\beta_{\lambda'}}
{\beta_{\lambda\lambda'}^2}\right)^{7/2}}{\left[1+  \frac{3}{2}\frac{m^2_\sigma}{m^2_{\Omega_q}}
\frac{p^2}{\beta_{\lambda\lambda'}^2}\right]^4}.\nonumber
\end{eqnarray}
\subsection{$1/2^-, |210001>$}
\subsubsection{Harmonic Oscillator Form Factors}
\begin{eqnarray}
F_1 &=&-I_H \frac{\alpha_{\lambda}}{6}\left(\frac{1}{m_q}+\frac{5}{m_Q}\right),\nonumber\\
F_2 &=&I_H\left\{\frac{2\ms}{\al}\left[1-\frac{m_\sigma}{\all^2}\left(\frac{2\alp^2}{m_q}-
\frac{\al^2}{m_Q}\right)\right]-\frac{\alpha_{\lambda}}{6m_q}\right\},\nonumber\\
F_3 &=&  I_H\frac{4\ms}{\al} \left[1+\frac{m_\sigma}{\all^2}\left(\frac{\alpha^2_{\lambda'}}{m_q}
-\frac{\alpha^2_{\lambda}}{2m_Q}\right)\right],\nonumber\\
G_1 &=&  I_H\left(\frac{2m_\sigma}{\alpha_{\lambda}}-\frac{5\alpha_{\lambda}}{6m_Q}\right),\nonumber\\
G_2 &=&  -I_H \left[\frac{2m_\sigma}{\alpha_{\lambda}}-\alpha_\lambda\left(\frac{1}{2m_q}
 +\frac{2}{3m_Q} \right)\right],\nonumber\\
G_3 &=& I_H \frac{2\alpha_{\lambda}}{3m_Q},\nonumber
\end{eqnarray}
where
\begin{eqnarray}
I_H = -\frac{1}{3}\left(\frac{\alpha_\lambda\alpha_{\lambda'}}{\alpha_{\lambda\lambda'}^2}\right)^{5/2}
\exp\left( -\frac{3 m^2_\sigma}{2m^2_{\Omega_q}}\frac{p^2}{\alpha_{\lambda\lambda'}^2}\right).\nonumber
\end{eqnarray}
\subsubsection{Sturmian Form Factors}
\begin{eqnarray}
F_1 &=& -I_S\frac{\beta_{\lambda\lambda'}}{12}\left(\frac{1}{m_q}
+\frac{5}{m_Q}\right),\nonumber\\
F_2 &=& I_S \left\{\frac{2\ms}{\bl}\left[1-\frac{m_\sigma}{\bll}\left(\frac{2\beta
_{\lambda'}}{m_q}-\frac{\beta_\lambda}{m_Q}\right)\right]-\frac{\beta_{\lambda\lambda'}}{12m_q}\right\}
,\nonumber\\
F_3 &=& I_S\frac{4\ms}{\bl}\left[1+\frac{m_\sigma}{\bll}\left(\frac{\beta_{\lambda'}}{m_q}- 
\frac{\beta_\lambda}{2m_Q}\right)\right],\nonumber\\
G_1 &=& I_S\left(\frac{2m_\sigma}{\beta_{\lambda}}-\frac{5\beta_{\lambda\lambda'}}{12m_Q}\right),\nonumber\\
G_2 &=& -I_S\left[\frac{2m_\sigma}{\beta_{\lambda}}-\beta_{\lambda\lambda'}\left(\frac{1}{4m_q}+
\frac{1}{3m_Q}\right)\right],\nonumber\\
G_3 &=& I_S \frac{\beta_{\lambda\lambda'}}{3m_Q},\nonumber
\end{eqnarray}
where
\begin{eqnarray}
I_S = -\frac{\sqrt{2}}{3}\frac{\left(\frac{\beta_\lambda\beta_{\lambda'}}
{\beta_{\lambda\lambda'}^2}\right)^{5/2}}{\left[1+  \frac{3}{2}\frac{m^2_\sigma}{m^2_{\Omega_q}}
\frac{p^2}{\beta_{\lambda\lambda'}^2}\right]^3}.\nonumber
\end{eqnarray}
\subsection{$3/2^-, |210001>$}
\subsubsection{Harmonic Oscillator Form Factors}
\begin{eqnarray}
F_1 &=&  -I_H \frac{m_\sigma}{\alpha_\lambda} \left[1
+\frac{m_\sigma}{\alpha^2_{\lambda\lambda'}}
\left(\frac{\alpha^2_{\lambda'}}{m_q}+\frac{\alpha^2_{\lambda}}{m_Q}\right)\right],\nonumber\\
F_2 &=&
I_H\left\{\frac{2\ms}{\al}\left[1-\frac{\ms}{2\all^2}\left(\frac{\alp^2}{m_q}-\frac{2\al^2}{m_Q}\right)
\right]-\frac{\al}{3m_q}\right\},\nonumber\\
F_3 &=&  I_H\left\{\frac{2\ms}{\al}\left[1+\frac{\ms}{2\all^2}\left(\frac{2\alpha^2_
{\lambda'}}{m_q}-\frac{\alpha^2_{\lambda}}{m_Q}\right)\right]-\frac{\al}{6}\left(\frac{2}{m_q}+
\frac{1}{m_Q}\right)\right\},\nonumber\\
F_4 &=& I_H \frac{\al}{3}\left(\frac{2}{m_q}+\frac{1}{m_Q}\right),\nonumber\\
G_1 &=&-I_H\left(\frac{m_\sigma}{\alpha_\lambda}+\frac{\alpha_\lambda}{6m_Q}\right),\nonumber\\
G_2 &=& I_H \frac{m_\sigma^2}{m_q}\frac{\alpha^2_{\lambda'}}{\alpha^2
_{\lambda\lambda'}\alpha_\lambda},\nonumber\\
G_3 &=& -I_H\frac{\alpha_{\lambda}}{m_Q}\left(\frac
{m_\sigma^2}{\alpha^2_{\lambda\lambda'}}-\frac{1}{6}\right),\,\,\,\,\,\,\,\,\,\,\,\,\,\,
G_4 = - I_H \frac{\alpha_\lambda}{3m_Q},\nonumber\\
I_H &=&-\frac{1}{\sqrt{3}}\left(\frac{\alpha_\lambda\alpha_{\lambda'}}
{\alpha_{\lambda\lambda'}^2}\right)^{5/2}\exp\left( -\frac{3m^2_\sigma}{2m^2_{\Omega_q}}
\frac{p^2}{\alpha_{\lambda\lambda'}^2}\right).\nonumber
\end{eqnarray}
\subsubsection{Sturmian Form Factors}
\begin{eqnarray}
F_1 &=&  -I_S \frac{m_\sigma}{\beta_\lambda} \left[1+\frac{m_\sigma}{\beta_
{\lambda\lambda'}}\left(\frac{\beta_{\lambda'}}{m_q}+\frac{\beta_{\lambda}}
{m_Q}\right)\right],\nonumber\\
F_2 &=&I_S\left\{\frac{2\ms}{\bl}\left[1-\frac{\ms}{\bll}\left(\frac{\beta_{\lambda'}}{2m_q}-
\frac{\beta_{\lambda}}{m_Q}\right)\right]-\frac{\bll}{6m_q}\right\},\nonumber\\
F_3 &=&I_S\left\{\frac{2\ms}{\bl}\left[1+\frac{\ms}{\bll}\left(\frac{\beta_{\lambda'}}{m_q}-
\frac{\beta_{\lambda}}{2m_Q}\right)\right]-\frac{\bll}{12}\left(\frac{2}{m_q}+\frac{1}{m_Q}\right)\right\},\nonumber\\
F_4 &=& I_S \frac{\bll}{6}\left(\frac{2}{m_q}+\frac{1}{m_Q}\right),\nonumber\\
G_1 &=& - I_S\left(\frac{m_\sigma}{\beta_\lambda}+\frac{\beta_{\lambda\lambda'}}
{12m_Q}\right),\nonumber\\
G_2 &=&I_S\frac{m^2_\sigma}{m_q}\frac{\beta_{\lambda'}}{\beta_\lambda\beta_{\lambda\lambda'}},\nonumber\\
G_3 &=& -I_S\frac{\beta_{\lambda\lambda'}}{m_Q}\left(\frac{m^2_\sigma}{\beta_{\lambda\lambda'}^2}-\frac{1}{12}\right),\,\,\,\,\,\,\,\,\,\,\,\,\,\,
G_4 =  - I_S\frac{\beta_{\lambda\lambda'}}{6m_Q} ,\nonumber\\
I_S &=& -\sqrt{\frac{2}{3}}\frac{\left(\frac{\beta_\lambda\beta_{\lambda'}}
{\beta_{\lambda\lambda'}^2}\right)^{5/2}}{\left[1+  \frac{3}{2}\frac{m^2_\sigma}{m^2_{\Omega_q}}
\frac{p^2}{\beta_{\lambda\lambda'}^2}\right]^3}.\nonumber
\end{eqnarray}
\pagebreak
\subsection{$1/2^-_1,|310001>$}
\subsubsection{Harmonic Oscillator Form Factors}
\begin{eqnarray}
F_1 &=&-I_H \frac{\alpha_{\lambda}}{3}\left(\frac{1}{m_q}-\frac{1}{m_Q}\right),\nonumber\\
F_2 &=& I_H\left\{\frac{\ms}{\al}\left[1+\frac{\ms}{\all^2}\left(\frac{\alpha^2_{\lambda'}}{m_q}
+\frac{\alpha^2_{\lambda}}{m_Q}\right)\right]-\frac{\al}{3m_Q}\right\},\nonumber\\
F_3 &=& - I_H \frac{\ms}{\al}\left[1+\frac{\ms}{\all^2}\left(\frac{\alpha^2_{\lambda'}}{m_q}+
\frac{\alpha^2_{\lambda}}{m_Q}\right)\right],\nonumber\\
G_1 &=&  -I_H\left(\frac{2m_\sigma}{\alpha_{\lambda}}-\frac{\alpha_{\lambda}}{3m_Q}\right),\nonumber\\
G_2 &=&  -I_H \left[\frac{m_\sigma}{\alpha_{\lambda}}-\frac{\alpha_{\lambda}}{3}\left(
\frac{3}{m_q}+\frac{1}{m_Q}\right)\right],\nonumber\\
G_3 &=& I_H \left(\frac{3m_\sigma}{\alpha_{\lambda}}+\frac{\alpha_{\lambda}}{3m_Q}\right),\nonumber
\end{eqnarray}
where
\begin{eqnarray}
I_H = -\frac{\sqrt{2}}{3}\left(\frac{\alpha_\lambda\alpha_{\lambda'}}{\alpha_{\lambda\lambda'}^2}\right)^{5/2}
\exp\left( -\frac{3 m^2_\sigma}{2m^2_{\Omega_q}}\frac{p^2}{\alpha_{\lambda\lambda'}^2}\right).\nonumber
\end{eqnarray}
\subsubsection{Sturmian Form Factors}
\begin{eqnarray}
F_1 &=& -I_S\frac{\beta_{\lambda\lambda'}}{6}\left(\frac{1}{m_q}
-\frac{1}{m_Q}\right),\nonumber\\
F_2 &=& I_S\left\{\frac{\ms}{\bl}\left[1+\frac{\ms}{\bll}\left(\frac{\beta_{\lambda'}}{m_q}
+\frac{\beta_{\lambda}}{m_Q}\right)\right]-\frac{\bll}{6m_q}\right\},\nonumber\\
F_3 &=& -  I_S\frac{\ms}{\bl}\left[1+\frac{m_\sigma}{\bll}\left(\frac{\beta_{\lambda'}}{m_q}+
\frac{\beta_{\lambda}}{m_Q}\right)\right],\nonumber\\
G_1 &=&-I_S\left(\frac{2m_\sigma}{\beta_{\lambda}}-\frac{\beta_{\lambda\lambda'}}{6m_Q}\right),\nonumber\\
G_2 &=&-I_S\left[\frac{m_\sigma}{\beta_{\lambda}}-\frac{\beta_{\lambda\lambda'}}{6}\left(
\frac{3}{m_q}+\frac{1}{m_Q}\right)\right],\nonumber\\
G_3 &=& I_S\left(\frac{3m_\sigma}{\beta_{\lambda}}+\frac{\beta_{\lambda\lambda'}}{6m_Q}\right),\nonumber
\end{eqnarray}
where
\begin{eqnarray}
I_S = -\frac{2}{3}\frac{\left(\frac{\beta_\lambda\beta_{\lambda'}}
{\beta_{\lambda\lambda'}^2}\right)^{5/2}}{\left[1+  \frac{3}{2}\frac{m^2_\sigma}{m^2_{\Omega_q}}
\frac{p^2}{\beta_{\lambda\lambda'}^2}\right]^3}.\nonumber
\end{eqnarray}
\subsection{$3/2^-_1, |310001>$}
\subsubsection{Harmonic Oscillator Form Factors}
\begin{eqnarray}
F_1 &=&  I_H \frac{2m_\sigma}{\alpha_\lambda} \left[1+\frac{m_\sigma}{\alpha^2_{\lambda\lambda'}}
\left(\frac{\alpha^2_{\lambda'}}{m_q}+\frac{\alpha^2_{\lambda}}{m_Q}\right)\right],\nonumber\\
F_2 &=&I_H\left\{\frac{2\ms}{\al}\left[1+\frac{\ms}{\all^2}\left(\frac{\alpha^2_{\lambda'}}{m_q}
+\frac{\alpha^2_{\lambda}}{m_Q}\right)\right]-\frac{5\al}{6m_q}\right\},\nonumber\\
F_3 &=& - I_H\left\{\frac{4\ms}{\al}\left[1+\frac{\ms}{\all^2}\left(\frac{\alpha^2_
{\lambda'}}{m_q}+\frac{\alpha^2_{\lambda}}{m_Q}\right)\right]+\frac{5\al}{6}\left(\frac{1}{m_q}-
\frac{1}{m_Q}\right)\right\},\nonumber\\
F_4 &=& I_H \frac{5\al}{3}\left(\frac{1}{m_q}-\frac{1}{m_Q}\right),\nonumber\\
G_1 &=&-I_H\left(\frac{4m_\sigma}{\alpha_\lambda}-\frac{\alpha_\lambda}{6m_Q}\right),\,\,\,\,\,\,\,\,\,\,\,\,\,\,
G_2 =-I_H\left(\frac{2\alpha_\lambda}{3m_Q}+\frac{2m_\sigma^2\alpha^2_{\lambda'}}{m_q
\alpha^2_{\lambda\lambda'}\alpha_\lambda}\right),\nonumber\\
G_3 &=& I_H \left[\frac{6\ms}{\al}\left(1+\frac{\al^2\ms}{3\all^2m_Q}\right)-\frac{5\al}{6m_Q}\right],\nonumber\\
G_4 &=& - I_H \left(\frac{12m_\sigma}{\alpha_{\lambda}}-\frac{5\alpha_\lambda}{3m_Q}\right),\nonumber\\
I_H &=&-\frac{1}{\sqrt{15}}\left(\frac{\alpha_\lambda\alpha_{\lambda'}}
{\alpha_{\lambda\lambda'}^2}\right)^{5/2}\exp\left( -\frac{3m^2_\sigma}{2m^2_{\Omega_q}}
\frac{p^2}{\alpha_{\lambda\lambda'}^2}\right).\nonumber
\end{eqnarray}
\subsubsection{Sturmian Form Factors}
\begin{eqnarray}
F_1 &=&  I_S\frac{2m_\sigma}{\beta_\lambda} \left[1+\frac{m_\sigma}{\beta_{\lambda\lambda'}}
\left(\frac{\beta_{\lambda'}}{m_q}+\frac{\beta_{\lambda}}{m_Q}\right)\right],\nonumber\\
F_2 &=&I_S\left\{\frac{2\ms}{\bl}\left[1+\frac{\ms}{\bll}\left(\frac{\beta_{\lambda'}}{m_q}+
\frac{\beta_{\lambda}}{m_Q}\right)\right]-\frac{5\bll}{12m_q}\right\},\nonumber\\
F_3 &=& - I_S\left\{\frac{4\ms}{\bl}\left[1+\frac{m_\sigma}{\bll}
\left(\frac{\beta_{\lambda'}}{m_q}+\frac{\beta_{\lambda}}{m_Q}\right)\right]+\frac{5\bll}{12}\left(
\frac{1}{m_q}-\frac{1}{m_Q}\right)\right\},\nonumber\\
F_4 &=& I_S\frac{5\bll}{6}\left(\frac{1}{m_q}-\frac{1}{m_Q}\right),\nonumber\\
G_1 &=&-I_S\left(\frac{4m_\sigma}{\beta_\lambda}-\frac{\bll}{12m_Q}\right),\,\,\,\,\,\,\,\,\,\,\,\,\,\,
G_2 = -I_S\left(\frac{\beta_{\lambda\lambda'}}{3m_Q}
+\frac{2m_\sigma^2\beta_{\lambda'}}{m_q\beta_{\lambda\lambda'}\beta_\lambda}\right),\nonumber\\
G_3 &=& I_S\left[\frac{6\ms}{\bl}\left(1+\frac{\bl\ms}{3\bll m_Q}\right)-\frac{5\bll}{12m_Q}\right],\nonumber\\
G_4 &=& - I_S\left(\frac{12m_\sigma}{\beta_{\lambda}}-\frac{5\beta_\lambda}{6m_Q}\right),\nonumber\\
I_S &=& -\sqrt{\frac{2}{15}}\frac{\left(\frac{\beta_\lambda\beta_{\lambda'}}
{\beta_{\lambda\lambda'}^2}\right)^{5/2}}{\left[1+  \frac{3}{2}\frac{m^2_\sigma}{m^2_{\Omega_q}}
\frac{p^2}{\beta_{\lambda\lambda'}^2}\right]^3}.\nonumber
\end{eqnarray}

\pagebreak
\subsection{$3/2^+, |300000>$}
\label{ffthpa}

\subsubsection{Harmonic Oscillator Form Factors}
\begin{eqnarray}
F_1 &=&I_H\left[1 +\frac{m_\sigma}{\alpha^2_{\lambda\lambda'}}\left(\frac{\alpha^2_{\lambda'}}{m_q}
+\frac{\alpha^2_{\lambda}}{m_Q}\right)\right],\nonumber\\
F_2 &=& 0,\nonumber\\
F_3 &=&-I_H \left[1 +\frac{m_\sigma}{\alpha^2_{\lambda\lambda'}}\left(\frac{\alpha^2_{\lambda'}}{m_q}
+\frac{\alpha^2_{\lambda}}{m_Q}\right)\right],\nonumber\\ 
F_4 &=& 2 I_H \left[1 +\frac{m_\sigma}{\alpha^2_{\lambda\lambda'}}\left(\frac{\alpha^2_{\lambda'}}{m_q}
+\frac{\alpha^2_{\lambda}}{m_Q}\right)\right],\nonumber\\
G_1 &=& 0,\nonumber\\
G_2 &=& -I_H\frac{m_\sigma}{m_q}\frac{\alpha^2_{\lambda'}}{\alpha^2_{\lambda\lambda'}},\nonumber\\
G_3 &=& I_H\left(1+\frac{m_\sigma}{m_Q}\frac{\alpha^2_{\lambda}}{\alpha^2_{\lambda\lambda'}}
\right),\nonumber\\
G_4 &=&-2I_H,\nonumber
\end{eqnarray}
where
\begin{center}
\begin{eqnarray}
I_H =\frac{1}{\sqrt{3}}\left(\frac{\alpha_\lambda\alpha_{\lambda'}}{\alpha_{\lambda\lambda'}^2}\right)^{3/2}
\exp\left( -\frac{3m^2_\sigma}{2m^2_{\Omega_q}}\frac{p^2}{\alpha_{\lambda\lambda'}^2}\right).\nonumber
\end{eqnarray}
\end{center}
\subsubsection{Sturmian Form Factors}
\begin{eqnarray}
F_1 &=&  I_S\left[1+\frac{m_\sigma}{\beta_{\lambda\lambda'}}\left(\frac{\beta_{\lambda'}}{m_q}
+\frac{\beta_{\lambda}}{m_Q}\right)\right],\nonumber\\
F_2 &=& 0,\nonumber\\
F_3 &=&-I_S\left[1 +\frac{m_\sigma}{\beta_{\lambda\lambda'}}\left(\frac{\beta_{\lambda'}}{m_q}
+\frac{\beta_{\lambda}}{m_Q}\right)\right],\nonumber\\ 
F_4 &=& 2 I_S\left[1 +\frac{m_\sigma}{\beta_{\lambda\lambda'}}\left(\frac{\beta_{\lambda'}}{m_q}
+\frac{\beta_{\lambda}}{m_Q}\right)\right],\nonumber\\
G_1 &=& 0,\nonumber\\
G_2 &=& -I_S\frac{m_\sigma}{m_q}\frac{\beta_{\lambda'}}{\beta_{\lambda\lambda'}}
,\nonumber\\
G_3 &=& I_S\left(1+\frac{m_\sigma}{m_Q}\frac{\beta_{\lambda}}{\beta_{\lambda\lambda'}}
\right),\nonumber\\
G_4 &=&-2I_S,\nonumber
\end{eqnarray}
where
\begin{eqnarray}
I_S = \frac{1}{\sqrt{3}}\frac{\left(\frac{\beta_\lambda\beta_{\lambda'}}
{\beta_{\lambda\lambda'}^2}\right)^{3/2}}{\left[1+  \frac{3}{2}\frac{m^2_\sigma}{m^2_{\Omega_q}}
\frac{p^2}{\beta_{\lambda\lambda'}^2}\right]^2}.\nonumber
\end{eqnarray}
\subsection{$3/2^+_1, |300010>$}
\label{ffthpb}
\subsubsection{Harmonic Oscillator Form Factors}
\begin{eqnarray}
F_1 &=& I_H\frac{1}{\al\alp}\left[3\left(\alpha^2_{\lambda}-\alpha^2_{\lambda'}\right)
+\frac{m_\sigma}{\alpha^2_{\lambda\lambda'}}\left(\frac{\alp^2(7\alpha^2_{\lambda}-
3\alpha^2_{\lambda'})}{m_q}+ \frac{\al^2
(3\alpha^2_{\lambda} -7\alpha^2_{\lambda'})}{m_Q}\right)\right],\nonumber\\
F_2 &=& 0,\nonumber\\
F_3 &=&-I_H\frac{1}{\al\alp}\left[3\left(\alpha^2_{\lambda}-\alpha^2_{\lambda'}\right)
+\frac{m_\sigma}{\alpha^2_{\lambda\lambda'}}\left(\frac{\alp^2(7\alpha^2_{\lambda}-
3\alpha^2_{\lambda'})}{m_q}+ \frac{\al^2
(3\alpha^2_{\lambda} -7\alpha^2_{\lambda'})}{m_Q}\right)\right],\nonumber\\
F_4 &=& I_H\frac{2}{\al\alp}\left[3\left(\alpha^2_{\lambda}-\alpha^2_{\lambda'}\right)
+\frac{m_\sigma}{\alpha^2_{\lambda\lambda'}}\left(\frac{\alp^2(7\alpha^2_{\lambda}-
3\alpha^2_{\lambda'})}{m_q}+ \frac{\al^2(3\alpha^2_{\lambda} -7\alpha^2_{\lambda'})}{m_Q}\right)\right],\nonumber\\
G_1 &=& 0,\nonumber\\
G_2 &=& -I_H\frac{m_\sigma}{m_q}\frac{\alpha_{\lambda'}(7\alpha^2_{\lambda}-3\alpha^2_{\lambda'})}
{\alpha_{\lambda}\alpha^2_{\lambda\lambda'}},\nonumber\\
G_3 &=&I_H\frac{1}{\al\alp}\left[3\left(\alpha^2_{\lambda}-\alpha^2_{\lambda'}\right)
+\frac{m_\sigma}{m_Q}\frac{\al^2(3\alpha^2_{\lambda}-7\alpha^2_{\lambda'})}
{\alpha^2_{\lambda\lambda'}}\right],\nonumber\\
G_4 &=&-I_H\frac{6(\alpha^2_{\lambda}-\alpha^2_{\lambda'})}{\alpha_{\lambda}\alpha_{\lambda}},\nonumber\\
I_H
&=&\frac{1}{6\sqrt{2}}\left(\frac{\alpha_\lambda\alpha_{\lambda'}}{\alpha_{\lambda\lambda'}^{2}}\right)^{5/2}
\exp\left( -\frac{3m^2_\sigma}{2m^2_{\Omega_q}}\frac{p^2}{\alpha_{\lambda\lambda'}^2}\right).\nonumber
\end{eqnarray}
\subsubsection{Sturmian Form Factors}
\begin{eqnarray}
F_1 &=&  I_S\frac{1}{2\bl\blp} \left[3\left(\beta^2_{\lambda}-\beta^2_{\lambda'}\right)
 +2m_\sigma\left(\frac{\blp(5\beta_{\lambda}-3\beta_{\lambda'})}{m_q}+ \frac{\bl(3\beta_{\lambda}
-5\beta_{\lambda'})}{m_Q}\right)\right],\nonumber\\
F_2 &=& 0,\nonumber\\
F_3 &=&-I_S\frac{1}{2\bl\blp}\left[3\left(\beta^2_{\lambda}-\beta^2_{\lambda'}\right)
 +2m_\sigma\left(\frac{\blp(5\beta_{\lambda}-3\beta_{\lambda'})}{m_q}+ \frac{\bl(3\beta_{\lambda}
-5\beta_{\lambda'})}{m_Q}\right)\right],\nonumber\\
F_4 &=& I_S\frac{1}{\bl\blp} \left[3\left(\beta^2_{\lambda}-\beta^2_{\lambda'}\right)
 +2m_\sigma\left(\frac{\blp(5\beta_{\lambda}-3\beta_{\lambda'})}{m_q}+ \frac{\bl(3\beta_{\lambda}
-5\beta_{\lambda'})}{m_Q}\right)\right],\nonumber\\ 
G_1 &=& 0,\nonumber\\
G_2 &=& -I_S\frac{m_\sigma}{m_q}\frac{(5\beta_{\lambda}-3\beta_{\lambda'})}{\beta_{\lambda}},\nonumber\\
G_3 &=&I_S\frac{1}{2\bl\blp}\left[3\left(\beta^2_{\lambda}-\beta^2_{\lambda'}\right)
+\frac{2m_\sigma}{m_Q}\bl(3\beta_{\lambda}-5\beta_{\lambda'})\right],\nonumber\\
G_4 &=&-I_S\frac{3(\beta^2_{\lambda}-\beta^2_{\lambda'})}{\beta_{\lambda}\beta_{\lambda'}},\nonumber\\
I_S &=& \frac{1}{6}\frac{\left(\frac{\beta_\lambda\beta_{\lambda'}}
{\beta_{\lambda\lambda'}^2}\right)^{5/2}}{\left[1+  \frac{3}{2}\frac{m^2_\sigma}{m^2_{\Omega_q}}
\frac{p^2}{\beta_{\lambda\lambda'}^2}\right]^3}.\nonumber
\end{eqnarray}
\subsection{$3/2^+_2, |220002>$}
\subsubsection{Harmonic Oscillator Form Factors}
\begin{eqnarray}
F_1 &=&  I_H\frac{ m_\sigma}{2}\left(\frac{1}{m_q}-\frac{7}{m_Q}\right),\nonumber\\
F_2 &=&I_H\frac{m_\sigma}{\al}\left\{\frac{6\ms}{\al}\left[1-\frac{m_\sigma}{\all^2}\left(\frac{2\alpha^2_{\lambda'}}
{m_q}-\frac{\alpha^2_{\lambda}}{m_Q}\right)\right]-\frac{3\al}{2m_q}\right\},\nonumber\\
F_3 &=& I_H\frac{m_\sigma}{\al}\left\{\frac{12\ms}{\al}\left[1+\frac{m_\sigma}{\all^2}\left(\frac{\alpha^2_{\lambda'}}
{m_q}-\frac{\alpha^2_{\lambda}}{2m_Q}\right)\right]-\al\left(\frac{2}{m_q}+\frac{1}{m_Q}\right)\right\},\nonumber\\
F_4 &=&2 I_H \ms\left(\frac{2}{m_q}+\frac{1}{m_Q}\right),\nonumber\\
G_1 &=&I_H\frac{3\ms}{\al}\left(\frac{2m_\sigma}{\alpha_{\lambda}}-\frac{3\al}{2m_Q}\right),\nonumber\\
G_2 &=&-I_H\frac{m_\sigma}{\al}\left[\frac{6m_\sigma}{\alpha_{\lambda}}-\al\left(\frac{5}{2m_q}+\frac{3}{m_Q}
\right)\right],\nonumber\\
G_3 &=& I_H \frac{4m_\sigma}{m_Q},\,\,\,\,\,\,\,\,\,\,\,\,\,\,
G_4 =  0,\nonumber
\end{eqnarray}
where
\begin{eqnarray}
I_H =\frac{1}{3\sqrt{5}}\left(\frac{\alpha_\lambda\alpha_{\lambda'}}
{\alpha_{\lambda\lambda'}^2}\right)^{7/2}\exp\left( -\frac{3m^2_\sigma}{2m^2_{\Omega_q}}
\frac{p^2}{\alpha_{\lambda\lambda'}^2}\right).\nonumber
\end{eqnarray}
\subsubsection{Sturmian Form Factors}
\begin{eqnarray}
F_1 &=&  I_S \frac{m_\sigma\bll}{2\bl}\left(\frac{1}{m_q}-\frac{7}{m_Q}\right),\nonumber\\
F_2 &=&I_S \frac{m_\sigma}{\bl}\left\{\frac{18\ms}{\bl}\left[1-\frac{m_\sigma}{\bll}
\left(\frac{2\beta_{\lambda'}}{m_q}-\frac{\beta_{\lambda}}{m_Q}\right)\right]-\frac{3\bll}{2m_q}\right\},\nonumber\\
F_3 &=& I_S \frac{m_\sigma}{\bl}\left\{\frac{36\ms}{\bl}\left[1+\frac{m_\sigma}{\beta_{\lambda\lambda'}}
\left(\frac{\beta_{\lambda'}}{m_q}-\frac{\beta_{\lambda}}{2m_Q}\right)\right]-\bll\left(\frac{2}{m_q}+
\frac{1}{m_Q}\right)\right\},\nonumber\\
F_4 &=& I_S \frac{2\ms\bll}{\bl}\left(\frac{2}{m_q}+\frac{1}{m_Q}\right),\nonumber\\
G_1 &=&I_S \frac{9m_\sigma}{\bl}\left(\frac{2m_\sigma}{\bl}
-\frac{\bll}{2m_Q}\right),\nonumber\\
G_2 &=&-I_S\frac{\ms}{\bl}\left[\frac{18\ms}{\bl}-
\bll \left(\frac{5}{2m_q}+\frac{3}{m_Q}\right)\right],\nonumber\\
G_3 &=& I_S  \frac{4m_\sigma\bll}{m_Q\bl},\,\,\,\,\,\,\,\,\,\,\,\,\,\,
G_4 = 0,\nonumber
\end{eqnarray}
where
\begin{eqnarray}
I_S = \frac{\sqrt{2}}{3\sqrt{15}}\frac{\left(\frac{\beta_\lambda\beta_{\lambda'}}
{\beta_{\lambda\lambda'}^2}\right)^{7/2}}{\left[1+  \frac{3}{2}\frac{m^2_\sigma}{m^2_{\Omega_q}}
\frac{p^2}{\beta_{\lambda\lambda'}^2}\right]^4}.\nonumber
\end{eqnarray}
\subsection{$3/2^+_3, |320002>$}
\subsubsection{Harmonic Oscillator Form Factors}
\begin{eqnarray}
F_1 &=&  I_H\ms\left(\frac{1}{m_q}-\frac{1}{m_Q}\right),\nonumber\\
F_2 &=&I_H\frac{3\ms}{\al}\left\{\frac{2\ms}{\al}\left[1+\frac{\ms}{\all^2}\left(\frac{\alpha^2_{\lambda'}}{m_q}
+\frac{\alpha^2_{\lambda}}{m_Q}\right)\right]-\frac{\al}{m_q}\right\},\nonumber\\
F_3 &=& -I_H\frac{2\ms}{\al}\left\{\frac{3\ms}{\al}\left[1+\frac{\ms}{\all^2}\left(\frac{\alpha^2_{\lambda'}}{m_q}
+\frac{\alpha^2_{\lambda}}{m_Q}\right)\right]+2\al\left(\frac{1}{m_q}-\frac{1}{m_Q}\right)\right\},\nonumber\\
F_4 &=&8 I_H\ms \left(\frac{1}{m_q}-\frac{1}{m_Q}\right),\nonumber\\
G_1 &=&- I_H\frac{3\ms}{\al}\left(\frac{4m_\sigma}{\alpha_{\lambda}}-\frac{\al}{m_Q}\right),\nonumber\\
G_2 &=&-I_H\frac{\ms}{\al}\left[\frac{6m_\sigma}{\alpha_{\lambda}}-\al\left(\frac{5}{m_q}+\frac{3}{m_Q}
\right)\right],\nonumber\\
G_3 &=& I_H \frac{\ms}{\al}\left(\frac{18m_\sigma}{\alpha_{\lambda}}-\frac{\al}{m_Q}\right),\,\,\,\,\,\,\,\,\,\,\,\,\,\,
G_4 = 0,\nonumber
\end{eqnarray}
where
\begin{eqnarray}
I_H =\frac{1}{3\sqrt{5}}\left(\frac{\alpha_\lambda\alpha_{\lambda'}}
{\alpha_{\lambda\lambda'}^2}\right)^{7/2}\exp\left( -\frac{3m^2_\sigma}{2m^2_{\Omega_q}}
\frac{p^2}{\alpha_{\lambda\lambda'}^2}\right).\nonumber
\end{eqnarray}
\subsubsection{Sturmian Form Factors}
\begin{eqnarray}
F_1 &=&  I_S \frac{\ms\bll}{\bl}\left(\frac{1}{m_q}-\frac{1}{m_Q}\right),\nonumber\\
F_2 &=&I_S \frac{3m_\sigma}{\bl}\left\{\frac{6\ms}{\bl}\left[1+\frac{m_\sigma}{\beta_{\lambda\lambda'}}
\left(\frac{\beta_{\lambda'}}{m_q}+\frac{\beta_{\lambda}}{m_Q}\right)\right]-
\frac{\bll}{m_q}\right\},\nonumber\\
F_3 &=&- I_S \frac{2m_\sigma}{\bl}\left\{\frac{9\ms}{\bl}\left[1+\frac{m_\sigma}{\beta_{\lambda\lambda'}}
\left(\frac{\beta_{\lambda'}}{m_q}+\frac{\bl}{m_Q}\right)\right]
+2\bll\left(\frac{1}{m_q}-\frac{1}{m_Q}\right)\right\},\nonumber\\
F_4 &=& I_S\frac{8\ms\bll}{\bl} \left(\frac{1}{m_q}-\frac{1}{m_Q}\right),\nonumber\\
G_1 &=&-I_S \frac{3\ms}{\bl}\left(\frac{12\ms}{\bl}
-\frac{\bll}{m_Q}\right),\nonumber\\
G_2 &=&-I_S\frac{\ms}{\bl}\left[\frac{18m_\sigma}{\bl}-
\bll\left(\frac{5}{m_q}+\frac{3}{m_Q}\right)\right],\nonumber\\
G_3 &=& I_S\frac{\ms}{\bl} \left(\frac{54\ms}{\bl}-\frac{\bll}{m_Q}\right),\,\,\,\,\,\,\,\,\,\,\,\,\,\,
G_4 =0,\nonumber
\end{eqnarray}
where
\begin{eqnarray}
I_S = \frac{1}{3}\sqrt{\frac{2}{15}}\frac{\left(\frac{\beta_\lambda\beta_{\lambda'}}
{\beta_{\lambda\lambda'}^2}\right)^{7/2}}{\left[1+  \frac{3}{2}\frac{m^2_\sigma}{m^2_{\Omega_q}}
\frac{p^2}{\beta_{\lambda\lambda'}^2}\right]^4}.\nonumber
\end{eqnarray}
\subsection{$5/2^-, |310001>$}
\label{fffhm}
\subsubsection{Harmonic Oscillator Form Factors}
\begin{eqnarray}
F_1 &=&  I_H \frac{m_\sigma}{\alpha_\lambda} \left[1+\frac{m_\sigma}{\alpha^2_{\lambda\lambda'}}
\left(\frac{\alpha^2_{\lambda'}}{m_q}+\frac{\alpha^2_{\lambda}}{m_Q}\right)\right],\nonumber\\
F_2 &=&0,\nonumber\\
F_3 &=& -I_H\frac{\ms}{\al}\left[1+\frac{m_\sigma}{\alpha^2_{\lambda\lambda'}}
\left(\frac{\alpha^2_{\lambda'}}{m_q}+\frac{\alpha^2_{\lambda}}{m_Q}\right)\right],\nonumber\\
F_4 &=& I_H\frac{2m_\sigma}{\alpha_{\lambda}}\left[1+\frac{m_\sigma}{\alpha^2_{\lambda\lambda'}}
\left(\frac{\alpha^2_{\lambda'}}{m_q}+\frac{\alpha^2_{\lambda}}{m_Q}\right)\right],\nonumber\\
G_1 &=&0,\nonumber\\
G_2 &=&-I_H\frac{m_\sigma^2\alpha^2_{\lambda'}}{m_q\alpha_{\lambda}\alpha^2_{\lambda\lambda'}},\nonumber\\
G_3 &=& I_H \frac{\ms}{\al}\left(1+\frac{m_\sigma\alpha_{\lambda}^2}{m_Q\alpha^2_
{\lambda\lambda'}}\right),\nonumber\\
G_4 &=&-I_H \frac{2m_\sigma}{\alpha_\lambda},\nonumber
\end{eqnarray}
where
\begin{eqnarray}
I_H =-\left(\frac{\alpha_\lambda\alpha_{\lambda'}}
{\alpha_{\lambda\lambda'}^2}\right)^{5/2}\exp\left( -\frac{3m^2_\sigma}{2m^2_{\Omega_q}}
\frac{p^2}{\alpha_{\lambda\lambda'}^2}\right).\nonumber
\end{eqnarray}
\subsubsection{Sturmian Form Factors}
\begin{eqnarray}
F_1 &=&  I_S\frac{m_\sigma}{\beta_\lambda} \left[1+\frac{m_\sigma}{\beta_{\lambda\lambda'}}
\left(\frac{\beta_{\lambda'}}{m_q}+\frac{\beta_{\lambda}}{m_Q}\right)\right],\nonumber\\
F_2 &=&0,\nonumber\\
F_3 &=& -I_S\frac{\ms}{\bl}\left[1+\frac{m_\sigma}{\beta_{\lambda\lambda'}}
\left(\frac{\beta_{\lambda'}}{m_q}+\frac{\beta_{\lambda}}{m_Q}\right)\right],\nonumber\\
F_4 &=& I_S\frac{2\ms}{\bl}\left[1+\frac{m_\sigma}{\beta_{\lambda\lambda'}}
\left(\frac{\beta_{\lambda'}}{m_q}+\frac{\beta_{\lambda}}{m_Q}\right)\right],\nonumber\\
G_1 &=&0,\nonumber\\
G_2 &=&-I_S\frac{m_\sigma^2\beta_{\lambda'}}{m_q\beta_{\lambda}\beta_{\lambda\lambda'}},\nonumber\\
G_3 &=& I_S\frac{\ms}{\bl}\left(1+\frac{m_\sigma\bl}{m_Q\beta_{\lambda\lambda'}}\right),\nonumber\\
G_4 &=& - I_S\frac{2m_\sigma}{\beta_\lambda},\nonumber
\end{eqnarray}
where
\begin{eqnarray}
I_S = -\sqrt{2}\frac{\left(\frac{\beta_\lambda\beta_{\lambda'}}
{\beta_{\lambda\lambda'}^2}\right)^{5/2}}{\left[1+  \frac{3}{2}\frac{m^2_\sigma}{m^2_{\Omega_q}}
\frac{p^2}{\beta_{\lambda\lambda'}^2}\right]^3}.\nonumber
\end{eqnarray}
\subsection{$5/2^+, |220002>$}
\subsubsection{Harmonic Oscillator Form Factors}
\begin{eqnarray}
F_1 &=& - I_H\frac{3m_\sigma^2}{\alpha^2_{\lambda}}\left[1+\frac{m_\sigma}{\alpha^2_{\lambda\lambda'}}
\left(\frac{\alpha^2_{\lambda'}}{m_q}+\frac{\alpha^2_{\lambda}}{m_Q}\right)\right],\nonumber\\
F_2 &=& I_H\frac{2\ms}{\al}\left\{\frac{3\ms}{\al}\left[1-\frac{\ms}{\all^2}\left(
\frac{\alpha^2_{\lambda'}}{2m_q}-\frac{\alpha^2_{\lambda}}{m_Q}\right)\right]-\frac{\al}{m_q}\right\},\nonumber\\
F_3 &=& I_H\frac{2\ms}{\al}\left\{\frac{3\ms}{\al}\left[1+\frac{\ms}{\all^2}\left(
\frac{\alpha^2_{\lambda'}}{m_q}-\frac{\alpha^2_{\lambda}}{2m_Q}\right)\right]-\al\left(
\frac{1}{m_q}+\frac{1}{2m_Q}\right)\right\},\nonumber\\
F_4 &=&2 I_H \ms\left(\frac{2}{m_q}+\frac{1}{m_Q}\right),\nonumber\\
G_1 &=&-I_H\frac{\ms}{\al}\left(\frac{3m_\sigma}{\alpha_{\lambda}}+\frac{\al}{m_Q}\right),\nonumber\\
G_2 &=&I_H\frac{3m_\sigma^3}{m_q}\frac{\alpha^2_{\lambda'}}{\alpha^2_{\lambda\lambda'}\alpha^2_{\lambda}},\nonumber\\
G_3 &=& I_H \frac{m_\sigma}{m_Q}\left(1-\frac{3m_\sigma^2}{\alpha^2_{\lambda\lambda'}}\right),\,\,\,\,\,\,\,\,\,\,\,\,\,\,
G_4 =  -I_H \frac{2m_\sigma}{m_Q},\nonumber
\end{eqnarray}
where
\begin{eqnarray}
I_H =\frac{1}{3\sqrt{2}}\left(\frac{\alpha_\lambda\alpha_{\lambda'}}
{\alpha_{\lambda\lambda'}^2}\right)^{7/2}\exp\left( -\frac{3m^2_\sigma}{2m^2_{\Omega_q}}
\frac{p^2}{\alpha_{\lambda\lambda'}^2}\right).\nonumber
\end{eqnarray}
\subsubsection{Sturmian Form Factors}
\begin{eqnarray}
F_1 &=& - I_S \frac{9m_\sigma^2}{\bl^2}\left[1+\frac{m_\sigma}{\bll}\left(
\frac{\beta_{\lambda'}}{m_q}+\frac{\beta_{\lambda}}{m_Q}\right)\right],\nonumber\\
F_2 &=& I_S\frac{2\ms}{\bl}\left\{\frac{9\ms}{\bl}\left[1-\frac{\ms}{\bll}\left(\frac{\beta_{\lambda'}}{2m_q}
-\frac{\beta_{\lambda}}{m_Q}\right)\right]-\frac{\bll}{m_q}\right\},\nonumber\\
F_3 &=& I_S\frac{2\ms}{\bl}\left\{\frac{9\ms}{\bl}\left[1+\frac{\ms}{\bll}\left(\frac{\beta_{\lambda'}}{m_q}
-\frac{\beta_{\lambda}}{2m_Q}\right)\right]-\bll\left(\frac{1}{m_q}+\frac{1}{2m_Q}\right)\right\},\nonumber\\
F_4 &=&I_S \frac{2\ms\bll}{\bl}\left(\frac{2}{m_q}+\frac{1}{m_Q}\right),\nonumber\\
G_1 &=&-I_S\frac{\ms}{\bl}\left(\frac{9\ms}{\bl}+\frac{\bll}{m_Q}\right),\nonumber\\
G_2 &=&I_S\frac{9m_\sigma^3\beta_{\lambda'}}{m_q\bl^2\bll},\nonumber\\
G_3 &=& I_S \frac{m_\sigma\bll}{m_Q\bl}\left(1-\frac{9m_\sigma^2}{\bll^2}\right),\,\,\,\,\,\,\,\,\,\,\,\,\,\,
G_4 =  -I_S \frac{2m_\sigma\bll}{m_Q\bl},\nonumber
\end{eqnarray}
where
\begin{eqnarray}
I_S = \frac{1}{3\sqrt{3}}\frac{\left(\frac{\beta_\lambda\beta_{\lambda'}}
{\beta_{\lambda\lambda'}^2}\right)^{7/2}}
{\left[1+  \frac{3}{2}\frac{m^2_\sigma}{m^2_{\Omega_q}}
\frac{p^2}{\beta_{\lambda\lambda'}^2}\right]^4}.\nonumber
\end{eqnarray}
\subsection{$5/2^+_1, |320002>$}
\subsubsection{Harmonic Oscillator Form Factors}
\begin{eqnarray}
F_1 &=&I_H\frac{3m_\sigma^2}{\alpha^2_{\lambda}}\left[1+\frac{m_\sigma}{\alpha^2_{\lambda\lambda'}}
\left(\frac{\alpha^2_{\lambda'}}{m_q}+\frac{\alpha^2_{\lambda}}{m_Q}\right)\right],\nonumber\\
F_2 &=&I_H\frac{\ms}{\al}\left\{\frac{3\ms}{\al}\left[1+\frac{\ms}{\all^2}
\left(\frac{\alpha^2_{\lambda'}}{m_q}+\frac{\alpha^2_{\lambda}}{m_Q}\right)\right]-\frac{\al}{4}
\left(\frac{1}{m_q}+\frac{6}{m_Q}\right)\right\},\nonumber\\
F_3 &=& -I_H\frac{\ms}{\al}\left\{\frac{6\ms}{\al}\left[1+\frac{\ms}{\all^2}
\left(\frac{\alpha^2_{\lambda'}}{m_q}+\frac{\alpha^2_{\lambda}}{m_Q}\right)\right]+\frac{\al}{4}
\left(\frac{1}{m_q}-\frac{1}{m_Q}\right)\right\},\nonumber\\
F_4 &=& I_H \frac{\ms}{2}\left(\frac{1}{m_q}-\frac{1}{m_Q}\right),\nonumber\\
G_1 &=&-I_H\frac{\ms}{\al}\left(\frac{6m_\sigma}{\alpha_{\lambda}}-\frac{7\al}{4m_Q}\right),\,\,\,\,\,\,\,\,\,\,\,\,\,\,
G_2 =-I_H\frac{3m_\sigma^3}{\alpha^2_{\lambda\lambda'}\alpha^2_{\lambda}}
\left(\frac{\alpha^2_{\lambda'}}{m_q}+\frac{2\alpha^2_{\lambda}}{m_Q}\right),\nonumber\\
G_3 &=&I_H\frac{\ms}{\al}\left[\frac{9m_\sigma}{\alpha_{\lambda}}+\frac{\al}{m_Q}\left(
\frac{9\ms^2}{\all^2}-\frac{7}{4}\right)\right],\nonumber\\
G_4 &=& -I_H \frac{\ms}{\al}\left(\frac{18m_\sigma}{\alpha_{\lambda}}-\frac{7\al}{2m_Q}\right),\nonumber
\end{eqnarray}
where
\begin{eqnarray}
I_H =\frac{2}{3\sqrt{7}}\left(\frac{\alpha_\lambda\alpha_{\lambda'}}
{\alpha_{\lambda\lambda'}^2}\right)^{7/2}\exp\left( -\frac{3m^2_\sigma}{2m^2_{\Omega_q}}
\frac{p^2}{\alpha_{\lambda\lambda'}^2}\right).\nonumber
\end{eqnarray}
\subsubsection{Sturmian Form Factors}
\begin{eqnarray}
F_1 &=&  I_S\frac{18m_\sigma^2}{\bl^2}\left[1+\frac{m_\sigma}
{\beta_{\lambda\lambda'}}\left(\frac{\beta_{\lambda'}}{m_q}+\frac{\beta_{\lambda}}{m_Q}\right)\right],\nonumber\\
F_2 &=&I_S\frac{m_\sigma}{\bl}\left\{\frac{18\ms}{\bl}\left[1+\frac{\ms}{\bll}
\left(\frac{\beta_{\lambda'}}{m_q}+\frac{\beta_{\lambda}}{m_Q}\right)\right]-\frac{\bll}{2}\left(
\frac{1}{m_q}+\frac{6}{m_Q}\right)\right\},\nonumber\\
F_3 &=& -I_S\frac{m_\sigma}{\bl}\left\{\frac{36\ms}{\bl}\left[1+\frac{\ms}{\bll}
\left(\frac{\beta_{\lambda'}}{m_q}+\frac{\beta_{\lambda}}{m_Q}\right)\right]+\frac{\bll}{2}\left(
\frac{1}{m_q}-\frac{1}{m_Q}\right)\right\},\nonumber\\
F_4 &=& I_S \frac{\ms\bll}{\bl}\left(\frac{1}{m_q}-\frac{1}{m_Q}\right),\nonumber\\
G_1 &=& -I_S\frac{\ms}{\bl}\left(\frac{36\ms}{\bl}-\frac{7\bl}{2m_Q}\right),\,\,\,\,\,\,\,\,\,\,\,\,\,\,
G_2 =-I_S\frac{18m_\sigma^3}{\bl^2\bll}\left(\frac{\beta_{\lambda'}}
{m_q}+\frac{2\beta_{\lambda}}{m_Q}\right),\nonumber\\
G_3 &=& I_S\frac{\ms}{\bl}\left[\frac{54m_\sigma}{\bl}+\frac{\bll}{m_Q}\left(
\frac{54\ms^2}{\bll^2}-\frac{7}{2}\right)\right],\nonumber\\
G_4 &=& -I_S\frac{\ms}{\bl}\left(\frac{108m_\sigma}{\bl}-\frac{7\bll}{m_Q}\right),\nonumber
\end{eqnarray}
where
\begin{eqnarray}
I_S = \frac{\sqrt{2}}{3\sqrt{21}}\frac{\left(\frac{\beta_\lambda\beta_{\lambda'}}
{\beta_{\lambda\lambda'}^2}\right)^{7/2}}{\left[1+  \frac{3}{2}
\frac{m^2_\sigma}{m^2_{\Omega_q}}\frac{p^2}{\beta_{\lambda\lambda'}^2}\right]^4}.\nonumber
\end{eqnarray}
\section{HQET Quark Model Form Factors}
\label{hqetformfactors}
\subsection{$1/2^-,\,\,j=0$}
\label{ffhmsinglet}
\subsubsection{Harmonic Oscillator Form Factors}
\begin{eqnarray}
F_1 &=&I_H \frac{\alpha_{\lambda}}{6}\left(\frac{1}{m_q}-\frac{3}{m_Q}\right),\nonumber\\
F_2 &=& I_H\frac{1}{6m_q\al}\left(\alpha^2_{\lambda}-\frac{12\alpha_{\lambda'}^2m_\sigma^2}{\alpha_{\lambda\lambda'}^2}\right),\nonumber\\
F_3 &=&  I_H\frac{2m_\sigma}{\al} \left(1+\frac{m_\sigma\alpha_{\lambda'}^2}{m_q\alpha_{\lambda\lambda'}^2}\right),\nonumber\\
G_1 &=& I_H \left(\frac{2m_\sigma}{\al}-\frac{\alpha_\lambda}{2m_Q}\right), \nonumber\\
G_2 &=&  -I_H\frac{\alpha_\lambda}{2m_q},\nonumber\\
G_3 &=& -I_H\frac{2m_\sigma}{\al},\nonumber
\end{eqnarray}
where
\begin{eqnarray}
I_H = \frac{1}{\sqrt{3}}
\left(\frac{\alpha_\lambda\alpha_{\lambda'}}{\alpha_{\lambda\lambda'}^2}\right)^{5/2}
\exp\left( -\frac{3 m^2_\sigma}{2m^2_{\Omega_q}}\frac{p^2}{\alpha_{\lambda\lambda'}^2}\right).\nonumber
\end{eqnarray}
\subsubsection{Sturmian Form Factors}
\begin{eqnarray}
F_1 &=& I_S\frac{\beta_{\lambda\lambda'}}{24}\left(\frac{1}{m_q}
-\frac{3}{m_Q}\right),\nonumber\\
F_2 &=& I_S \frac{1}{24m_q}\left(\beta_{\lambda\lambda'}-\frac{24\beta_{\lambda'}m_\sigma^2}
{\beta_\lambda\beta_{\lambda\lambda'}}\right),\nonumber\\
F_3 &=&I_S \frac{m_\sigma}{\beta_\lambda}\left(1+\frac{m_\sigma\beta_{\lambda'}}{m_q\beta_{\lambda\lambda'}}\right),\nonumber\\
G_1 &=& I_S\left(\frac{m_\sigma}{\beta_{\lambda}}-\frac{\beta_{\lambda\lambda'}}{8m_Q}\right),\nonumber\\
G_2 &=& -I_S \frac{\beta_{\lambda\lambda'}}{8m_q},\nonumber\\
G_3 &=& -I_S \frac{m_\sigma}{\beta_\lambda},\nonumber
\end{eqnarray}
where
\begin{eqnarray}
I_S = \frac{2\sqrt{2}}{\sqrt{3}}\frac{\left(\frac{\beta_\lambda\beta_{\lambda'}}
{\beta_{\lambda\lambda'}}\right)^{5/2}}{\left[1+  \frac{3}{2}\frac{m^2_\sigma}{m^2_{\Omega_q}}
\frac{p^2}{\beta_{\lambda\lambda'}^2}\right]^3}.\nonumber
\end{eqnarray}
\subsection{$1/2^-,\,\, j=1$}
\label{ffhmsl1}
\subsubsection{Harmonic Oscillator Form Factors}
\begin{eqnarray}
F_1 &=&  I_H\frac{\alpha_\lambda}{6}\left(\frac{1}{m_q}+\frac{1}{m_Q}\right),\nonumber\\
F_2 &=&I_H\left\{-\frac{m_\sigma}{\al}\left[1-\frac{\ms}{\all^2}\left(\frac{\alp^2}{m_q}-\frac{\al^2}{m_Q}\right)\right]+
\frac{\alpha_\lambda}{6m_q}\right\},\nonumber\\
F_3 &=&-I_H\frac{m_\sigma}{\al}\left[1+\frac{\ms}{\all^2}\left(\frac{\alp^2}{m_q}-\frac{\al^2}{m_Q}\right)\right],\nonumber\\
G_1 &=&I_H\frac{\al}{6m_Q},\nonumber\\
G_2&=&I_H\left[\frac{m_\sigma}{\al}-\frac{\al}{6}\left(\frac{3}{m_q}+\frac{2}{m_Q}\right)\right],\nonumber\\
G_3 &=&- I_H \left(\frac{m_\sigma}{\alpha_\lambda}+\frac{\al}{3m_Q}\right),\nonumber
\end{eqnarray}
where
\begin{eqnarray}
I_H =\sqrt{\frac{2}{3}}\left(\frac{\alpha_\lambda\alpha_{\lambda'}}
{\alpha_{\lambda\lambda'}^2}\right)^{5/2}\exp\left( -\frac{3m^2_\sigma}{2m^2_{\Omega_q}}
\frac{p^2}{\alpha_{\lambda\lambda'}^2}\right).\nonumber
\end{eqnarray}
\subsubsection{Sturmian Form Factors}
\begin{eqnarray}
F_1 &=&I_S\frac{\beta_{\lambda\lambda'}}{6}\left(\frac{1}{m_q}+\frac{1}{m_Q}\right),\nonumber\\
F_2 &=&I_S\left\{-\frac{2m_\sigma}{\beta_\lambda}\left[1-\frac{m_\sigma}{\beta_{\lambda\lambda'}}
\left(\frac{\beta_{\lambda'}}{m_q}-\frac{\beta_\lambda}{m_Q}\right)\right]+\frac{\beta_{\lambda\lambda'}}{6m_q}
\right\},\nonumber\\
F_3 &=& -I_S\frac{2m_\sigma}{\beta_\lambda}\left[1+\frac{m_\sigma}{\beta_{\lambda\lambda'}}
\left(\frac{\beta_{\lambda'}}{m_q}-\frac{\beta_\lambda}{m_Q}\right)\right],\nonumber\\
G_1 &=& I_S\frac{\beta_{\lambda\lambda'}}{6m_Q},\nonumber\\
G_2 &=&I_S\left[\frac{2m_\sigma}{\bl}-\frac{\bll}{6}\left(\frac{3}{m_q}+\frac{2}{m_Q}\right)\right],\nonumber\\
G_3 &=&- I_S\left(\frac{2m_\sigma}{\bl}+\frac{\bll}{3m_Q}\right),\nonumber
\end{eqnarray}
where
\begin{eqnarray}
I_S = \frac{1}{\sqrt{3}}\frac{\left(\frac{\beta_\lambda\beta_{\lambda'}}
{\beta_{\lambda\lambda'}^2}\right)^{5/2}}{\left[1+  \frac{3}{2}\frac{m^2_\sigma}{m^2_{\Omega_q}}
\frac{p^2}{\beta_{\lambda\lambda'}^2}\right]^4}.\nonumber
\end{eqnarray}
\subsection{$3/2^+,\,\, j=1$}
\label{ffthpsl1}
\subsubsection{Harmonic Oscillator Form Factors}
\begin{eqnarray}
F_1 &=&  I_H\frac{\ms}{6}\left(\frac{1}{m_q}+\frac{5}{m_Q}\right),\nonumber\\
F_2 &=&-I_H\frac{\ms}{2m_q}\left(1-\frac{12m_\sigma^2\alpha^2_{\lambda'}}
{\alpha^2_\lambda\alpha^2_{\lambda\lambda'}}
\right),\nonumber\\
F_3 &=& -I_H\frac{\ms}{\al}\left[\frac{6m_\sigma}{\alpha_{\lambda}}\left(1+\frac{m_\sigma\alpha^2_{\lambda'}}
{m_q\alpha^2_{\lambda\lambda'}}\right)+\frac{\alpha_\lambda}{3}\left(\frac{2}{m_q}-\frac{5}{m_Q}\right)\right],\nonumber\\
F_4 &=&I_H\frac{2\ms}{3} \left(\frac{2}{m_q}-\frac{5}{m_Q}\right),\nonumber\\
G_1 &=&-I_H\frac{\ms}{\al}\left(\frac{6m_\sigma}{\alpha_{\lambda}}-\frac{5\alpha_\lambda}{2m_Q}\right),\nonumber\\
G_2 &=&I_H\frac{5\ms}{6m_q},\nonumber\\
G_3 &=& I_H\frac{\ms}{\al} \left(\frac{6m_\sigma}{\alpha_{\lambda}}-\frac{5\alpha_\lambda}{3m_Q}\right),\,\,\,\,\,\,\,\,\,\,\,\,\,\,
G_4 =  0,\nonumber
\end{eqnarray}
where
\begin{eqnarray}
I_H =\sqrt{\frac{1}{10}}\left(\frac{\alpha_\lambda\alpha_{\lambda'}}
{\alpha_{\lambda\lambda'}^2}\right)^{7/2}\exp\left( -\frac{3m^2_\sigma}{2m^2_{\Omega_q}}
\frac{p^2}{\alpha_{\lambda\lambda'}^2}\right).\nonumber
\end{eqnarray}
\subsubsection{Sturmian Form Factors}
\begin{eqnarray}
F_1 &=&  I_S \frac{\beta_{\lambda\lambda'}\ms}{108\bl}\left(\frac{1}{m_q}+\frac{5}{m_Q}\right),\nonumber\\
F_2&=&-I_S\frac{\ms\bll}{36m_q\bl}\left(1-\frac{36\ms^2\blp}{\bl\bll^2}\right),\nonumber\\
F_3 &=& -I_S\frac{\ms}{\bl} \left[\frac{\ms}{\bl}\left(1+
\frac{m_\sigma\beta_{\lambda'}}{m_q\beta_{\lambda\lambda'}}\right)
+\frac{\beta_{\lambda\lambda'}}{54}\left(\frac{2}{m_q}-
\frac{5}{m_Q}\right)\right],\nonumber\\
F_4 &=& I_S\frac{\beta_{\lambda\lambda'}\ms}{27\bl} \left(\frac{2}{m_q}-\frac{5}{m_Q}\right),\nonumber\\
G_1 &=&-I_S\frac{\ms}{\bl}\left(\frac{m_\sigma}{\beta_{\lambda}}-\frac{5\beta_{\lambda\lambda'}}{36m_Q}\right),\nonumber\\
G_2 &=&I_S\frac{5\ms\beta_{\lambda\lambda'}}{108m_q\bl},\nonumber\\
G_3 &=& I_S\frac{\ms}{\bl}\left(\frac{m_\sigma}{\beta_{\lambda}}-\frac{5\beta_{\lambda\lambda'}}{54m_Q}\right),\,\,\,\,\,\,\,\,\,\,\,\,\,\,
G_4 = 0,\nonumber
\end{eqnarray}
where
\begin{eqnarray}
I_S = \frac{6\sqrt{3}}{\sqrt{5}}\frac{\left(\frac{\beta_\lambda\beta_{\lambda'}}
{\beta^2_{\lambda\lambda'}}\right)^{7/2}}{\left[1+  \frac{3}{2}\frac{m^2_\sigma}{m^2_{\Omega_q}}
\frac{p^2}{\beta_{\lambda\lambda'}^2}\right]^4}.\nonumber
\end{eqnarray}
\subsection{$3/2^+,\,\,j=2 $}
\label{ffthpsl2}
\subsubsection{Harmonic Oscillator Form Factors}
\begin{eqnarray}
F_1 &=& I_H \frac{\ms}{4}\left(\frac{1}{m_q}-\frac{3}{m_Q}\right),\nonumber\\
F_2 &=&I_H\left\{\frac{2\ms^2}{\al^2}\left[1+\frac{\ms}{2\all^2}\left(\frac{2\al^2}{m_Q}-\frac{\alp^2}{m_q}\right)\right]
-\frac{3\ms}{4m_q}\right\},\nonumber\\
F_3 &=&I_H\left\{\frac{\ms^2}{\al^2}\left[1+\frac{\ms}{\all^2}\left(\frac{\alp^2}{m_q}-\frac{2\al^2}{m_Q}\right)\right]
+\ms\left(\frac{1}{2m_Q}-\frac{1}{m_q}\right)\right\},\nonumber\\
F_4 &=&I_H\ms  \left(\frac{2}{m_q}-\frac{1}{m_Q}\right),\nonumber\\
G_1 &=&-I_H\ms \left(\frac{m_\sigma}{\alpha^2_{\lambda}}+\frac{1}{4m_Q}\right),\nonumber\\
G_2 &=&I_H\frac{\ms}{\al}\left[-\frac{2m_\sigma}{\alpha_{\lambda}}+\frac{\al}{4}\left(\frac{5}{m_q}+\frac{4}{m_Q}\right)\right],\nonumber\\
G_3 &=& I_H\frac{\ms}{\al} \left(\frac{3m_\sigma}{\alpha_{\lambda}}+\frac{\al}{2m_Q}\right),\,\,\,\,\,\,\,\,\,\,\,\,\,\,
G_4 = 0,\nonumber
\end{eqnarray}
where
\begin{eqnarray}
I_H =\sqrt{\frac{2}{5}}\left(\frac{\alpha_\lambda\alpha_{\lambda'}}
{\alpha_{\lambda\lambda'}^2}\right)^{7/2}\exp\left( -\frac{3m^2_\sigma}{2m^2_{\Omega_q}}
\frac{p^2}{\alpha_{\lambda\lambda'}^2}\right).\nonumber
\end{eqnarray}
\subsubsection{Sturmian Form Factors}
\begin{eqnarray}
F_1 &=& I_S\frac{\ms\bll}{24\bl} \left(\frac{1}{m_q}-\frac{3}{m_Q}\right),\nonumber\\
F_2 &=&I_S\frac{\ms}{\bl}\left\{\frac{\ms}{\bl}\left[1+\frac{\ms}{2\bll}\left(\frac{2\bl}{m_Q}-\frac{\blp}{m_q}\right)
\right]-\frac{\bll}{8m_q}\right\},\nonumber\\
F_3 &=&I_S\frac{\ms}{\bl}\left\{\frac{\ms}{2\bl}\left[1+\frac{\ms}{\bll}\left(\frac{\blp}{m_q}-\frac{2\bl}{m_Q}\right)
\right]+\frac{\bll}{12}\left(\frac{1}{m_Q}-\frac{2}{m_q}\right)\right\},\nonumber\\
F_4 &=& I_S\frac{\ms\bll}{6\bl} \left(\frac{2}{m_q}-\frac{1}{m_Q}\right),\nonumber\\
G_1 &=&-I_S\frac{\ms}{2\bl}\left(\frac{m_\sigma}{\beta_{\lambda}}+\frac{\bll}{12m_Q}\right),\nonumber\\
G_2 &=&I_S\frac{\ms}{\bl}\left[-\frac{\ms}{\bl}+\frac{\bll}{24}\left(\frac{5}{m_q}+\frac{4}{m_Q}\right)\right],\nonumber\\
G_3 &=& I_S\frac{\ms}{2\bl} \left(\frac{3\ms}{\bl}+\frac{\bll}{6m_Q}\right),\,\,\,\,\,\,\,\,\,\,\,\,\,\,
G_4 =0,\nonumber
\end{eqnarray}
where
\begin{eqnarray}
I_S = \frac{4\sqrt{3}}{\sqrt{5}}\frac{\left(\frac{\beta_\lambda\beta_{\lambda'}}
{\beta_{\lambda\lambda'}^2}\right)^{7/2}}{\left[1+  \frac{3}{2}\frac{m^2_\sigma}{m^2_{\Omega_q}}
\frac{p^2}{\beta_{\lambda\lambda'}^2}\right]^4}.\nonumber
\end{eqnarray}
\subsection{$3/2^-, \,\,j=1$}
\label{ffthmsl1}
\subsubsection{Harmonic Oscillator Form Factors}
\begin{eqnarray}
F_1 &=&  -I_H \frac{m_\sigma}{\alpha_\lambda} \left[1+\frac{m_\sigma}{\alpha^2_{\lambda\lambda'}}
\left(\frac{\alpha^2_{\lambda'}}{m_q}+\frac{\alpha^2_{\lambda}}{m_Q}\right)\right],\nonumber\\
F_2 &=& I_H\frac{\al}{m_q}\left(\frac{1}{6}-\frac{m_\sigma^2\alp^2}{\al^2\all^2}\right),\nonumber\\
F_3 &=&  I_H\left\{\frac{2\ms}{\al}\left[1+\frac{\ms}{2\all^2}\left(\frac{\al^2}{m_Q}+\frac{2\alp^2}{m_q}\right)\right]+
\frac{\al}{6}\left(\frac{1}{m_q}-\frac{2}{m_Q}\right)\right\},\nonumber\\
F_4 &=& I_H\frac{\al}{3} \left(\frac{2}{m_Q}-\frac{1}{m_q}\right),\nonumber\\
G_1 &=&I_H\left(\frac{m_\sigma}{\alpha_\lambda}-\frac{\alpha_\lambda}{9m_Q}\right),\,\,\,\,\,\,\,\,\,\,\,\,\,\,
G_2 = 2I_H \left(\frac{\al}{9m_Q}+\frac{\ms^2\alp^2}{2m_q\al\all^2}\right),\nonumber\\
G_3 &=& I_H\left[-\frac{2\ms}{\al}\left(1+\frac{\ms\al^2}{2m_Q\all^2}\right)+\frac{\al}{3m_Q}\right],\nonumber\\
G_4 &=& 2I_H\left(\frac{2\ms}{\al}-\frac{\al}{3m_Q}\right),\nonumber
\end{eqnarray}
where
\begin{eqnarray}
I_H =-\frac{1}{\sqrt{2}}\left(\frac{\alpha_\lambda\alpha_{\lambda'}}
{\alpha_{\lambda\lambda'}^2}\right)^{5/2}\exp\left( -\frac{3m^2_\sigma}{2m^2_{\Omega_q}}
\frac{p^2}{\alpha_{\lambda\lambda'}^2}\right).\nonumber
\end{eqnarray}
\subsubsection{Sturmian Form Factors}
\begin{eqnarray}
F_1 &=&  -I_S \frac{m_\sigma}{\beta_\lambda} \left[1+\frac{m_\sigma}{\beta_
{\lambda\lambda'}}\left(\frac{\beta_{\lambda'}}{m_q}+\frac{\beta_{\lambda}}
{m_Q}\right)\right],\nonumber\\
F_2 &=&I_S\frac{\bll}{m_q}\left(\frac{1}{12}-\frac{m_\sigma^2\blp}{\bl\bll^2}\right),\nonumber\\
F_3&=&I_S\left\{\frac{2m_\sigma}{\beta_\lambda}\left[1+\frac{\ms}{2\bll}\left(\frac{2\blp}{m_q}+
\frac{\bl}{m_Q}\right)\right]+\frac{\beta_{\lambda\lambda'}}{12}\left(\frac{1}{m_q}-\frac{2}{m_Q}\right)\right\},\nonumber\\
F_4 &=& I_S \frac{\beta_{\lambda\lambda'}}{6}\left(\frac{2}{m_Q}-\frac{1}{m_q}\right),\nonumber\\
G_1 &=&I_S\left(\frac{m_\sigma}{\beta_\lambda}-\frac{\beta_{\lambda\lambda'}}{18m_Q}\right),\,\,\,\,\,\,\,\,\,\,\,\,\,\,
G_2 =I_S\left(\frac{m^2_\sigma\blp}{m_q\bl\bll}+\frac{\bll}{9m_Q}\right),\nonumber\\
G_3 &=& I_S\left[\frac{-2\ms}{\bl}+\frac{1}{m_Q}\left(\frac{\bll}{6}-\frac{\ms^2}{\bll}\right)\right],\nonumber\\
G_4 &=&I_S\left(\frac{4\ms}{\bl}-\frac{\bll}{3m_Q}\right),\nonumber
\end{eqnarray}
where
\begin{eqnarray}
I_S = \frac{\left(\frac{\beta_\lambda\beta_{\lambda'}}
{\beta_{\lambda\lambda'}^2}\right)^{5/2}}{\left[1+  \frac{3}{2}\frac{m^2_\sigma}{m^2_{\Omega_q}}
\frac{p^2}{\beta_{\lambda\lambda'}^2}\right]^3}.\nonumber
\end{eqnarray}
\subsection{$3/2^-, \,\, j=2$}
\label{ffthmsl2}
\subsubsection{Harmonic Oscillator Form Factors}
\begin{eqnarray}
F_1 &=&  I_H \frac{m_\sigma}{\alpha_\lambda} \left[1+\frac{m_\sigma}{\alpha^2_{\lambda\lambda'}}
\left(\frac{\alpha^2_{\lambda'}}{m_q}+\frac{\alpha^2_{\lambda}}{m_Q}\right)\right],\nonumber\\
F_2 &=&I_H\left\{\frac{2m_\sigma}{\alpha_\lambda} \left[-2+\frac{m_\sigma}{2\alpha^2_{\lambda\lambda'}}
\left(\frac{\alpha^2_{\lambda'}}{m_q}-\frac{4\alpha^2_{\lambda}}{m_Q}\right)\right]+
\frac{5\al}{6m_q}\right\},\nonumber\\
F_3 &=&I_H\left\{\frac{2m_\sigma}{\alpha_\lambda} \left[-1+\frac{m_\sigma}{2\alpha^2_{\lambda\lambda'}}
\left(\frac{3\al^2}{m_Q}-\frac{2\alp^2}{m_q}\right)\right]+
\frac{5\al}{6m_q}\right\} ,\nonumber\\
F_4 &=& -I_H\frac{5\al}{3m_q},\nonumber\\
G_1 &=&I_H\left(\frac{3m_\sigma}{\alpha_\lambda}+\frac{2\alpha_\lambda}{9m_Q}\right),\nonumber\\
G_2 &=&I_H\left(\frac{2\alpha_\lambda}{9m_Q}-\frac{m_\sigma^2\alpha^2_{\lambda'}}{m_q
\alpha^2_{\lambda\lambda'}\alpha_\lambda}\right),\nonumber\\
G_3 &=& -I_H\frac{2\ms}{\al} \left(1-\frac{m_\sigma\al^2}{2m_Q\all^2}\right),\,\,\,\,\,\,\,\,\,\,\,\,\,\,
G_4 = I_H\frac{4\ms}{\al},\nonumber
\end{eqnarray}
where
\begin{eqnarray}
I_H =\frac{1}{\sqrt{10}}\left(\frac{\alpha_\lambda\alpha_{\lambda'}}
{\alpha_{\lambda\lambda'}^2}\right)^{5/2}\exp\left( -\frac{3m^2_\sigma}{2m^2_{\Omega_q}}
\frac{p^2}{\alpha_{\lambda\lambda'}^2}\right).\nonumber
\end{eqnarray}
\subsubsection{Sturmian Form Factors}
\begin{eqnarray}
F_1 &=&  I_S\frac{m_\sigma}{\beta_\lambda} \left[1+\frac{m_\sigma}{\beta_{\lambda\lambda'}}
\left(\frac{\beta_{\lambda'}}{m_q}+\frac{\beta_{\lambda}}{m_Q}\right)\right],\nonumber\\
F_2 &=&I_S\left\{\frac{2m_\sigma}{\beta_{\lambda}}\left[-2+\frac{\ms}{2\bll}
\left(\frac{\blp}{m_q}-\frac{4\bl}{m_Q}\right)\right]+\frac{5\bll}{12m_q}\right\},\nonumber\\
F_3 &=&I_S\left\{\frac{2m_\sigma}{\beta_{\lambda}}\left[-1+\frac{\ms}{2\bll}
\left(\frac{3\bl}{m_Q}-\frac{2\blp}{m_q}\right)\right]+\frac{5\bll}{12m_q}\right\},\nonumber\\
F_4 &=& -I_S\frac{5\bll}{6m_q},\nonumber\\
G_1 &=&I_S\left(\frac{3m_\sigma}{\beta_\lambda}+\frac{\bll}{9m_Q}\right),\nonumber\\
G_2 &=& I_S\left(\frac{\beta_{\lambda\lambda'}}{9m_Q}-\frac{\ms^2\blp}{m_q\bl\bll}\right),\nonumber\\
G_3 &=& -I_S\frac{2\ms}{\bl}\left(1-\frac{\ms\bl}{2m_Q\bll}\right),\,\,\,\,\,\,\,\,\,\,\,\,\,\,
G_4 = I_S\frac{4\ms}{\bl},\nonumber
\end{eqnarray}
where
\begin{eqnarray}
I_S = \frac{1}{\sqrt{5}}\frac{\left(\frac{\beta_\lambda\beta_{\lambda'}}
{\beta_{\lambda\lambda'}^2}\right)^{5/2}}{\left[1+  \frac{3}{2}\frac{m^2_\sigma}{m^2_{\Omega_q}}
\frac{p^2}{\beta_{\lambda\lambda'}^2}\right]^3}.\nonumber
\end{eqnarray}
\subsection{$5/2^+,\,\,j=2$}
\label{fffhpsl2}
\subsubsection{Harmonic Oscillator Form Factors}
\begin{eqnarray}
F_1 &=& I_H\frac{\ms^2}{\al^2}\left[1+\frac{\ms}{\all^2}\left(\frac{\alp^2}{m_q}+\frac{\al^2}{m_Q}\right)\right],\nonumber\\
F_2 &=&I_H\ms
\left[\frac{1}{6}\left(\frac{1}{m_q}-\frac{2}{m_Q}\right)+\frac{\ms^2\alp^2}{m_q\al^2\all^2}\right],\nonumber\\
F_3 &=& I_H\frac{\ms}{\al}\left\{-\frac{2\ms}{\al}\left[1+\frac{\ms}{2\all^2}\left(\frac{2\alp^2}{m_q}+\frac{\al^2}{m_Q}\right)\right]
+\frac{\al}{6}\left(\frac{1}{m_q}+\frac{1}{m_Q}\right)\right\},\nonumber\\
F_4 &=&-I_H\frac{\ms}{3} \left(\frac{1}{m_q}+\frac{1}{m_Q}\right),\nonumber\\
G_1 &=&-I_H\frac{\ms}{\al}\left(\frac{m_\sigma}{\alpha_{\lambda}}-\frac{\al}{2m_Q}\right),\,\,\,\,\,\,\,\,\,\,\,\,\,\,
G_2 =-I_H\frac{\ms^3}{\al^2\all^2}\left(\frac{\alp^2}{m_q}+\frac{4\al^2}{3m_Q}\right),\nonumber\\
G_3 &=& I_H\frac{\ms}{\al}\left[\frac{2\ms}{\al}\left(1+\frac{7\ms\al^2}{6m_Q\all^2}\right)-\frac{\al}{2m_Q}\right],\nonumber\\
G_4 &=&  I_H\ms \left(-\frac{4\ms}{\al^2}+ \frac{1}{m_Q}\right),\nonumber
\end{eqnarray}
where
\begin{eqnarray}
I_H =\left(\frac{\alpha_\lambda\alpha_{\lambda'}}
{\alpha_{\lambda\lambda'}^2}\right)^{7/2}\exp\left( -\frac{3m^2_\sigma}{2m^2_{\Omega_q}}
\frac{p^2}{\alpha_{\lambda\lambda'}^2}\right).\nonumber
\end{eqnarray}
\subsubsection{Sturmian Form Factors}
\begin{eqnarray}
F_1 &=& I_S\frac{\ms^2}{\bl^2}\left[1+\frac{\ms}{\bll}\left(\frac{\blp}{m_q}+\frac{\bl}{m_Q}\right)\right],\nonumber\\
F_2 &=&I_S\frac{\ms}{\bl}\left[\frac{\bll}{18}\left(\frac{1}{m_q}-\frac{2}{m_Q}\right)+\frac{\ms^2\blp}
{m_q\bl\bll}\right],\nonumber\\
F_3 &=& I_S\frac{\ms}{\bl}\left\{-\frac{2\ms}{\bl}\left[1+\frac{\ms}{2\bll}\left(\frac{2\blp}{m_q}+
\frac{\bl}{m_Q}\right)\right]+\frac{\bll}{18}\left(\frac{1}{m_q}+\frac{1}{m_Q}\right)\right\},\nonumber\\
F_4 &=&-I_S\frac{\ms\bll}{9\bl}  \left(\frac{1}{m_q}+\frac{1}{m_Q}\right),\nonumber\\
G_1 &=&-I_S\frac{\ms}{\bl}\left(\frac{m_\sigma}{\beta_{\lambda}}-\frac{\bll}{6m_Q}\right),\,\,\,\,\,\,\,\,\,\,\,\,\,\,
G_2 =-I_S\frac{\ms^3}{\bl^2\bll}\left(\frac{\blp}{m_q}+\frac{4\bl}{3m_Q}\right),\nonumber\\
G_3 &=& I_S\frac{\ms}{\bl}\left[
\frac{2m_\sigma}{\bl}\left(1+\frac{7m_\sigma\bl}{6m_Q\bll}\right)-
\frac{\bll}{6m_Q}\right],\nonumber\\
G_4 &=& I_S\frac{\ms}{\bl}\left(-\frac{4\ms}{\bl}+\frac{\bll}{3m_Q}\right),\nonumber
\end{eqnarray}
where
\begin{eqnarray}
I_S = \sqrt{6}\frac{\left(\frac{\beta_\lambda\beta_{\lambda'}}
{\beta_{\lambda\lambda'}^2}\right)^{7/2}}
{\left[1+  \frac{3}{2}\frac{m^2_\sigma}{m^2_{\Omega_q}}
\frac{p^2}{\beta_{\lambda\lambda'}^2}\right]^4}.\nonumber
\end{eqnarray}
\subsection{$5/2^+,\,\,j=3$}
\label{fffhpsl3}
\subsubsection{Harmonic Oscillator Form Factors}
\begin{eqnarray}
F_1 &=& -I_H\frac{\ms^2}{\al^2}\left[1+\frac{\ms}{\all^2}\left(\frac{\alp^2}{m_q}+\frac{\al^2}{m_Q}\right)\right] ,\nonumber\\
F_2 &=&I_H\frac{\ms}{\al}\left\{\frac{2m_\sigma}{\alpha_\lambda} \left[3+\frac{m_\sigma}{2\alpha^2_{\lambda\lambda'}}
\left(\frac{6\al^2}{m_Q}-\frac{\alp^2}{m_q}\right)\right]-
\frac{\al}{3}\left(\frac{5}{m_q}+\frac{2}{m_Q}\right)\right\},\nonumber\\
F_3 &=& I_H\frac{\ms}{\al}\left\{\frac{2m_\sigma}{\al} \left[1+\frac{m_\sigma}{2\alpha^2_{\lambda\lambda'}}
\left(\frac{2\alp^2}{m_q}-\frac{5\al^2}{m_Q}\right)\right]-
\frac{\al}{3}\left(\frac{5}{m_q}+\frac{2}{m_Q}\right)\right\},\nonumber\\
F_4 &=& I_H\frac{2\ms}{3} \left(\frac{2}{m_Q}+\frac{5}{m_q}\right),\nonumber\\
G_1 &=&-I_H\frac{5\ms^2}{\al^2},\,\,\,\,\,\,\,\,\,\,\,\,\,\,
G_2 =I_H\frac{\ms^3}{\al^2\all^2}\left(\frac{\alp^2}{m_q}-\frac{8\al^2}{3m_Q}\right),\nonumber\\
G_3 &=& I_H\frac{\ms^2}{\al^2}\left(4+\frac{5\ms\al^2}{3m_Q\all^2}\right),\nonumber\\
G_4 &=& -I_H\frac{8\ms^2}{\al^2},\nonumber
\end{eqnarray}
where
\begin{eqnarray}
I_H =\frac{1}{\sqrt{14}}\left(\frac{\alpha_\lambda\alpha_{\lambda'}}
{\alpha_{\lambda\lambda'}^2}\right)^{7/2}\exp\left( -\frac{3m^2_\sigma}{2m^2_{\Omega_q}}
\frac{p^2}{\alpha_{\lambda\lambda'}^2}\right).\nonumber
\end{eqnarray}
\subsubsection{Sturmian Form Factors}
\begin{eqnarray}
F_1 &=& -I_S\frac{\ms^2}{\bl^2}\left[1+\frac{\ms}{\bll}\left(\frac{\blp}{m_q}+\frac{\bl}{m_Q}\right)\right],\nonumber\\
F_2 &=&I_S\frac{\ms}{\bl}\left\{\frac{2\ms}{\bl}\left[3+\frac{\ms}{2\bll}\left(\frac{6\bl}{m_Q}-
\frac{\blp}{m_q}\right)\right]-\frac{\bll}{9}\left(\frac{5}{m_q}+\frac{2}{m_Q}\right)\right\},\nonumber\\
F_3 &=&I_S\frac{\ms}{\bl}\left\{\frac{2\ms}{\bl}\left[1+\frac{\ms}{2\bll}\left(\frac{2\blp}{m_q}-
\frac{5\bl}{m_Q}\right)\right]-\frac{\bll}{9}\left(\frac{5}{m_q}+\frac{2}{m_Q}\right)\right\} ,\nonumber\\
F_4 &=&I_S\frac{2\ms\bll}{9\bl}  \left(\frac{2}{m_Q}+\frac{5}{m_q}\right),\nonumber\\
G_1 &=&-I_S\frac{5\ms^2}{\bl^2},\,\,\,\,\,\,\,\,\,\,\,\,\,\,
G_2 =I_S\frac{\ms^3}{\bl^2\bll}\left(\frac{\blp}{m_q}-\frac{8\bl}{3m_Q}\right),\nonumber\\
G_3 &=& I_S\frac{\ms^2}{\bl^2}\left(4+\frac{5\ms\bl}{3m_Q\bll}\right),\nonumber\\
G_4 &=& -I_S\frac{8\ms^2}{\bl^2},\nonumber
\end{eqnarray}
where
\begin{eqnarray}
I_S = \sqrt{\frac{3}{7}}\frac{\left(\frac{\beta_\lambda\beta_{\lambda'}}
{\beta_{\lambda\lambda'}^2}\right)^{7/2}}{\left[1+  \frac{3}{2}
\frac{m^2_\sigma}{m^2_{\Omega_q}}\frac{p^2}{\beta_{\lambda\lambda'}^2}\right]^4}.\nonumber
\end{eqnarray}

\newif\ifmultiplepapers
\def\beginpapers{\multiplepaperstrue}
\def\endpapers{\multiplepapersfalse}  
\def\journal#1&#2(#3)#4{\rm #1~{\bf #2}\unskip, \rm  #4 (19#3)}
\def\trjrnl#1&#2(#3)#4{\rm #1~{\bf #2}\unskip, \rm #4 (19#3)}
\def\baps{\journal {Bull.} {Am.} {Phys.} {Soc.}&}
\def\jap{\journal J. {Appl.} {Phys.}&}
\def\prl{\journal {Phys.} {Rev.} {Lett.}&}
\def\pl{\journal {Phys.} {Lett.}&}
\def\pr{\journal {Phys.} {Rev.}&}
\def\np{\journal {Nucl.} {Phys.}&}
\def\rmp{\journal {Rev.} {Mod.} {Phys.}&}
\def\jmp{\journal J. {Math.} {Phys.}&}
\def\rmm{\journal {Revs.} {Mod.} {Math.}&}
\def\jetp{\journal {J.} {Exp.} {Theor.} {Phys.}&}
\def\sjetp{\trjrnl {Sov.} {Phys.} {JETP}&}
\def\dokl{\journal {Dokl.} {Akad.} Nauk USSR&}
\def\spd{\trjrnl {Sov.} {Phys.} {Dokl.}&}
\def\tmf{\journal {Theor.} {Mat.} {Fiz.}&}
\def\snp{\trjrnl {Sov.} J. {Nucl.} {Phys.}&}
\def\hpa{\journal {Helv.} {Phys.} Acta&}
\def\yf{\journal {Yad.} {Fiz.}&}
\def\zp{\journal Z. {Phys.}&}
\def\anp{\journal {Adv.} {Nucl.} {Phys.}&}
\def\ap{\journal {Ann.} {Phys.}&}
\def\am{\journal {Ann.} {Math.}&}
\def\nc{\journal {Nuo.} {Cim.}&}
\def\etal{{\sl et al.}}
\def\pre{\journal {Phys.} {Rep.}&}
\def\pca{\journal Physica (Utrecht)&}
\def\prs{\journal {Proc.} R. {Soc.} London &}
\def\jcp{\journal J. {Comp.} {Phys.}&}
\def\pna{\journal {Proc.} {Nat.} {Acad.}&}
\def\jpg{\journal J. {Phys.} G (Nuclear Physics)&}
\def\fort{\journal {Fortsch.} {Phys.}&}
\def\jfa{\journal {J.} {Func.} {Anal.}&}
\def\cmp{\journal {Comm.} {Math.} {Phys.}&}


\newpage
\begin{thebibliography}{99}
\normalsize
\baselineskip = 0.2 in
\parskip 0pt
\bibitem{MWS} M.Pervin, W. Roberts, S. Capstick,  Phys.\ Rev.\ C {\bf 72}, 035201
(2005).
\bibitem{HQET} {N. Isgur and M. Wise, 
               Phys. Lett. {\bf B232} (1989) 113; 
	       Phys.Lett. {\bf B237} (1990) 527; \\
               B. Grinstein, 
	       Nucl. Phys. {\bf B339} (1990) 253; \\
               H. Georgi, 
	       Phys. Lett. {\bf B240} (1990) 447; \\
	       A. Falk, H. Georgi, B. Grinstein and M. Wise,
		 Nucl. Phys. {\bf B343} (1990) 1; \\
	       A. Falk and B. Grinstein,
		 Phys. Lett. {\bf B247} (1990) 406; \\
	       T. Mannel, W. Roberts and Z. Ryzak, 
	       Nucl. Phys. {\bf B368} (1992) 204.}
\bibitem{omegab1} C.~G.~Boyd and D.~E.~Brahm, Phys.\ Lett.\ B {\bf 254}, 468 (1991).
\bibitem{omegab2} M.~Sutherland,
		  Z.\ Phys.\ C {\bf 63}, 111 (1994).
\bibitem{omegab3}Q.~P.~Xu,
		Phys.\ Rev.\ D {\bf 48}, 5429 (1993).
\bibitem{cleooc} K.~K.~Gan  [CLEO Collaboration], arXiv:hep-ex/0202012.
\bibitem{argus} H.~Albrecht {\it et al.}  [ARGUS Collaboration], 
		 Phys.\ Lett.\ B {\bf 277}, 209 (1992).
\bibitem{belle} K.~Abe et al., Lepton-Photon 2003, BELLE-CONF-0333 (2003).

\bibitem{Falk} A. F. Falk, Nucl.\ Phys.\ B {\bf 378}, 79 (1992).
\bibitem{KP} B.~D.~Keister and W.~N.~Polyzou,
		J.\ Comput.\ Phys.\  {\bf 134}, 231 (1997).
\bibitem{godfrey} T.~Barnes, S.~Godfrey and E.~S.~Swanson,
                  Phys.\ Rev.\ D {\bf 72}, 054026 (2005)
\bibitem{ISGW} N. Isgur, D. Scora, B. Grinstein and M. Wise, 
		Phys. Rev. D {\bf 39},799 (1989).
\bibitem{ISGW1} N. Isgur and D. Scora, Phys. Rev. D {\bf 52}, 2783 (1989).
\bibitem{Hayne:1981zy}
  C.~Hayne and N.~Isgur,
  Phys.\ Rev.\ D {\bf 25}, 1944 (1982).
\bibitem{Foster:1983kn}
  F.~Foster and G.~Hughes,
  Rept.\ Prog.\ Phys.\  {\bf 46}, 1445 (1983).
\bibitem{CLEO} G.~D.~Crawford {\it et al.}  [CLEO Collaboration],
		 Phys.\ Rev.\ Lett.\  {\bf 75}, 624 (1995).
\bibitem{delphi} J.~Abdallah {\it et al.}  [DELPHI Collaboration],
Phys.\ Lett.\ B {\bf 585}, 63 (2004).
\bibitem{pdgxi} See article by C.G.~Wohl (pp.~977) in S.~Eidelman {\it
et al.}  [Particle Data Group],
Phys.\ Lett.\ B {\bf 592}, 1 (2004).
\end{thebibliography}
\end{document}